\documentclass[amsfonts,amsmath,prd,preprint,nofootinbib]{revtex4}

\usepackage{amssymb,amsmath,amsfonts,amsbsy}
\usepackage{color}
\usepackage{enumerate}
\usepackage{graphicx}
\usepackage{pifont}
\usepackage{import}
\usepackage{float}
\usepackage{physics}
\usepackage{subcaption}

\usepackage{cancel}

\RequirePackage[colorlinks,citecolor=red,urlcolor=red,linkcolor=red]{hyperref}

\usepackage{bm}

\usepackage[normalem]{ulem}

\begin{document}

\title{Orbital dynamics and spin-precession around a circular chiral vorton}

\author{S.~M.~Holme}
\author{H.~S.~Ramadhan\footnote{Corresponding author: hramad@sci.ui.ac.id}}
\affiliation{\textit{Departemen Fisika, FMIPA, Universitas Indonesia, Depok, 16424, Indonesia}}
\author{I.~Nurul Huda}
\affiliation{School of Astronomy and Space Science,\\ Key Laboratory of Modern Astronomy and Astrophysics (Ministry of Education), Nanjing University, Nanjing, People's Republic of China}
\affiliation{Research Center for Computing, National Research and Innovation Agency, Bogor, Indonesia.}
\author{Leonardus~B.~Putra}
\affiliation{Mathematical Institute, University of Oxford, Radcliffe Observatory, Andrew Wiles Building, Woodstock Rd, Oxford OX2 6GG}

\begin{abstract}
Vortons are of interest in high-energy physics as possible dark matter candidates and as probes of Grand Unified Theories. Using the recently derived weak-field metric for a chiral vorton, we study the dynamics of test particles by analyzing both timelike and null geodesics. We identify several classes of trajectories, including bound precessing orbits, circular orbits, toroidal and crown-type oscillations, as well as unbound scattering paths. Poincaré surfaces of section reveal transitions between regular and chaotic motion that depend sensitively on the vorton tension $G\mu$ and initial conditions. We further compute the Lense–Thirring and general spin-precession frequencies for gyroscopes along Killing trajectories. The resulting precession profiles exhibit several distinct features not present in Kerr black holes but reminiscent of Kerr naked singularities, such as: divergences near the ring core and multi-minima structures. These dynamical and precessional signatures may offer potential observational pathways for detecting vortons.
\end{abstract}

\maketitle

\section{Introduction} \label{sec:introduction}

The scientific fascination with ring-like celestial structures can be traced back more than four centuries, beginning with Galileo’s pioneering observation of Saturn’s distinctive rings~\cite{miner, tiscareno}. In the nineteenth century, Gauss established that the gravitational potential generated by a circular ring can be expressed in terms of elliptic integrals~\cite{Kellog}. Later, Maxwell rigorously demonstrated that Saturn’s rings could not remain stable if composed of a uniform solid or fluid body~\cite{maxwell}.  
With the advent of General Relativity (GR), the gravitational properties of ring-like configurations become richer and more complex. The Bach–Weyl solution, for instance, represents the relativistic analogue of a thin circular ring in vacuum~\cite{Bach:1922lav}. This metric provides an exact singular solution describing a ring-like gravitational source within Weyl’s class of static spacetimes~\cite{Hons, Semerak:2016dqd}. Ring singularities appear in the Kerr solution at $r=0$ and $\theta=\pi/2$~\cite{Kerr:1963ud}. By generalizing Weyl metric in higher-dimensional gravity~\cite{Emparan:2001wk}, Emparan and Reall discovered black ring solutions: asymptotically-flat vacuum solutions with $S^1\times S^2$ event-horizon topology~\cite{Emparan:2001wn}.

A viable, though still hypothetical, cosmological object with $S^1$ topology is a cosmic string loop~\cite{Vilenkin:1984ib, CSBookVilenkin:2000jqa}. Cosmic strings are one-dimensional topological defects that may have formed during symmetry-breaking phase transitions in the early universe when the associated vacuum manifold is non simply-connected. The large tension along the string causes a closed loop to behave as an oscillating relativistic ring, which can serve as an efficient source of gravitational radiation. To prevent the loop from collapsing, external mechanism(s) must be introduced~\cite{Hughes_PhysRevD.47.468}. The possibility that a cosmic string can be superconducting was first suggested by Witten~\cite{Witten:1984eb}. When such a current circulates along the loop, the resulting angular momentum can counterbalance the string tension, allowing for a stationary equilibrium configuration known as a {\it vorton}~\cite{DAVISShellardVorton1989209}. Due to their long-term stability, vortons have been proposed as viable dark matter candidates~\cite{Auclair_2021_vortondarkmatter}.  

The dynamics around ring-like mass distributions have long been studied in both Newtonian and relativistic frameworks. In the Newtonian regime, the orbital motion and stability of test particles around a rigid circular ring have been analyzed extensively by Elipe and collaborators~\cite{elipe}, who identified equilibrium points, bounded orbits, and regions of instability depending on the ring’s mass distribution and the particle’s energy. In general relativity, the spacetime generated by a ring-like source influences the motion of both massive and massless particles. The first investigation of photon trajectories and gravitational lensing by a cosmic string loop was carried out by De Laix and Vachaspati~\cite{de_Laix_1996_lensingcosmicstringloop}. A vorton provides a natural model of a stationary, non-oscillating ring. Its weak-field gravitational potential and corresponding lensing signatures were recently studied by two of the present authors~\cite{putra2024gravitationalfieldlensingcircular}. To the best of our knowledge, no detailed analysis has yet been published on the trajectories or orbital dynamics of massive test particles around a cosmic string loop or vorton\footnote{Timelike geodesics around a straight (infinitely long) cosmic string have been investigated, for example, in~\cite{Hartmann_2010,hartmann2011detection,Hartmann_2012}.}

In addition to particle and photon trajectories, another key probe of spacetime geometry is the precession of gyroscopes~\cite{Schiff:1960gi}. Gyroscopic motion provides a sensitive diagnostic of local curvature and frame-dragging effects, allowing one to distinguish between different gravitational sources even when their external fields appear similar~\cite{ Chakraborty:2016ipk, Chakraborty_2017_distinguishingKerrandNaked, Karmakar:2017lho}. In general relativity, two principal forms of gyroscopic precession are recognized: geodetic (or de Sitter) precession, which arises from the curvature produced by mass-energy~\cite{deSitter:1916zz}, and Lense–Thirring (LT) precession, originating from rotational frame-dragging~\cite{Lense:1918zz}. While such effects have been extensively analyzed around compact objects such as black holes and neutron stars, the corresponding analysis for extended ring-like configurations remains unexplored. The closed geometry and internal current of a vorton can induce distinctive frame-dragging patterns along and around the ring.

The objectives of this work are twofold. First, we derive the timelike geodesic equations from the vorton metric presented in~\cite{putra2024gravitationalfieldlensingcircular} and solve for the trajectories of massive test particles. Second, following the methodology of~\cite{Chakraborty_2017_distinguishingKerrandNaked}, we compute the precession frequencies associated with both geodetic and Lense–Thirring effects in the vorton spacetime. Due to the circular topology and intrinsic angular momentum of the vorton, these frequencies are expected to exhibit unique signatures, potentially differing from those of spherically symmetry or other axially rotating objects. Identifying such distinctive features is a key goal of this study, as they could provide indirect observational evidence for the existence of vortons through the orbital behavior and precessional dynamics of nearby stars or other celestial bodies.

\section{Vorton Geodesics}

The most general vorton ansatz that incorporates angular momentum along the string and circular topology of the loop with is an axially symmetric metric possessing two Killing vectors $\partial_t$ and $\partial_\phi$ \cite{Mc_Manus_1993}
\begin{equation}
ds^2=-e^{2\nu}dt^2+e^{-2\nu}r^2(d\phi-Adt)^2+e^{-2\nu}(dr^2+dz^2).\label{metricansatzcyllindrical}
\end{equation}
In~~\cite{putra2024gravitationalfieldlensingcircular} we obtain the solutions in the order of $\mathcal{O}\left(G\mu\right)$ as
\begin{equation} 
\label{eq:numetric}
    \nu(r,z) = -4\sqrt{2}G\mu \sqrt{\frac{R}{r}\sinh{\left[ \frac{1}{2} \ln{\left( \frac{(r+R)^2+z^2}{(r-R)^2+z^2} \right)} \right]}}\frac{K\left(\tanh{\left[ \frac{1}{4} \ln{\left( \frac{(r+R)^2+z^2}{(r-R)^2+z^2} \right)} \right]} \right)}{\cosh{\left[ \frac{1}{4} \ln{\left( \frac{(r+R)^2+z^2}{(r-R)^2+z^2} \right)} \right]}},
\end{equation}
and
\begin{eqnarray}\label{eq:Ametric}
    A(r,z) &=& \frac{16G\mu R}{r\sqrt{(r+R)^2+z^2}}\biggr[ \left(\frac{(r+R)^2+z^2}{2rR}-1 \right) K\left( \frac{4rR}{(r+R)^2+z^2} \right)\nonumber\\
    &&-\frac{(r+R)^2+z^2}{2rR} E\left( \frac{4rR}{(r+R)^2+z^2} \right) \biggr],
\end{eqnarray}
with 
\begin{equation}
K(a)\equiv\int_{0}^{\pi/2} \frac{d\theta}{\sqrt{1-a\sin^2{\theta}}},\ \ \ \ \     E(a)\equiv\int_{0}^{\pi/2} \sqrt{1-a\sin^2{\theta}}d\theta,
\end{equation}
the complete elliptical integral of the first and second kind, respectively. Both $\nu(r,z)$ and $A(r,z)$ functions diverge near the string core at $r=1$, $z=0$, as expected from the delta-function source associated with the thin-string approximation. This divergence can be regularized by adopting a more realistic field-theoretic treatment that includes the finite core radius $\delta$. 

The corresponding geodesic equations,
\begin{equation}
    \ddot{x}^\mu + \Gamma^\mu_{\alpha\beta}\dot{x}^\alpha \dot{x}^\beta = 0, 
\end{equation}
are
\begin{eqnarray}
    \ddot{t} + e^{-4\nu}\Big[\dot{z} \left(r^2 \dot{\phi}A_{,z} + 2 e^{4\nu} \dot{t} \nu_{,z} \right) - r^2 A \dot{t}\left(\dot{z} A_{,z} + \dot{r} A_{,r} \right)
     + \dot{r} \left(r^2 \dot{\phi} A_{,r} + 2 e^{4\nu} \dot{t} \nu_{,r} \right) \Big] &=& 0,\nonumber\\ 
     \nonumber\\
     \ddot{r} + \dot{z}^2 \nu_{,r} + r\dot{\phi}^2 \left( r\nu_{,r} - 1\right) + r\dot{t}\dot{\phi}\left[ r A_{,r} + A\left(2-2r \nu_{,r}  \right) \right]\nonumber\\
     +\dot{t}\left[e^{4\nu} \nu_{,r} - r^2 A A_{,r} + rA^2\left(r\nu_{,r} - 1   \right) \right] - \dot{r}\left(2\dot{z}\nu_{,z} + \dot{r} \nu_{,r} \right)  &=& 0,\nonumber\\ 
     \nonumber\\
     \ddot{\phi} + \dot{z}\dot{\phi} \left(e^{-4\nu} r^2 A A_{,z} - 2\nu_{,z}  \right) 
     + \dot{t}\dot{z}\left[2A_{,z} + 4A \nu_{,z} - \left(1 + e^{-4\nu} r^2 A^2   \right)    \right] \nonumber\\
     +\frac{\dot{r}\dot{\phi}}{r}\left(2+ e^{-4\nu}r^3A A_{,r} -2r \nu_{,r}  \right)
     -\frac{\dot{r}\dot{t}}{r}\left[ r A_{,r} + e^{-4\nu}r^3 A^2 A_{,r} + 2A\left(1-2r\nu_{,r} \right)\right]  &=&0,\nonumber\\
     \nonumber\\
     \ddot{z} + \dot{r}^2\nu_{,z} + r^2\dot{\phi}^2 \nu_{,z} + r^2 \dot{t}\dot{\phi} \left(A_{,z} - 2A\nu_{,z} \right) - \dot{z} \left(\dot{z} \nu_{,z} + 2\dot{r} \nu_{,r} \right)& \notag\\ 
+\dot{t}^2\left(e^{4\nu}\nu_{,z} + r^2 A^2 \nu_{,z}-r^2AA_{,z}   \right)  &=&0,\label{geodesicz}
\end{eqnarray}
where the Affine parameter is taken to be the proper time $\tau$. These equations are exact. However, since the metric is valid only in the weak-field regime, the geodesic equations can be consistently expanded up to first order in $\mathcal{O}(G\mu)$. Using the invariant line element (or geodesic Lagrangian),
\begin{equation}
\label{invariancemetricform}
\mathcal{L}=g_{\mu\nu}\dot{x}^\mu\dot{x}^\nu = \epsilon,
\end{equation}
where $\epsilon=-1$ for timelike and $\epsilon=0$ for null geodesics, we can eliminate $\dot{t}$ and express the $\phi$-component in first-order form as
\begin{eqnarray}
\dot{t} &=& \frac{ g_{t\phi}L + g_{\phi\phi} E }{g_{t\phi}^2-g_{\phi\phi}g_{tt}}
= e^{-2\nu} \left(E - LA\right), \label{tdotfull}\nonumber\\
\dot{\phi} &=& \frac{g_{t\phi}E + g_{tt}L}{ g_{\phi\phi}g_{tt} - g_{t\phi}^2}
=\frac{L}{r^2}e^{2\nu} - e^{-2\nu}\left(EA + Lr^2 A^2 \right). \label{phiLfull}
\end{eqnarray}
Substituting these expressions into the full geodesic equations and retaining only terms up to $\mathcal{O}(G\mu)$ yields the reduced system, 
\begin{eqnarray}
    \ddot{r}  -L A_{,r}\Sigma -\frac{L^2}{r^3}\left(1+4\nu \right)-\frac{2}{r}LA \left(E +\Sigma\right)&\notag\\
    +\nu_{,r}\left(\frac{2}{r^2}L^2-\epsilon+\dot{r}^2+2\dot{z}^2\right)-2\dot{r}\dot{z}\nu_{,z}&=& 0, \label{geodesicfinalr}\\
    \dot{\phi}- \frac{L}{r^2}(1+2\nu) -EA &=& 0, \label{geodesicfinalphi}\\
    \ddot{z}-LA_{,z}\Sigma  +\nu_{,z}\left(\frac{2}{r^2}L^2-\epsilon+2\dot{r}^2\right) -2\dot{r}\dot{z}\nu_{,r} &=&0,\label{geodesicfinalz}
\end{eqnarray}
where $\Sigma \equiv \sqrt{\dot{r}^2 + L^2/r^2 + \dot{z}^2 - \epsilon}$.

For purely radial motion, $L=0$, Eq.~\eqref{geodesicfinalphi} simplifies to
\begin{equation}
\dot{\phi} = EA.
\label{nonzeroL}
\end{equation}
Thus, a (null or timelike) test particle with nonzero energy experiences rotational frame dragging due to the vorton’s angular momentum. When the initial angular velocity vanishes, $\dot{\phi}=0$, the corresponding angular momentum is
\begin{equation}
L_0 = -\frac{EA r^2}{(1 + 2\nu)},
\end{equation}
defining the condition for a {\it Zero Angular Momentum Observer} (ZAMO)~\cite{Bardeen:1972fi, Frolov:2014dta, Costa:2025dzf}, i.e. a particle co-rotating with the local inertial frame induced by the vorton’s rotation.

Studying the dynamics around a vorton requires solving Eqs.~\eqref{geodesicfinalr}–\eqref{geodesicfinalz}, which generally admit no closed-form analytical solutions and must therefore be treated numerically. To obtain meaningful trajectories, one must first identify the regions where bound orbits may exist and determine suitable initial conditions. This preliminary analysis can be performed by constructing the effective potential $V_{\text{eff}}$. From Eq.~\eqref{invariancemetricform} we can express
\begin{eqnarray} 
    \dot{r}^2 + \dot{z}^2 &=& -\frac{1}{g_{rr}}\left(g_{tt} \dot{t}^2 + 2g_{t\phi}\dot{t}\dot{\phi}+g_{\phi\phi}\dot{\phi}^2 - \epsilon\right)\nonumber\\
&=&\left(1 - e^{-4\nu}r^2A^2\right)\left(E - LA\right)^2 +2r^2A \dot{t}\dot{\phi} \notag\\
    & &- r^2\left[ \frac{L}{r^2}e^{2\nu} +   e^{-2\nu}\left(EA - Lr^2 A^2 \right)\right]^2 + e^{2\nu}\epsilon.  \label{boundcond}  
\end{eqnarray}
Taking, for simplicity, the planar condition ($z=0$), the effective potential up to $\mathcal{O}(G\mu)$ is 
\begin{equation}
    \dot{r}^2 = \mathcal{E}-V_{\text{eff}}(r), \label{Orbitalequation}
\end{equation}
where $\mathcal{E} \equiv (E^2 + \epsilon)$, and
\begin{equation}
    V_{\text{eff}}(r) = \frac{L^2}{r^2}(1+4\nu)+2LEA - 2\nu\epsilon \label{effpotent}.
\end{equation}
Eq.~\eqref{effpotent} serves as a diagnostic tool to guide initial condition selection and to study the influence of parameters $L,\, E$, and $G\mu$. The first term corresponds to the centrifugal barrier, while the second term, $2LEA$, represents a Coriolis-like contribution arising from the rotational frame-dragging induced by the vorton’s angular momentum potential $A$. 
\begin{figure}[h] 
\centering
    \begin{subfigure}[t]{0.32\textwidth}
        \centering
        \includegraphics[width=\textwidth]{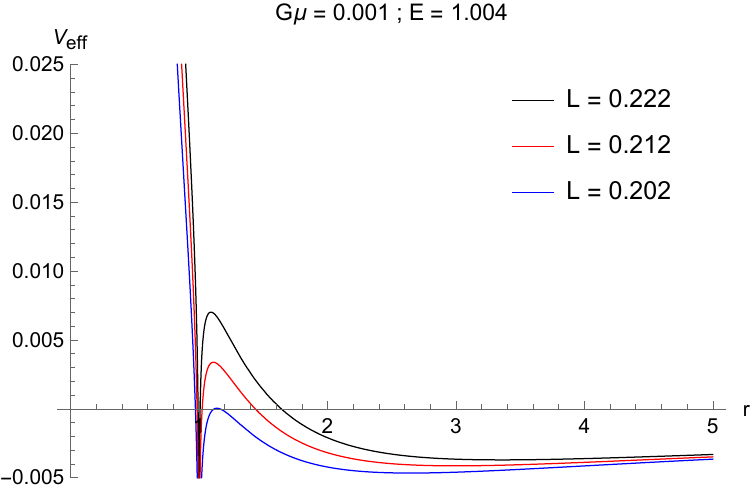}
        \caption{}
        \label{fig:veffgenerallyclose}
    \end{subfigure}
    \hfill
    \begin{subfigure}[t]{0.32\textwidth}
        \centering
        \includegraphics[width=\textwidth]{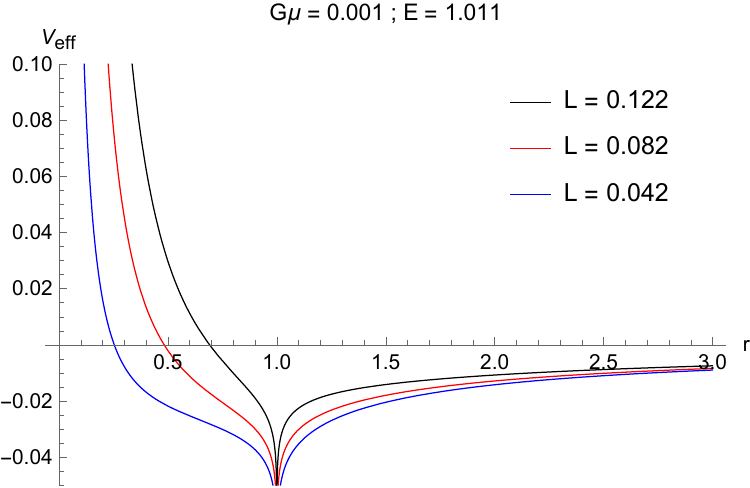}
        \caption{}
        \label{fig:Veffnearvorton}
    \end{subfigure}
    \hfill
    \begin{subfigure}[t]{0.32\textwidth}
        \centering
        \includegraphics[width=\textwidth]{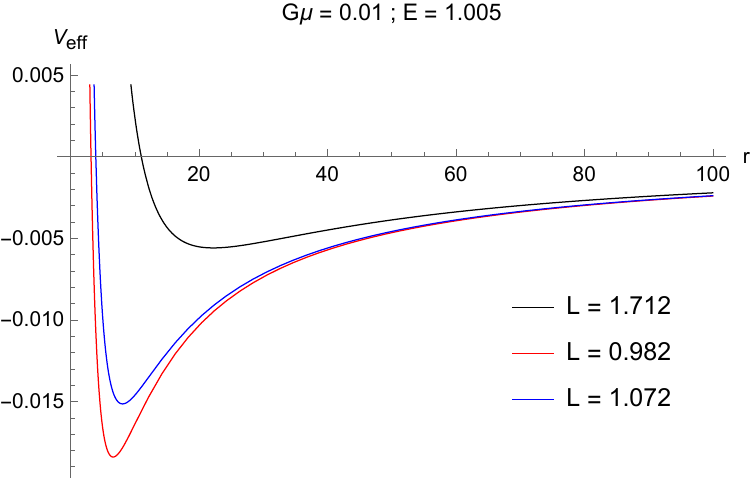}
        \caption{}
        \label{fig:Vefffaraway}
    \end{subfigure}
\caption{Radial profile of $V_{\text{eff}}(r)$ for several values of the angular momentum $L$.}
\label{fig:Veffplot}
\end{figure}

The behaviour of $V_{\text{eff}}(r)$ is illustrated in Fig.~\ref{fig:Veffplot}. The potential diverges at both $r=0$ and $r=R$. The divergence at $r=0$, although the vorton itself is absent there, originates from the centrifugal repulsion that prevents test particles with nonzero angular momentum from approaching the coordinate center. The singularity at $r=R$, on the other hand, is a consequence of the thin-string approximation which can be smoothed out in a full field-theoretic treatment. 

Circular orbits around the vorton can be determined by examining the stationary points of the effective potential,
\begin{equation}
\left.\frac{\partial V_{\text{eff}}(r)}{\partial r}\right|_{r = r_c} = 0,
\end{equation}
where $r_c$ is the radius of the circular orbit. Substituting Eq.~\eqref{effpotent} yields
\begin{equation}
L\left(\frac{2L}{r^2}\nu_{,r} + E A_{,r} - \frac{L}{r^3}(1 + 4\nu)\right) - \epsilon \nu_{,r} = 0, \label{circularorbitcondition}
\end{equation}
evaluated at $r=r_c$. The stability of these orbits can be examined by analyzing the second derivative of the effective potential. The marginally stable (innermost stable circular) orbit, $r_{ISCO}$, satisfies
\begin{equation}
\left.\frac{\partial^{2} V_{\text{eff}}(r)}{\partial r^{2}}\right|_{r = r_{c}} = 0,
\end{equation}
which, upon substitution, gives
\begin{equation}
L\left(\frac{3L}{r^{4}}(1 + 4\nu) + E A_{,rr} + \frac{2L}{r^{2}}\nu_{,rr} - \frac{8L}{r^{3}}\nu_{,r}\right) - \epsilon \nu_{,rr} = 0,
\end{equation}
which is again, evaluated at $r=r_c$.

\begin{figure}[h] 
\centering
        \includegraphics[width=0.5\textwidth]{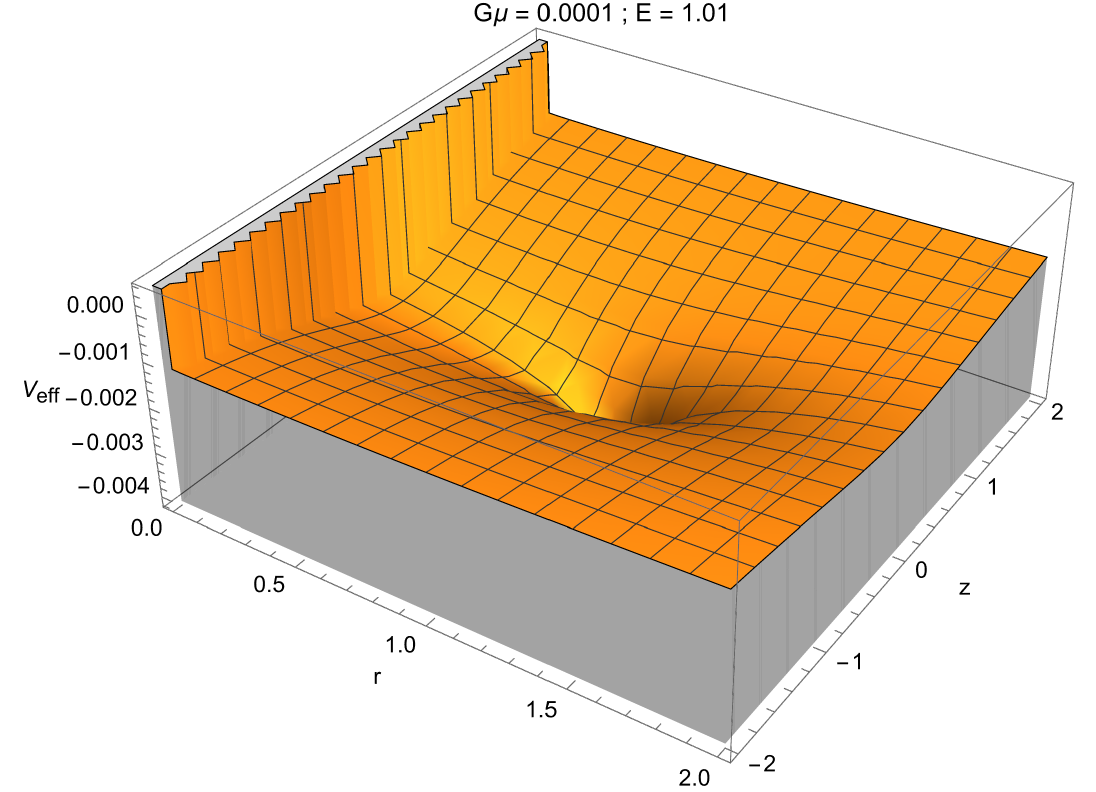}
\caption{The landscape of effective potential surface $V_{\text{eff}}(r,z)$, showing the bound region for test-particle motion for $G\mu = 0.0001$ and $L=0.0001$.}
\label{fig:Vplotznonzero}
\end{figure}

The inclusion of $z$-dependence breaks the planar symmetry and allows the trajectories to be off-equatorial. The $V_{\text{eff}}(r,z)$ landscape, as shown in Fig.~\ref{fig:Vplotznonzero}, reveals regions of confinement that suggest the existence of toroidal or quasi-periodic orbits near the vorton core. For moderate values of $L$, $V_{\text{eff}}(r,z)$ admits bounded trajectories whose radial turning points depend sensitively on $G\mu$ and the vorton radius $R$. Increasing $L$ forces the toroidal orbits to be tighter around the vorton ring, while a too large $L$ eliminates the bound states completely, resulting in scattering trajectories. At large distance the influence of the singularity at $r=R$ diminishes, and $V_{\text{eff}}(r,z)$ asymptotically approaches the point-mass potential.

\section{Numerical Results}

We numerically integrate Eqs.~\eqref{geodesicfinalr}-\eqref{geodesicfinalz} using Mathematica. In addition to the initial conditions ($r_0, \dot{r}_0, z_0, \dot{z}_0$ and $\phi_0$), the parameters that can be varied are $L$, $E$ and $G\mu$. We solve the trajectories around vorton for both timelike ($\epsilon=-1$) and null ($\epsilon=0$) test particles.

\subsection{Timelike Geodesics} \label{timelikegeodesics}

\subsubsection{Equatorial Orbits}

A prominent feature of the spacetime surrounding a vorton, as with any stationary and rotating mass distribution, is the frame-dragging effect, analogous to the Lense–Thirring effect, in which the local inertial frames are dragged by the circulating current (counterclockwise in this case). A test particle with zero initial angular momentum acquire a nonzero angular velocity around the vorton.  
\begin{figure}[h]  
    \begin{subfigure}[b]{0.4\textwidth}
        \centering
        \includegraphics[width=\textwidth]{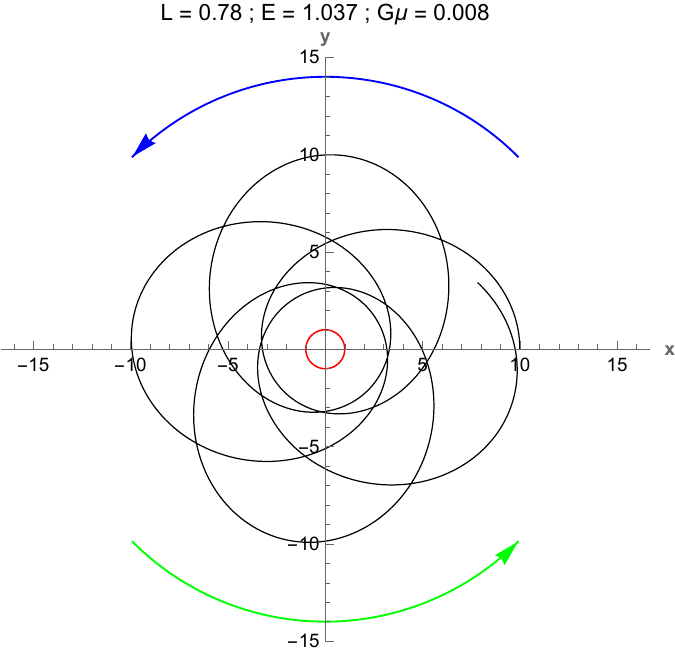}
        \caption{}
        \label{fig:framedraggingdemostratea}
    \end{subfigure}
    \begin{subfigure}[b]{0.4\textwidth}
        \centering
        \includegraphics[width=\textwidth]{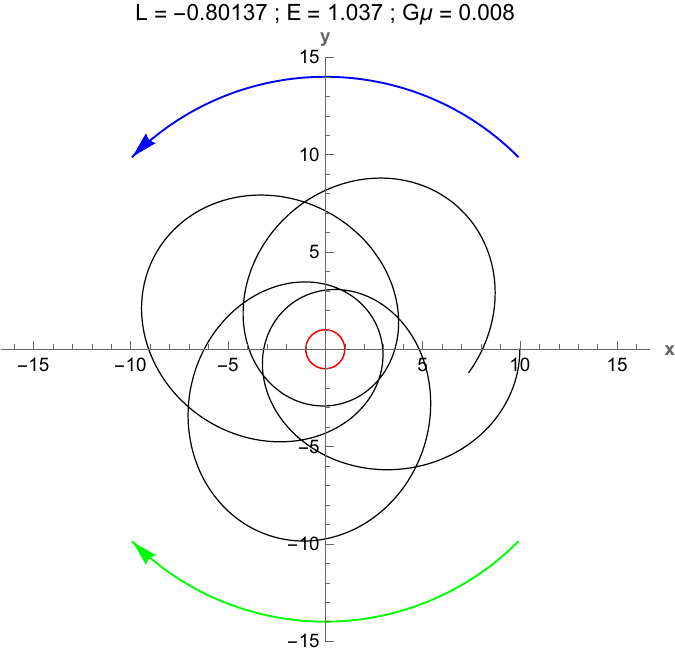}
        \caption{}
        \label{fig:framedraggingdemostrateb}
    \end{subfigure}
\caption{Precessing timelike orbits around a chiral vorton (red) with counter-clockwise frame dragging (blue arrow): (a) prograde and (b) retrograde initial angular velocities.}
\label{fig:framedraggingdemonstrate}
\end{figure}
In Fig.~\ref{fig:framedraggingdemonstrate}, two particles are shown orbiting a chiral vorton with $G\mu = 0.008$. The difference lies in their initial angular velocities $u_{\phi 0}$: one begins with a prograde (counterclockwise) motion, while the other starts in a retrograde (clockwise) direction. The prograde motion corresponds to the direction of the vorton’s current, whose local inertial frame angular velocity is given by
\begin{equation}
\Omega(r, z) = A(r, z).
\end{equation}
For the prograde orbit, we set $L_{po} = 0.78$, resulting in $u_{\phi 0} = 7.74 \times 10^{-3} \text{rad/s}$. To construct a retrograde orbit with the same magnitude of initial angular velocity, we use Eq.~\eqref{geodesicfinalphi} to determine the corresponding angular momentum $L_{ro}$
\begin{equation}
    u_{\phi0}= \frac{L_{ro}}{r_0^2}(1+2\nu(r0,z0)) -EA(r0,z0).  
\end{equation}
This yields $L_{ro} = -0.80137$, as shown in Fig.~\ref{fig:framedraggingdemostrateb}. 
As can be seen in Fig.~\ref{fig:framedraggingdemonstrate}, the prograde orbit exhibits a smaller periastron shift compared to the retrograde orbit. This occurs because the retrograde trajectory reaches a smaller minimum radius $r_-$ (see Table~\ref{tab:closestapproachfigdemonstrateframedrag}), causing the particle to experience a stronger gravitational “slingshot” effect. 

\begin{table}[H]
    \centering
    \begin{tabular}{|c|c|c|}
\hline
    \textbf{Orbit} & \( L\) & \( r_- \) \\ \hline
    Prograde & 0.78000 & 3.03985  \\
     Retrograde & -0.80137 & 2.90942 \\
\hline
\end{tabular}
    \caption{Closest approach $r_-$ for Fig.~\ref{fig:framedraggingdemonstrate}}
    \label{tab:closestapproachfigdemonstrateframedrag}
\end{table}
\begin{figure}[H] 
\centering
    \begin{subfigure}[t]{0.32\textwidth}
        \centering
    \includegraphics[width=\textwidth]{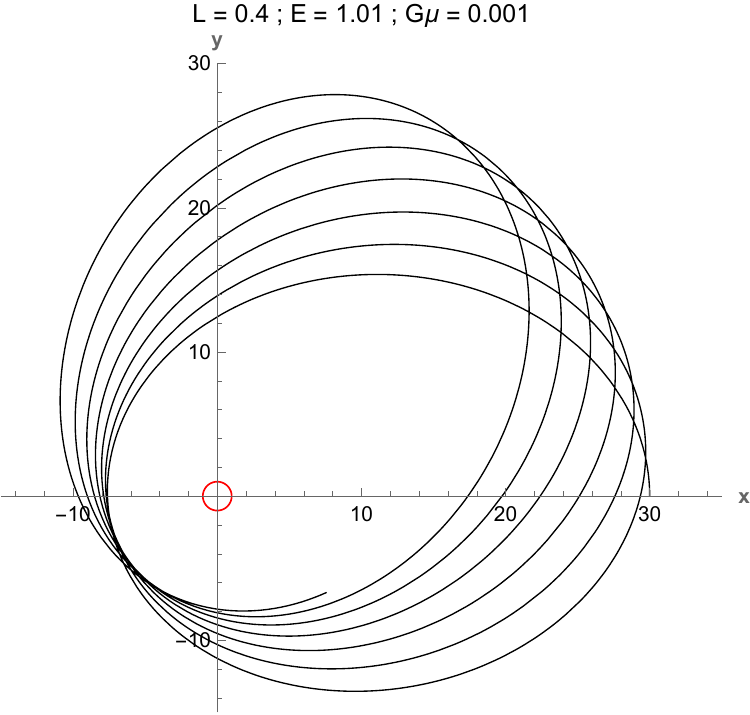}
        \caption{}
        \label{fig:precessingellipsea}
    \end{subfigure}
    \hfill
    \begin{subfigure}[t]{0.32\textwidth}
        \centering
    \includegraphics[width=\textwidth]{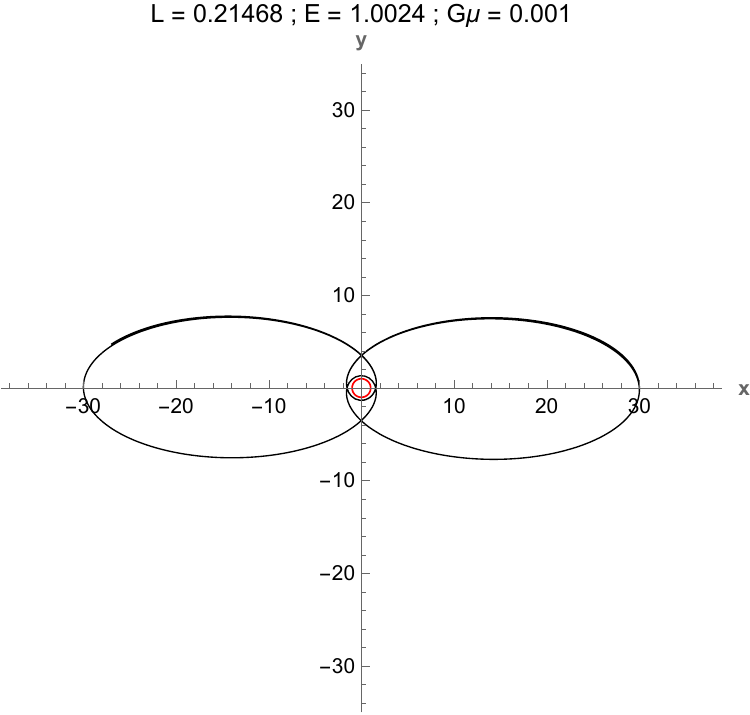}
        \caption{}
        \label{fig:precessingellipseb}
    \end{subfigure}
    \hfill
    \begin{subfigure}[t]{0.32\textwidth}
        \centering
    \includegraphics[width=\textwidth]{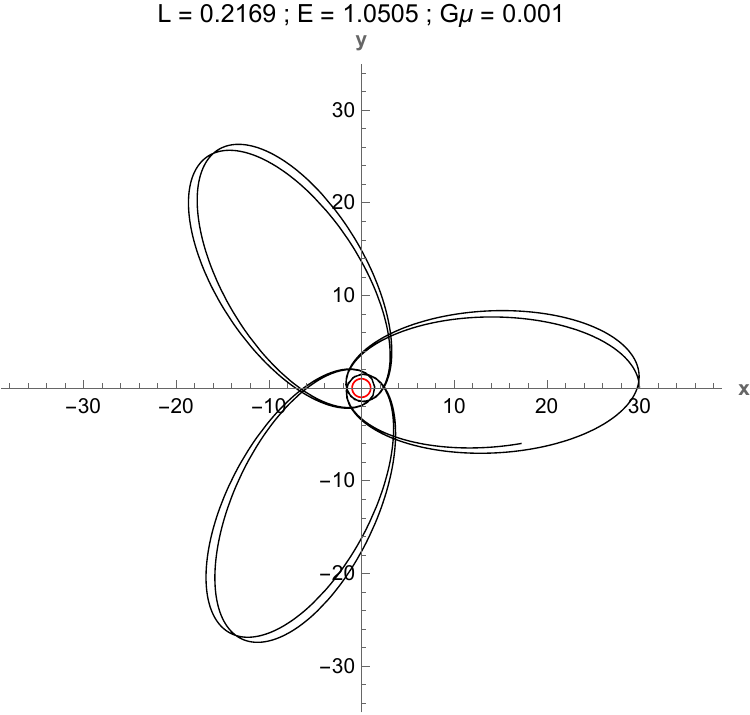}
        \caption{}
        \label{fig:precessingellipsec}
    \end{subfigure}
    \medskip
    \begin{subfigure}[t]{0.32\textwidth}
        \centering
    \includegraphics[width=\textwidth]{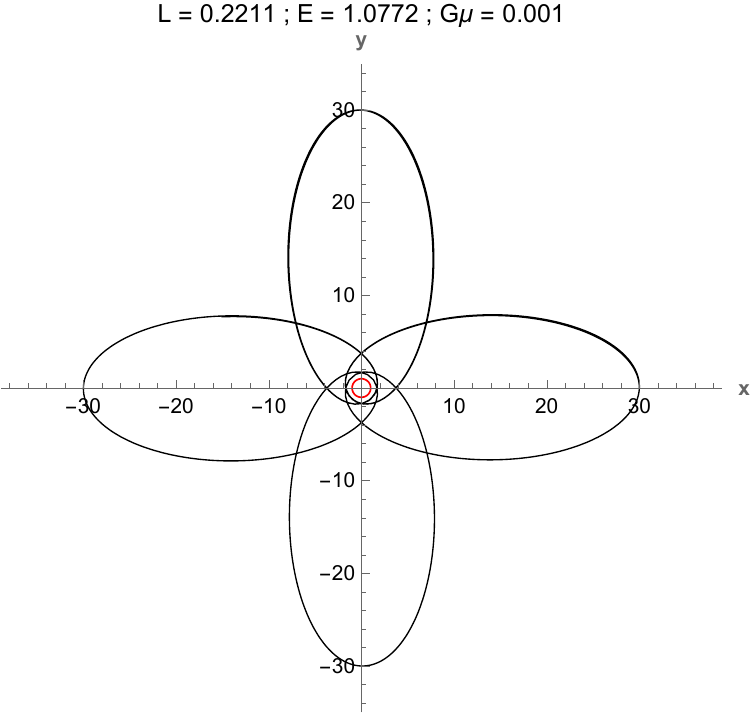}
        \caption{}
        \label{fig:precessingellipsed}
    \end{subfigure}
    \hfill
    \begin{subfigure}[t]{0.32\textwidth}
        \centering
    \includegraphics[width=\textwidth]{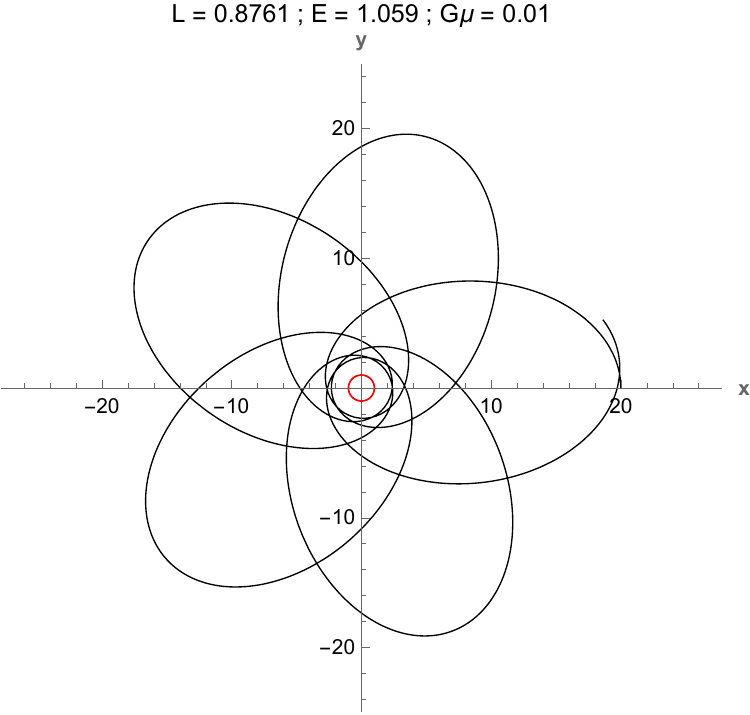}
        \caption{}
        \label{fig:precessingellipsee}
    \end{subfigure}
    \hfill
    \begin{subfigure}[t]{0.32\textwidth}
        \centering
    \includegraphics[width=\textwidth]{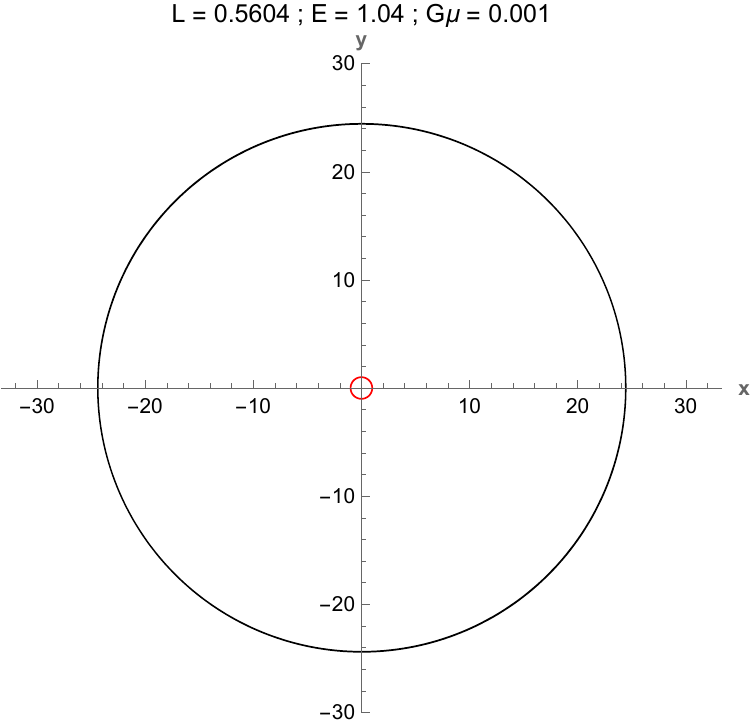}
        \caption{}
        \label{fig:circularorbit}
    \end{subfigure}
    \medskip
    \begin{subfigure}[t]{0.32\textwidth}
        \centering
    \includegraphics[width=\textwidth]{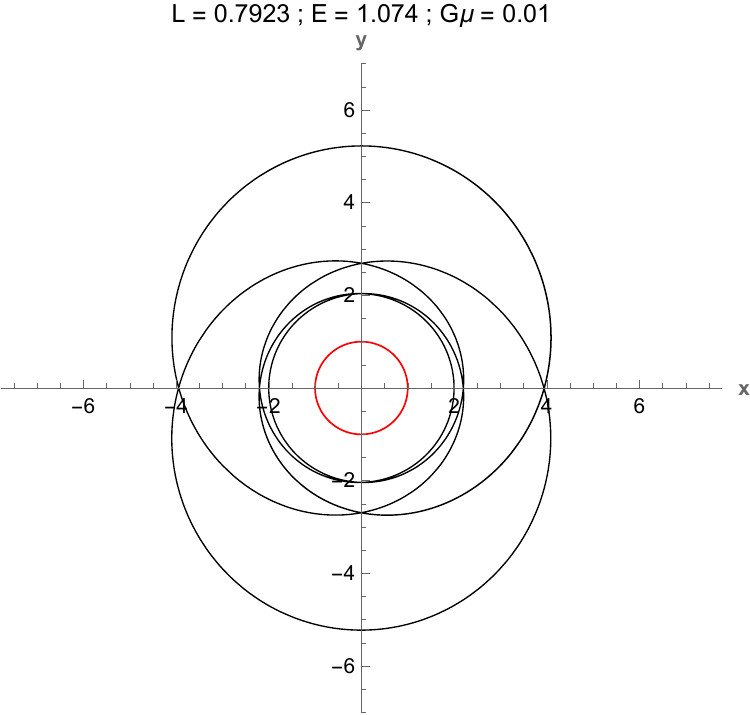}
        \caption{}
        \label{fig:precessingellipseg}
    \end{subfigure}
    \hfill
    \begin{subfigure}[t]{0.32\textwidth}
        \centering
    \includegraphics[width=\textwidth]{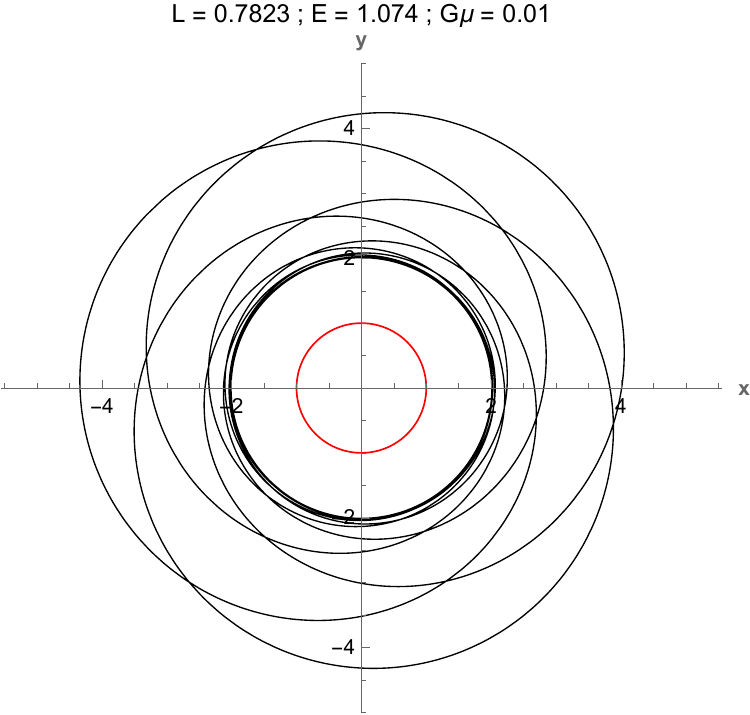}
        \caption{}
        \label{fig:precessingellipseh}
    \end{subfigure}
    \hfill
    \begin{subfigure}[t]{0.32\textwidth}
        \centering
    \includegraphics[width=\textwidth]{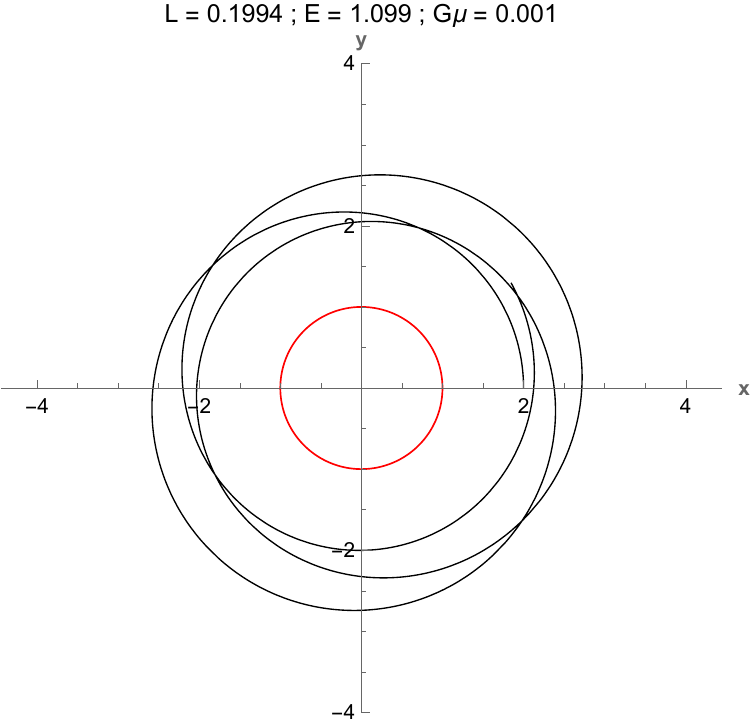}
        \caption{}
        \label{fig:precessingellipsei}
    \end{subfigure}
\caption{Bound precessing-ellipse trajectories of a massive test particle ($\epsilon = -1$). A circular orbit is shown for comparison in panel (f).}
\label{fig:precessingellipse_circular}
\end{figure}

We now examine bound orbits around the vorton. The effective potential, $V_{\text{eff}}$, serves as a guide to choose initial conditions and determine suitable values of $E$ and $L$ that yield bound trajectories. A few representative examples of these orbits are shown in Fig.~\ref{fig:precessingellipse_circular}. In the process of finding circular orbits, we find that some values of $r_c$ obtained from Eq.~\eqref{circularorbitcondition} do not actually produce circular trajectories when used as initial conditions. This discrepancy occurs for $r_c$ close to the vorton core. At large distance, however, the circular motions are recovered. This finding indicates that orbits predicted from $V_{\text{eff}}$ are valid only for weak-field condition.

\subsubsection{Toroidal Orbits}

An interesting aspect of studying geodesics in an extended mass distribution with $S^1$ topology, such as a vorton, is the possible existence of toroidal orbits. As suggested by Fig.~\ref{fig:Vplotznonzero}, the behavior of $V_{\text{eff}}(r,z)$ near the singular ring at $r=R$ implies that a particle could execute an orbit looping both inside ($r>R$) and outside ($r<R$) the ring,  encircling the vorton in a toroidal path. In cylindrical coordinates we explore such orbits using two approaches. In the first, the particle is initialized on the $xy$-plane with a nonzero velocity component in the $z$. In the second, the particle starts slightly off the $xy$-plane near the vorton and let the spacetime curvature induces the toroidal motion. 

\begin{figure}[H] 
\centering
    \begin{subfigure}[t]{0.32\textwidth}
        \centering
        \includegraphics[width=\textwidth]{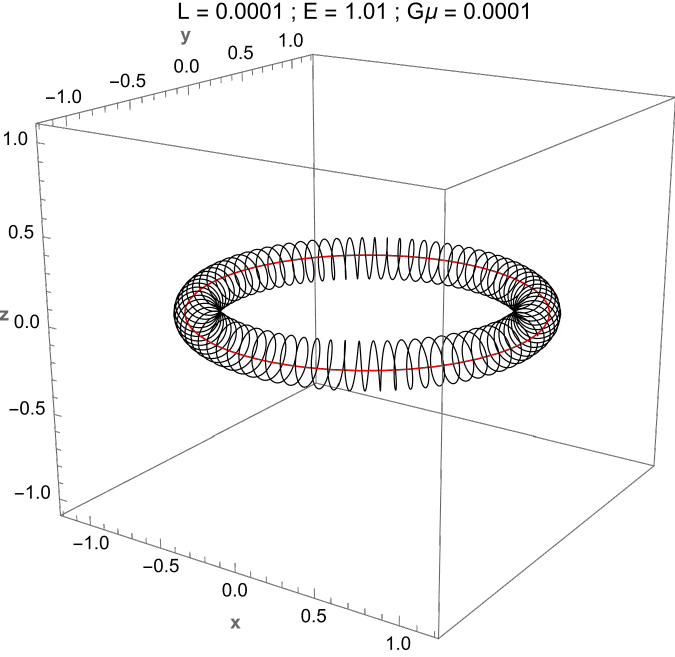}
        \label{fig:Toroidal3Dzdot_1}
    \end{subfigure}
    \hfill
    \begin{subfigure}[t]{0.32\textwidth}
        \centering
        \includegraphics[width=\textwidth]{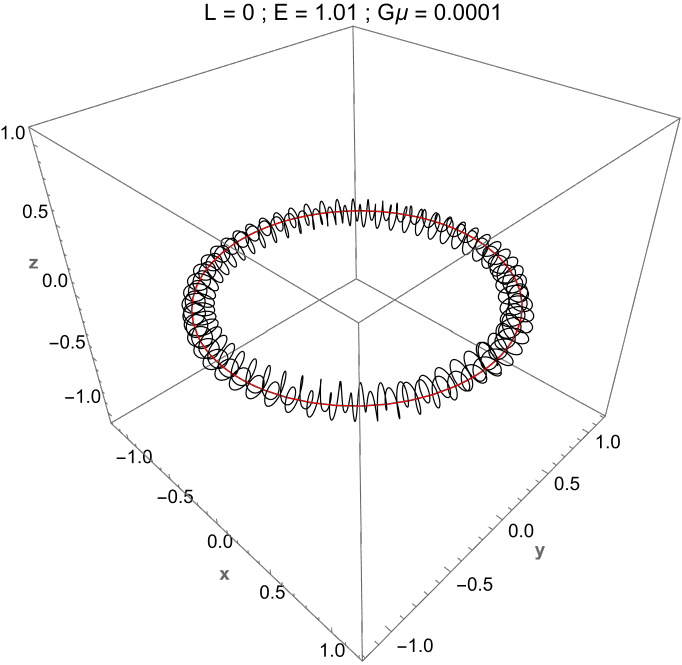}
        \label{fig:Toroidal3Dzdot_3}
    \end{subfigure}
    \hfill
    \begin{subfigure}[t]{0.32\textwidth}
        \centering
        \includegraphics[width=\textwidth]{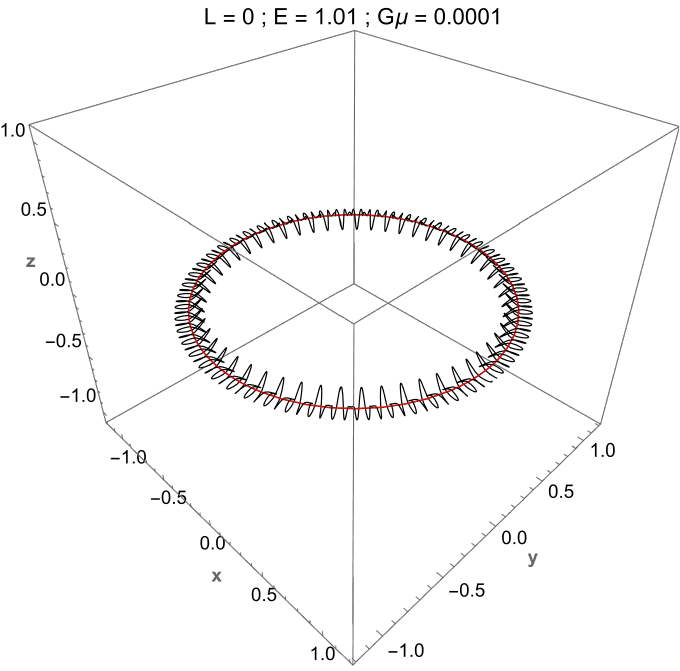}
        \label{fig:Toroidal3Dzdot_5}
    \end{subfigure}
    
    \medskip
    \begin{subfigure}[t]{0.32\textwidth}
        \centering
        \includegraphics[width=\textwidth]{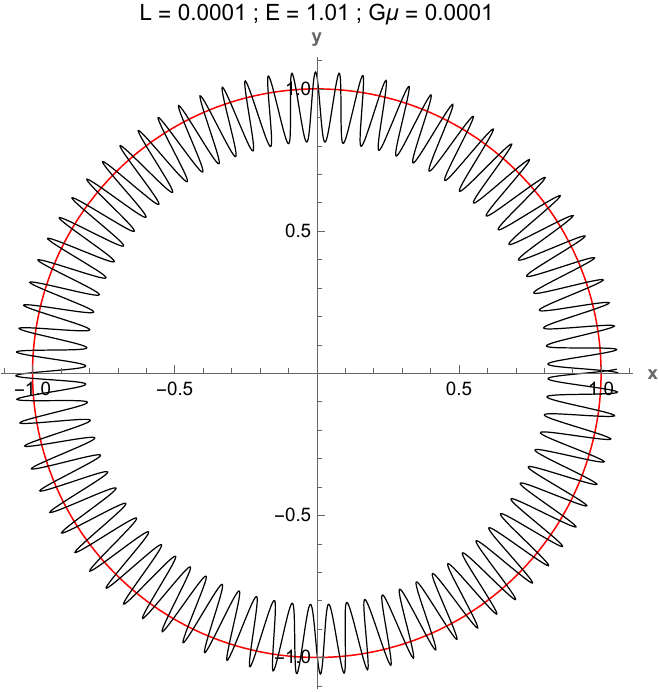}
        \label{fig:ToroidalXYzdot__1}
    \end{subfigure}
    \hfill
    \begin{subfigure}[t]{0.32\textwidth}
        \centering
        \includegraphics[width=\textwidth]{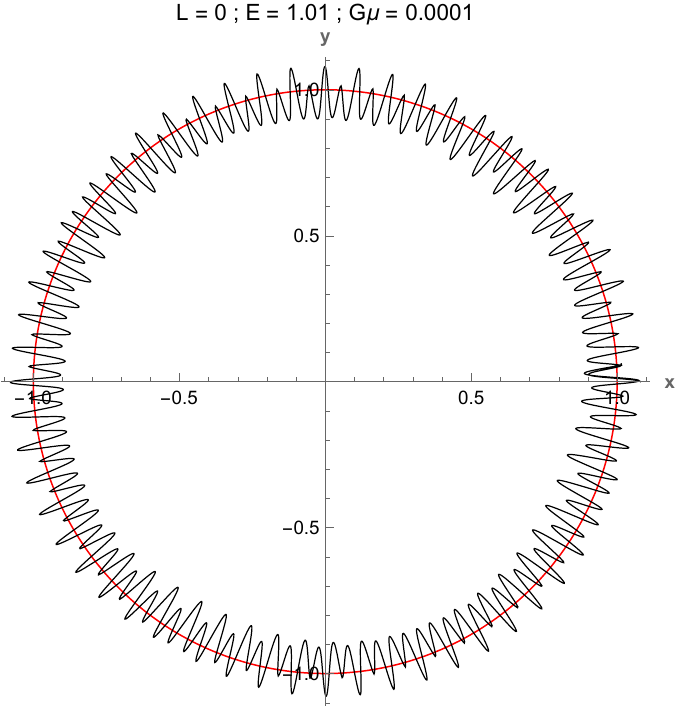}
        \label{fig:ToroidalXYzdot_3}
    \end{subfigure}
    \hfill
    \begin{subfigure}[t]{0.32\textwidth}
        \centering
        \includegraphics[width=\textwidth]{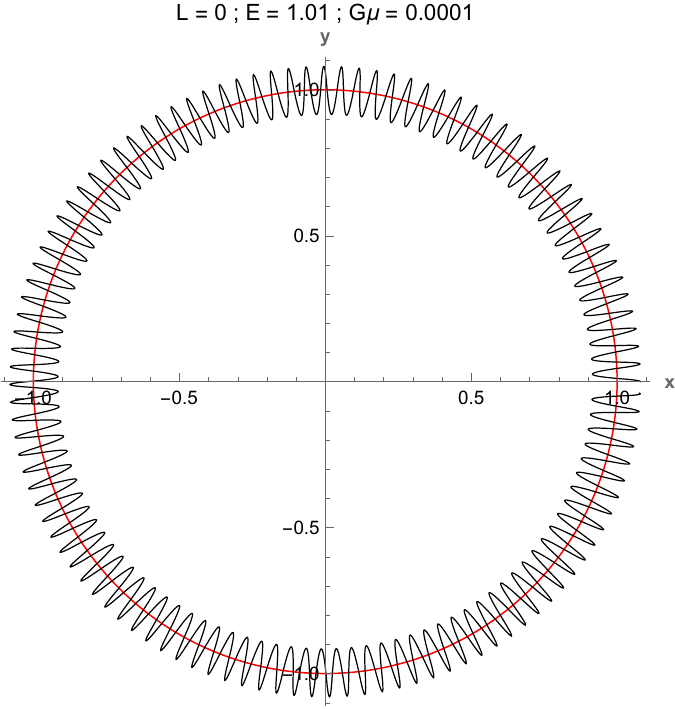}
        \label{fig:ToroidalXYzdot_5}
    \end{subfigure}
    
    \medskip
    \begin{subfigure}[t]{0.32\textwidth}
        \centering
        \includegraphics[width=\textwidth]{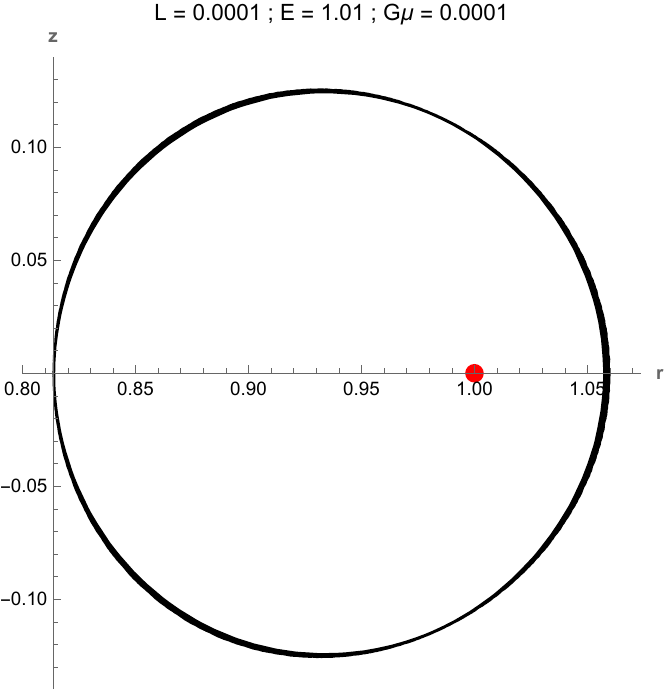}
        \caption{}
        \label{fig:ToroidalRZzdot_1}
    \end{subfigure}
    \hfill
    \begin{subfigure}[t]{0.32\textwidth}
        \centering
        \includegraphics[width=\textwidth]{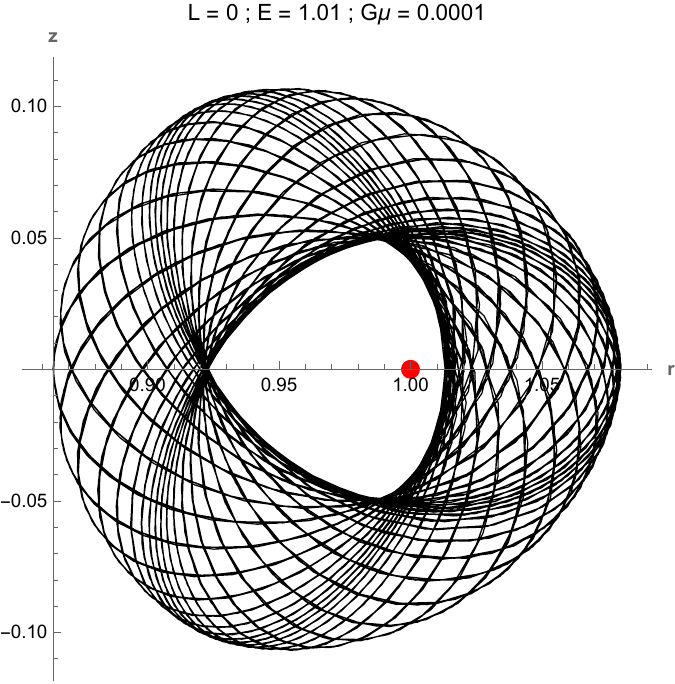}
        \caption{}
        \label{fig:ToroidalRZzdot_3}
    \end{subfigure}
    \hfill
    \begin{subfigure}[t]{0.32\textwidth}
        \centering
        \includegraphics[width=\textwidth]{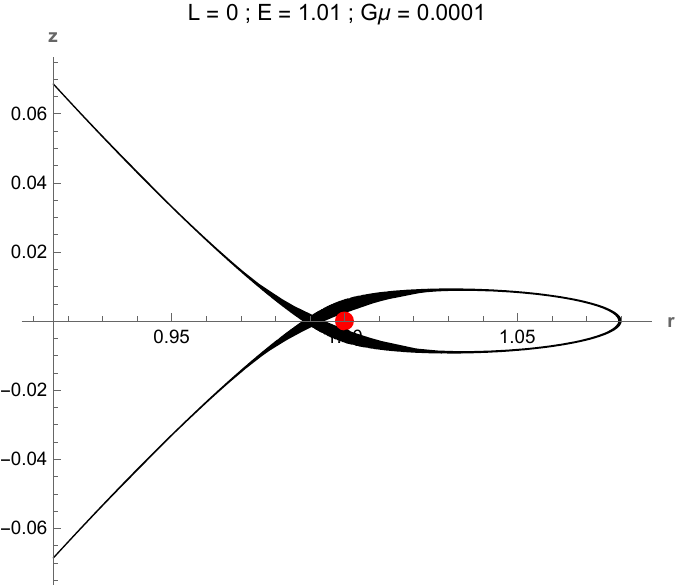}
        \caption{}
        \label{fig:ToroidalRZzdot_5}
    \end{subfigure}
\caption{Toroidal orbits of a massive particle around a vorton. Initial velocity is applied along the $z$-axis for a particle starting in the $xy$ plane. each column corresponds to a single orbits: top row – 3D view; middle row – top-down view in the $xy$ plane; bottom row – radial and vertical ($r$–$z$) trajectory, with the particle’s trajectory around the vorton superimposed.}
\label{fig:Toroidal_Orbits_zdot_1}
\end{figure}

\begin{figure}[H] 
\centering
    \begin{subfigure}[t]{0.32\textwidth}
        \centering
        \includegraphics[width=\textwidth]{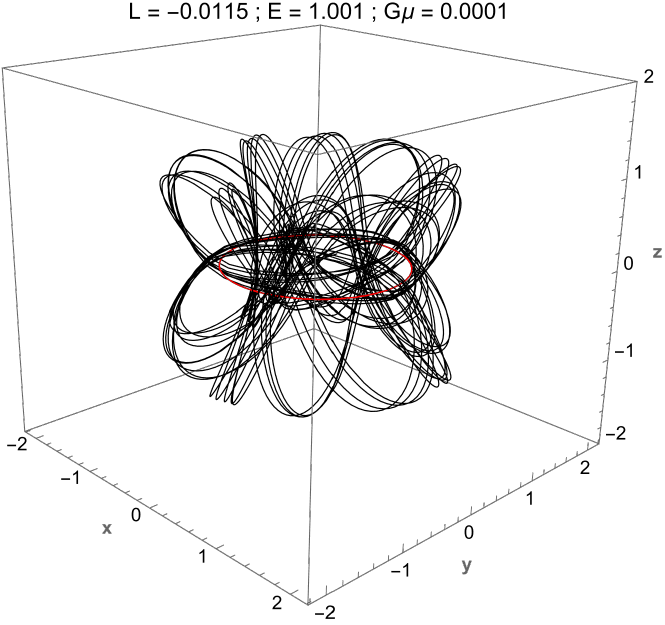}
        \label{fig:Toroidal3Dzdot_9}
    \end{subfigure}
    \hfill
    \begin{subfigure}[t]{0.32\textwidth}
        \centering
        \includegraphics[width=\textwidth]{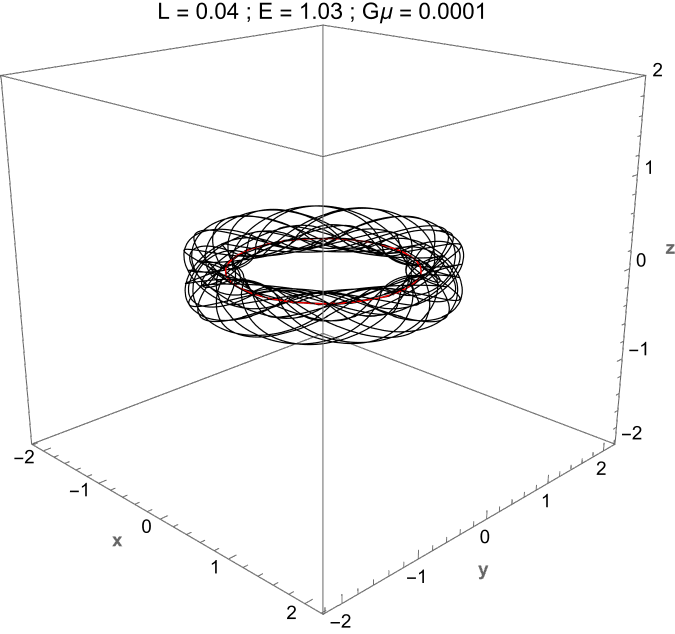}
        \label{fig:Toroidal3Dzdot_8}
    \end{subfigure}
    \hfill
    \begin{subfigure}[t]{0.32\textwidth}
        \centering
        \includegraphics[width=\textwidth]{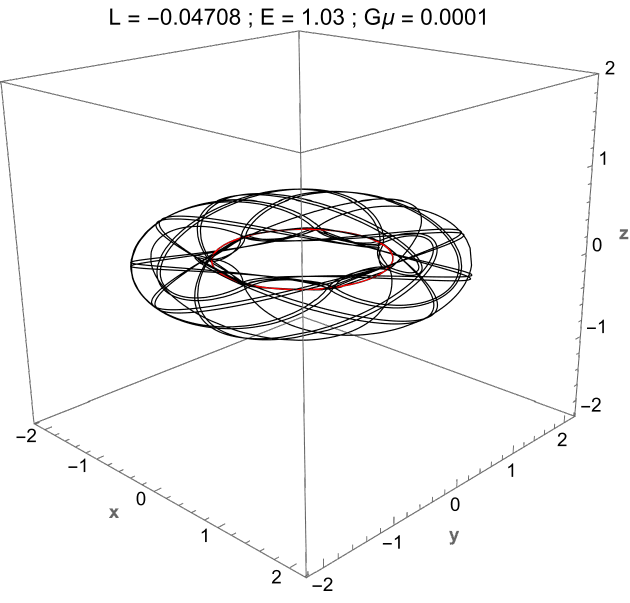}
        \label{fig:Toroidal3Dzdot_7}
    \end{subfigure}
    \medskip
    \begin{subfigure}[t]{0.32\textwidth}
        \centering
        \includegraphics[width=\textwidth]{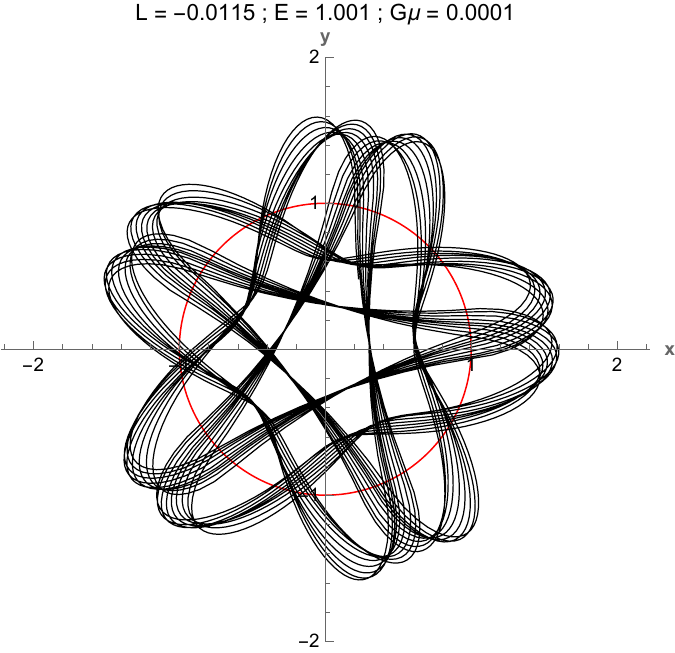}
        \label{fig:ToroidalXYzdot__9}
    \end{subfigure}
    \hfill
    \begin{subfigure}[t]{0.32\textwidth}
        \centering
        \includegraphics[width=\textwidth]{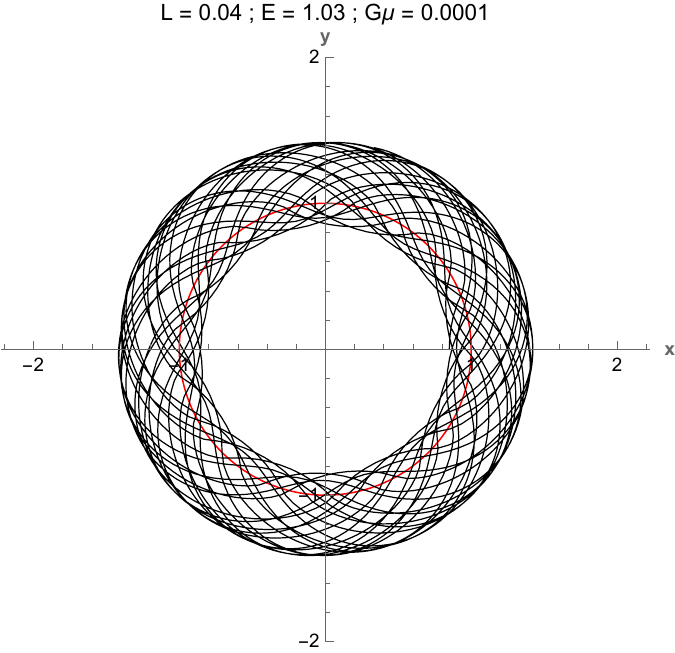}
        \label{fig:ToroidalXYzdot_8}
    \end{subfigure}
    \hfill
    \begin{subfigure}[t]{0.32\textwidth}
        \centering
        \includegraphics[width=\textwidth]{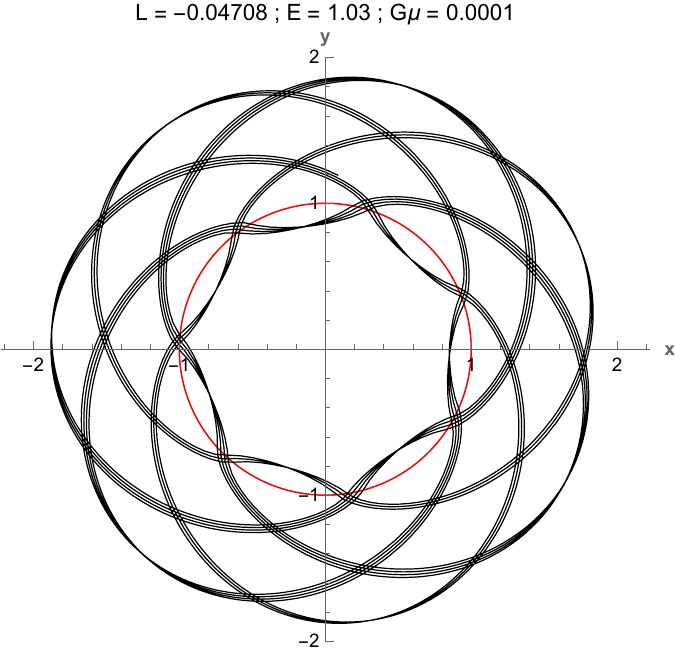}
        \label{fig:ToroidalXYzdot_7}
    \end{subfigure}
    
    \medskip
    \begin{subfigure}[t]{0.32\textwidth}
        \centering
        \includegraphics[height=\textwidth]{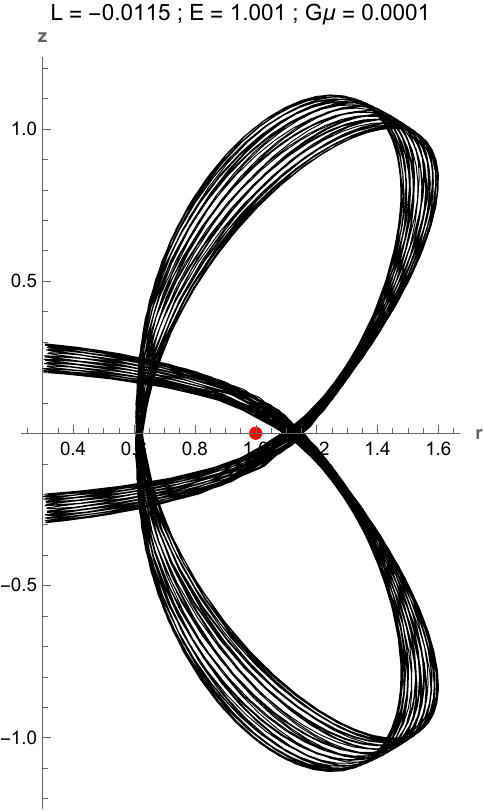}
        \caption{}
        \label{fig:ToroidalRZzdot_9}
    \end{subfigure}
    \hfill
    \begin{subfigure}[t]{0.32\textwidth}
        \centering
        \includegraphics[width=\textwidth]{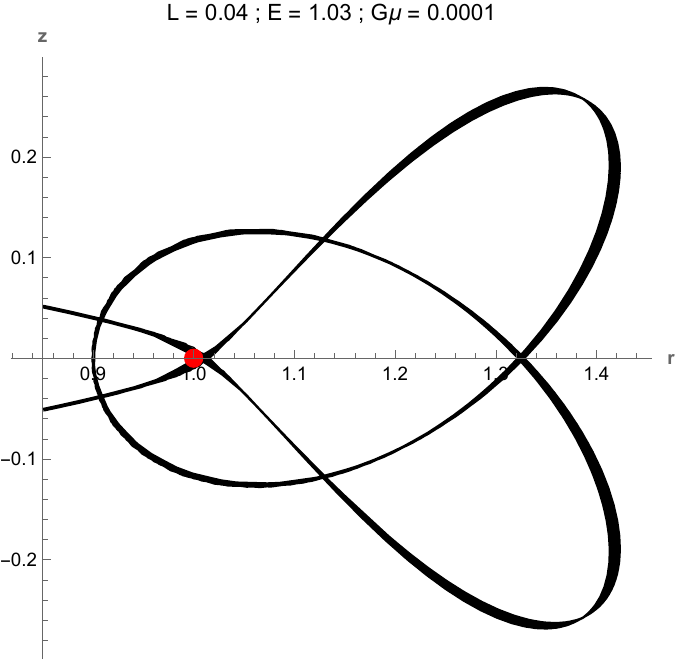}
        \caption{}
        \label{fig:ToroidalRZzdot_8}
    \end{subfigure}
    \hfill
    \begin{subfigure}[t]{0.32\textwidth}
        \centering
        \includegraphics[width=\textwidth]{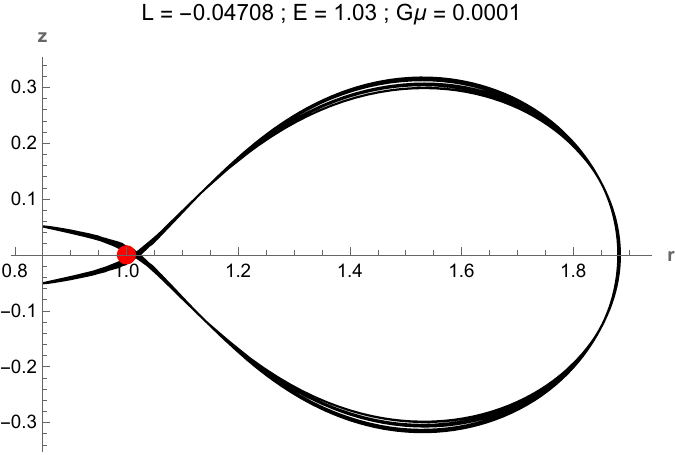}
        \caption{}
        \label{fig:ToroidalRZzdot_7}
    \end{subfigure}
\caption{Toroidal orbits of a massive particle around a vorton, starting off the $xy$ plane with zero initial velocity. Columns: top – 3D view; middle – top-down $xy$ view; bottom – $r$–$z$ trajectory with orbit path superimposed.}
\label{fig:Toroidal_Orbits_z_1}
\end{figure}

We present several examples of toroidal orbits obtained using the two methods described earlier, as shown in Figs.~\ref{fig:Toroidal_Orbits_zdot_1} and~\ref{fig:Toroidal_Orbits_z_1}. Some trajectories form distinct toroidal structures which, when looking from its $r-z$ projection shows precessing elliptical paths (Fig.~\ref{fig:ToroidalRZzdot_1}). All the orbits shown exhibit varying degrees of precession and slight shifts after completing a revolution around the vorton. Interestingly, a test particle with $L = 0$ still obeys a bound orbit, as seen in Figs.~\ref{fig:ToroidalRZzdot_3}-\ref{fig:ToroidalRZzdot_5}. This is due to the frame-dragging effect that induces an effective angular velocity $EA(r,z)$ for an initially non-orbitting particle. We also show orbits with $L < 0$ (Figs.~\ref{fig:ToroidalRZzdot_9}-\ref{fig:ToroidalRZzdot_7}) representing retrograde motion, or prograde motion weaker than the local frame rotation ($L_0 < L < 0$). Some trajectories appear enclosing the vorton but, upon closer inspection, merely cross regions inside and outside the ring ($r < R$ and $r>R$) without fully encapsulating it (Figs.~\ref{fig:ToroidalRZzdot_7}-\ref{fig:ToroidalRZzdot_5}). These particular orbits transition from $r < R$ and $r>R$ but do not encapsulate the vorton.

\begin{figure}[H] 
\centering
    \begin{subfigure}[c]{0.3\textwidth}
        \centering
        \includegraphics[width=\textwidth]{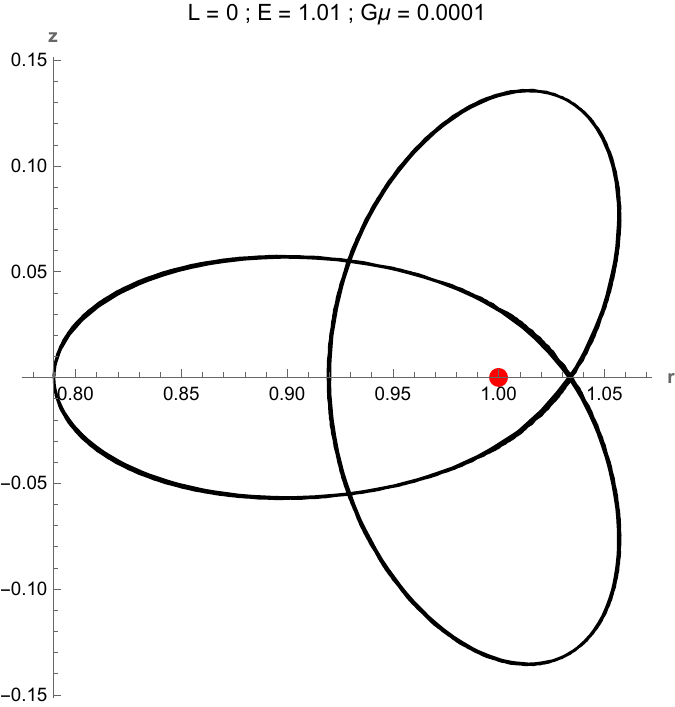}
        \label{fig:RZ_Toroidal_10}
        \caption{}
    \end{subfigure}
    \hfill
     \begin{subfigure}[c]{0.31\textwidth}
        \centering
        \includegraphics[width=\textwidth]{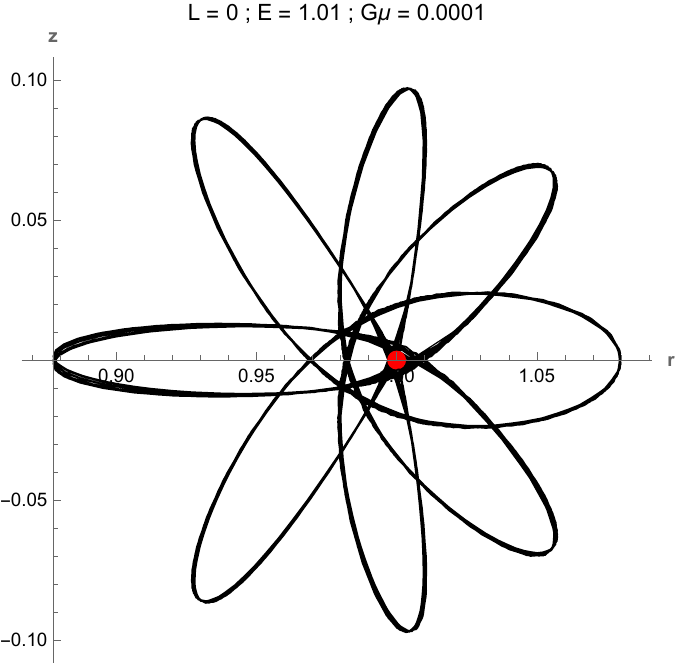}
        \label{fig:RZ_Toroidal_11}
        \caption{}
    \end{subfigure}
    \hfill
    \begin{subfigure}[c]{0.32\textwidth}
        \centering
        \includegraphics[width=\textwidth]{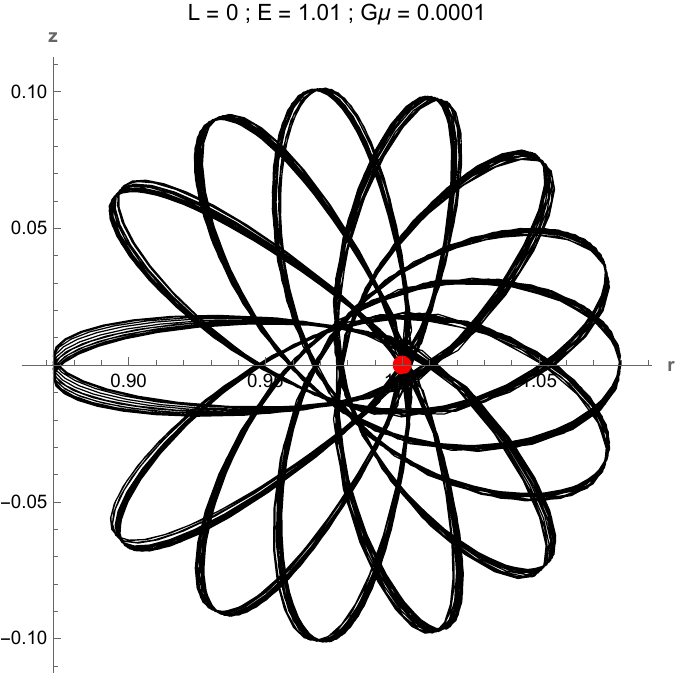}
        \caption{}
        \label{fig:RZ_Toroidal_12}
    \end{subfigure}
\caption{Toroidal orbits superimposed in the $r$–$z$ plane for $G\mu = 0.0001$, $L = 0$, and $E = 1.01$. Closer orbits (panels b and c) exhibit stronger precession, with larger shifts per period.}
\label{fig:RZ_Toroidal}
\end{figure}

Several toroidal orbits we obtained exhibit dynamical features resembling Newtonian trajectories around a solid circular ring \cite{elipe}. In particular, Fig.~\ref{fig:RZ_Toroidal} shows trajectories with multi-lobed structures similar to those reported in Fig. 9 of Ref.~\cite{elipe}, where the particle drifts near the ring and forms repeating lobe patterns. It should be emphasized, however, that orbits passing very close to the vorton’s core, $R = 1$, can no longer be described by our geodesic equations since in this regime the weak-field approximation breaks down. Nonetheless, our solutions serve as a general-relativistic analogue of 
Newtonian trajectories around a solid circular ring.

\begin{figure}[H] 
\centering
    \begin{subfigure}[c]{0.22\textwidth}
        \centering
        \includegraphics[width=\textwidth]{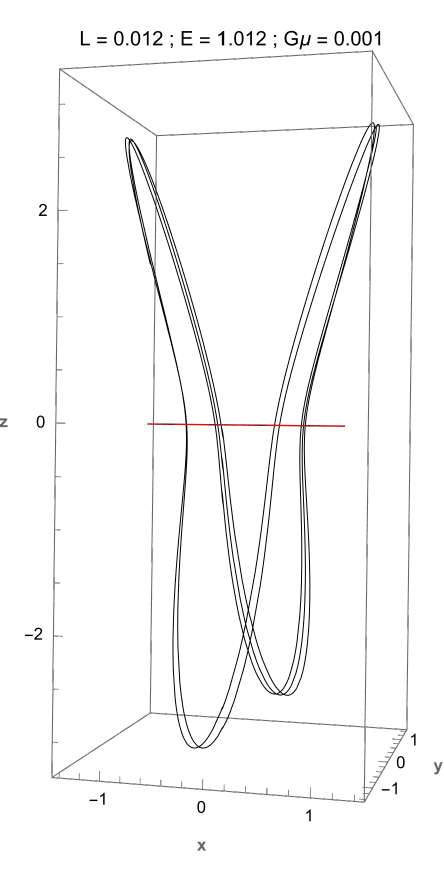}
        \label{fig:Crown3d2}
        \caption{}
    \end{subfigure}
    \hfill
     \begin{subfigure}[c]{0.3\textwidth}
        \centering
        \includegraphics[width=\textwidth]{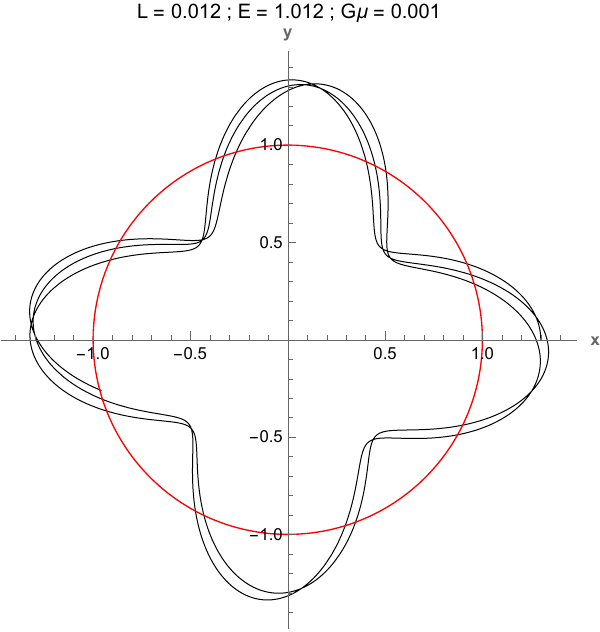}
        \label{fig:Crownxy2}
        \caption{}
    \end{subfigure}
    \hfill
    \begin{subfigure}[c]{0.2\textwidth}
        \centering
        \includegraphics[width=\textwidth]{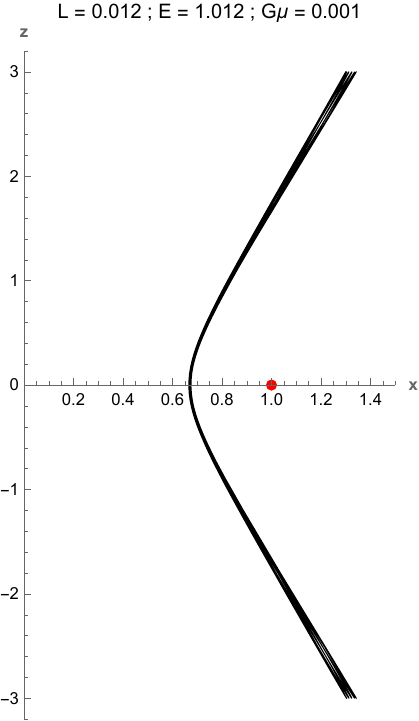}
        \caption{}
        \label{fig:Crownrz2}
    \end{subfigure}
    
    \medskip
    \begin{subfigure}[c]{0.32\textwidth}
        \centering
        \includegraphics[width=\textwidth]{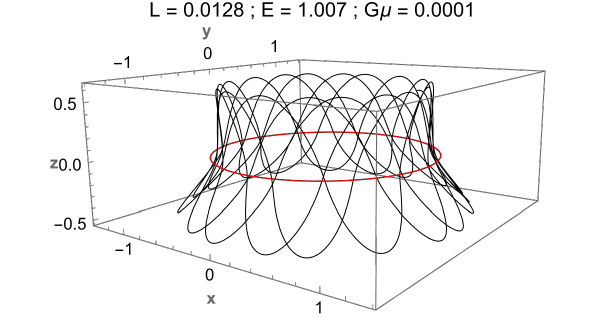}
        \label{fig:Crown3d3}
        \caption{}
    \end{subfigure}
    \hfill
    \begin{subfigure}[c]{0.3\textwidth}
        \centering
        \includegraphics[width=\textwidth]{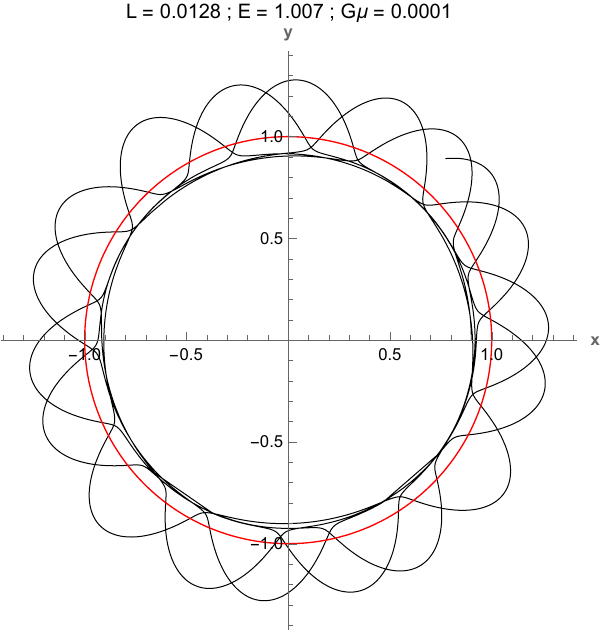}
        \label{fig:Crownxy3}
        \caption{}
    \end{subfigure}
    \hfill
    \begin{subfigure}[c]{0.2\textwidth}
        \centering
        \includegraphics[width=\textwidth]{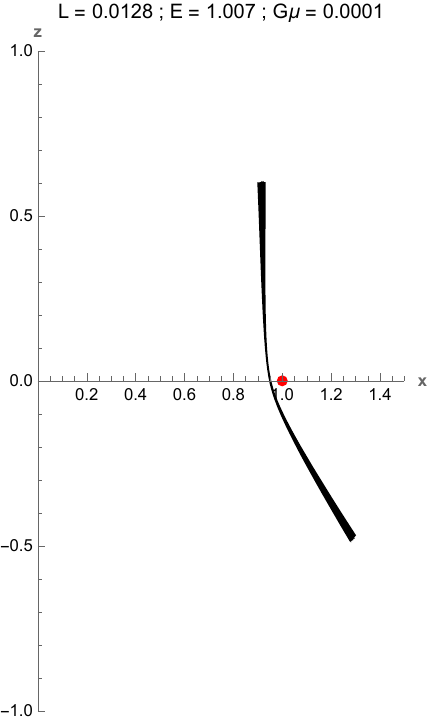}
        \caption{}
        \label{fig:Crownrz3}
    \end{subfigure}
\caption{Crown-type orbits around the chiral vorton. Each row shows a single trajectory from three perspectives: a 3D view (left), the projection onto the $xy$-plane (middle), and a superimposed $rz$-plot (right) illustrating the regularity of the oscillatory motion.}
\label{fig:crownorbit}
\end{figure}

\begin{figure}[H] 
\centering
    \begin{subfigure}[t]{0.30\textwidth}
        \centering
        \includegraphics[width=\textwidth]{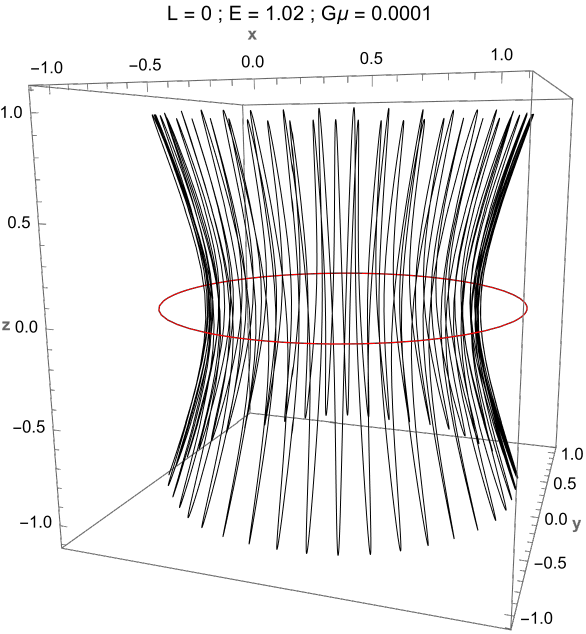}
        \label{fig:Crown3d4}
    \end{subfigure}
    \hfill
    \begin{subfigure}[t]{0.30\textwidth}
        \centering
        \includegraphics[width=\textwidth]{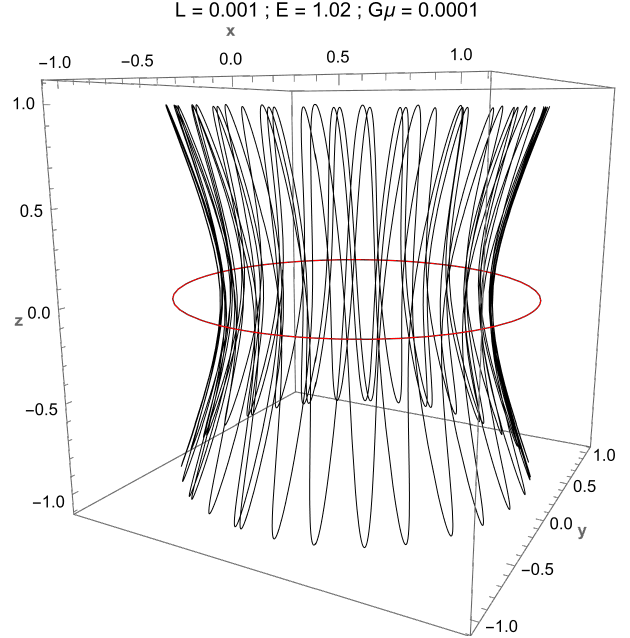}
        \label{fig:Crown3d5}
    \end{subfigure}
    \hfill
    \begin{subfigure}[t]{0.32\textwidth}
        \centering
        \includegraphics[width=\textwidth]{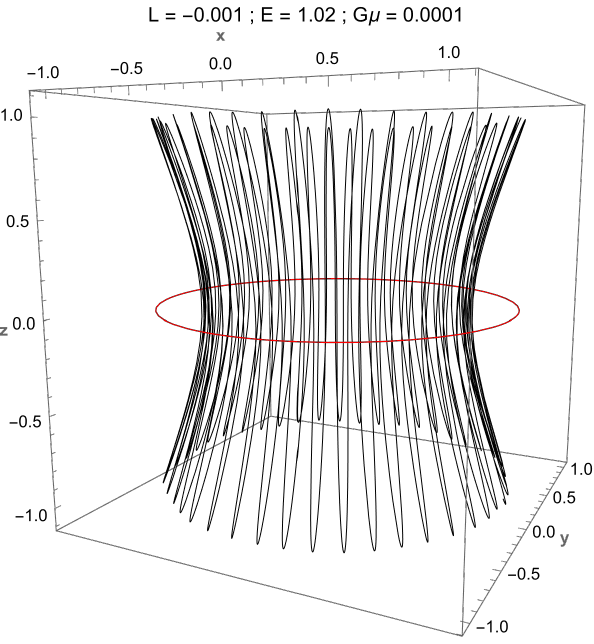}
        \label{fig:Crown3d6}
    \end{subfigure}
    
    \medskip
    \begin{subfigure}[t]{0.32\textwidth}
        \centering
        \includegraphics[width=\textwidth]{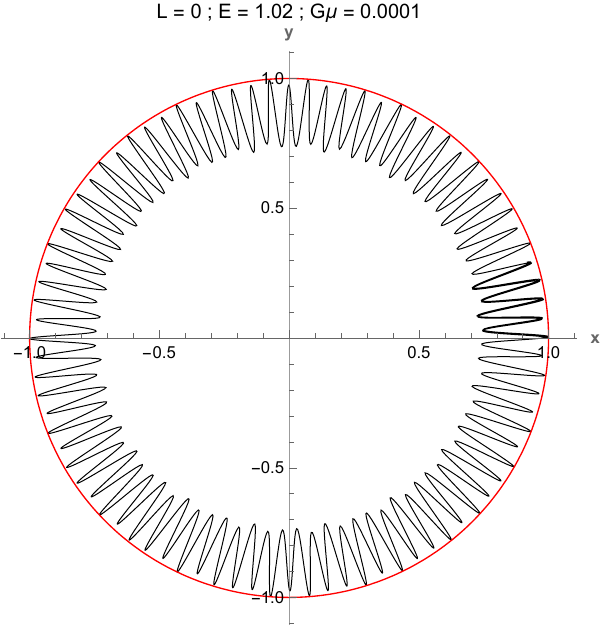}
        \label{fig:Crownxy4}
    \end{subfigure}
    \hfill
    \begin{subfigure}[t]{0.32\textwidth}
        \centering
        \includegraphics[width=\textwidth]{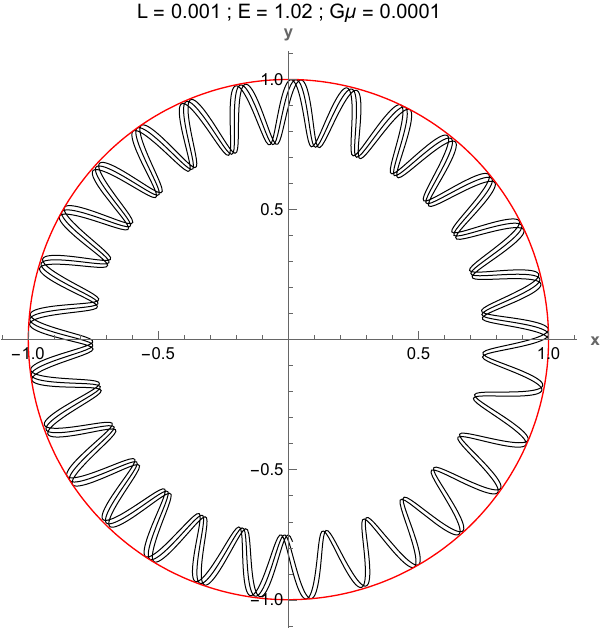}
        \label{fig:Crownxy5}
    \end{subfigure}
    \hfill
    \begin{subfigure}[t]{0.32\textwidth}
        \centering
        \includegraphics[width=\textwidth]{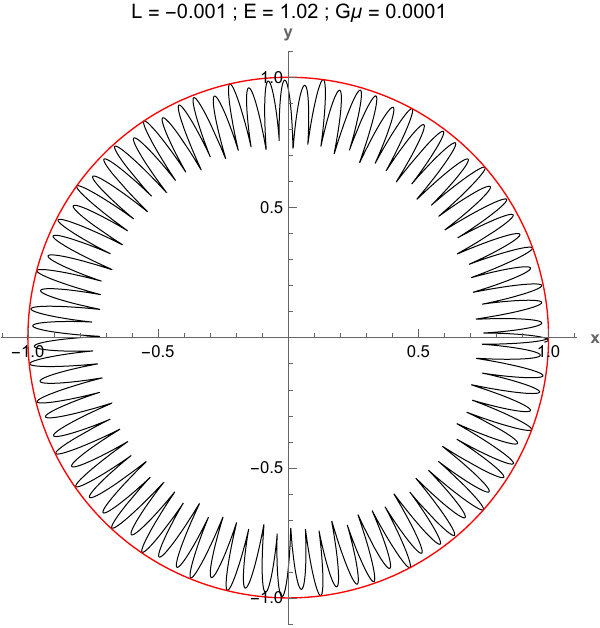}
        \label{fig:Crownxy6}
    \end{subfigure}
    
    \medskip
    \begin{subfigure}[t]{0.2\textwidth}
        \centering
        \includegraphics[width=\textwidth]{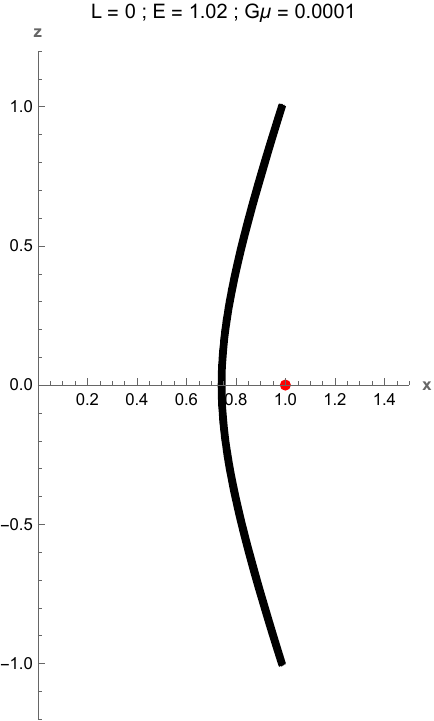}
        \caption{}
        \label{fig:Crownrz4}
    \end{subfigure}
    \hfill
    \begin{subfigure}[t]{0.2\textwidth}
        \centering
        \includegraphics[width=\textwidth]{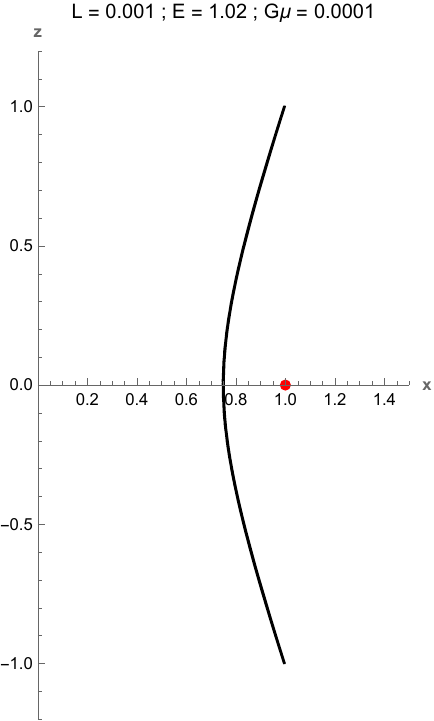}
        \caption{}
        \label{fig:Crownrz5}
    \end{subfigure}
    \hfill
    \begin{subfigure}[t]{0.2\textwidth}
        \centering
        \includegraphics[width=\textwidth]{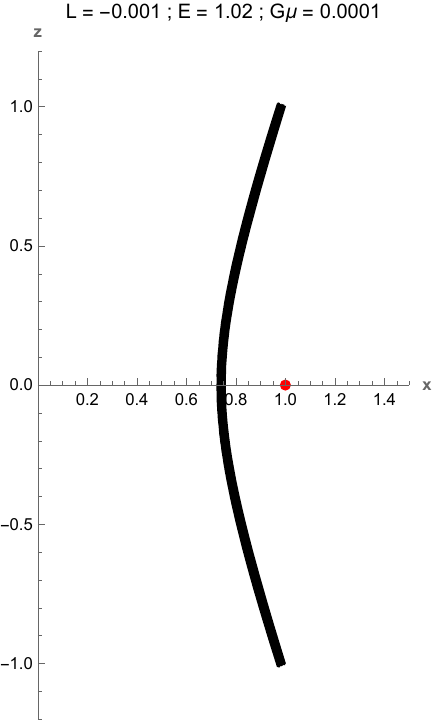}
        \caption{}
        \label{fig:Crownrz6}
    \end{subfigure}
\caption{Crown orbits around the vorton for different $L$ with identical initial conditions. Each trajectory (column) is shown in 3D (top row), in the $xy$-plane (middle row), and via the superimposed $r-z$ profile (bottom row).}
\label{fig:crownorbitLeff}
\end{figure}

We also found a class of trajectories in which a test particle oscillates through the vorton’s ring structure, moving alternately inside and outside the core, without intersecting its own path during a single oscillation. We refer to these trajectories as {\it crown orbits} due to their distinctive, thorny crown-like geometry. They are presented in Figs.~\ref{fig:crownorbit}-\ref{fig:crownorbitLeff}, where it can be observed that the particle oscillates both radially and vertically through the vorton plane. As can be seen in Fig.~\ref{fig:crownorbitLeff}, varying the angular momentum parameter $L=0, 0.001, -0.001$ leads to notable differences in the orbital dynamics rate around vorton. For $L=0$, the particle still acquires angular velocity due to the frame-dragging effect of the vorton. This effect is amplified for $L=0.001$, whereas for $L=-0.001$ the rotation is counteracted, leading to slower motion.

\begin{figure}[H] 
\centering
    \begin{subfigure}[t]{0.3\textwidth}
        \centering
        \includegraphics[width=\textwidth]{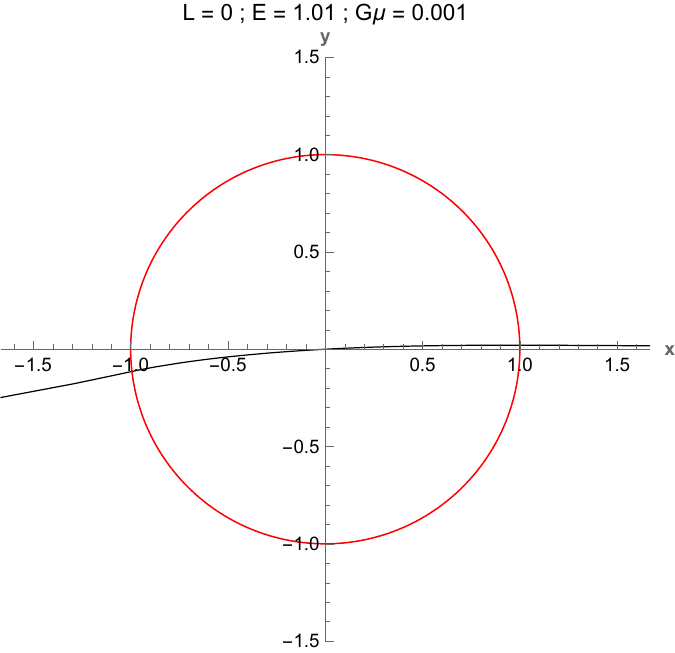}
        \caption{}
        \label{fig:scatter_over_XY}
    \end{subfigure}
    \hfill
    \begin{subfigure}[t]{0.3\textwidth}
        \centering
        \includegraphics[width=\textwidth]{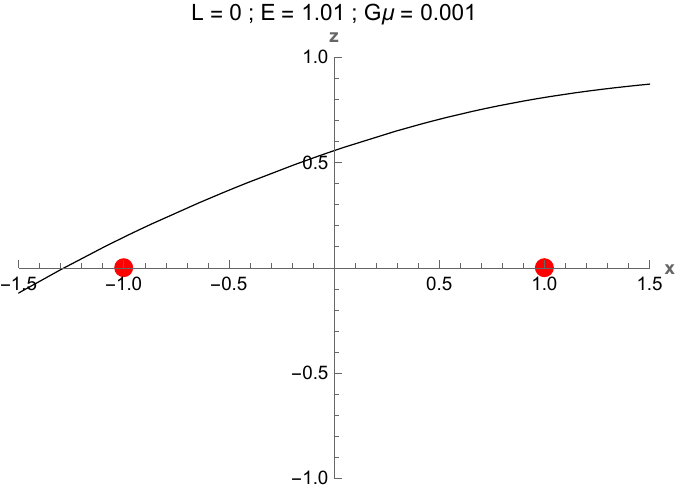}
        \caption{}
        \label{fig:scatter_over_RZ}
    \end{subfigure}
    \hfill
    \begin{subfigure}[t]{0.3\textwidth}
        \centering
        \includegraphics[width=\textwidth]{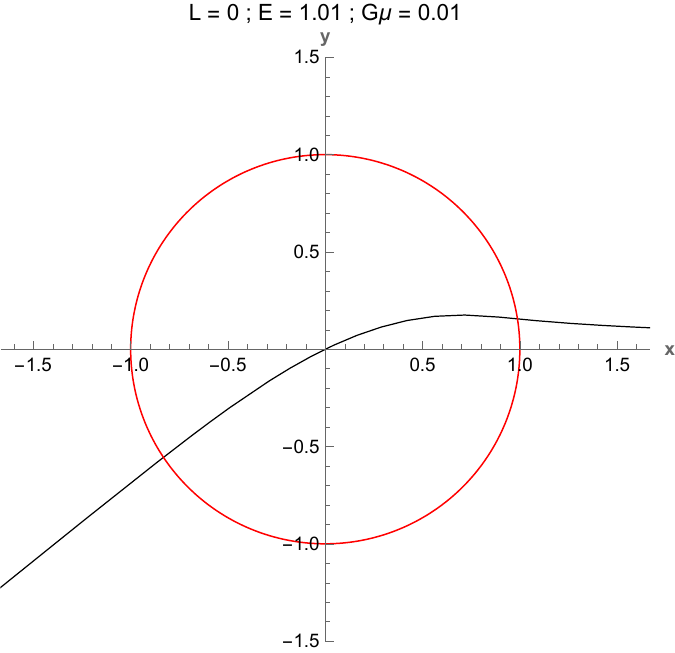}
        \caption{}
        \label{fig:scatter_upover_XY}
    \end{subfigure}
    \hfill
    \begin{subfigure}[t]{0.3\textwidth}
        \centering
        \includegraphics[width=\textwidth]{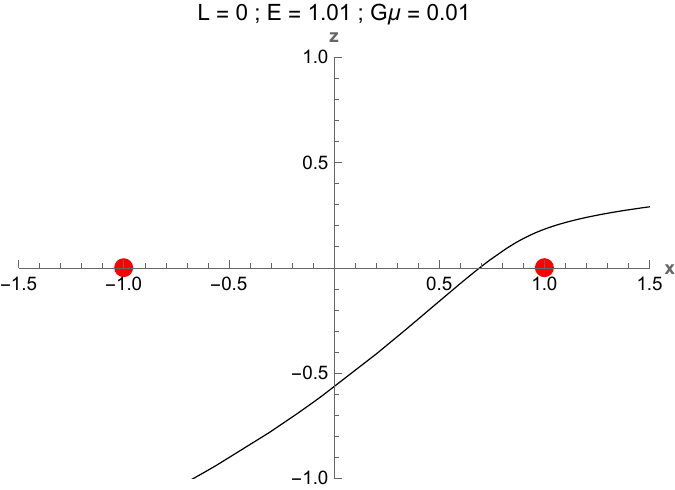}
        \caption{}
        \label{fig:scatter_upover_RZ}
    \end{subfigure}
    \hfill
    \begin{subfigure}[t]{0.3\textwidth}
        \centering
        \includegraphics[width=\textwidth]{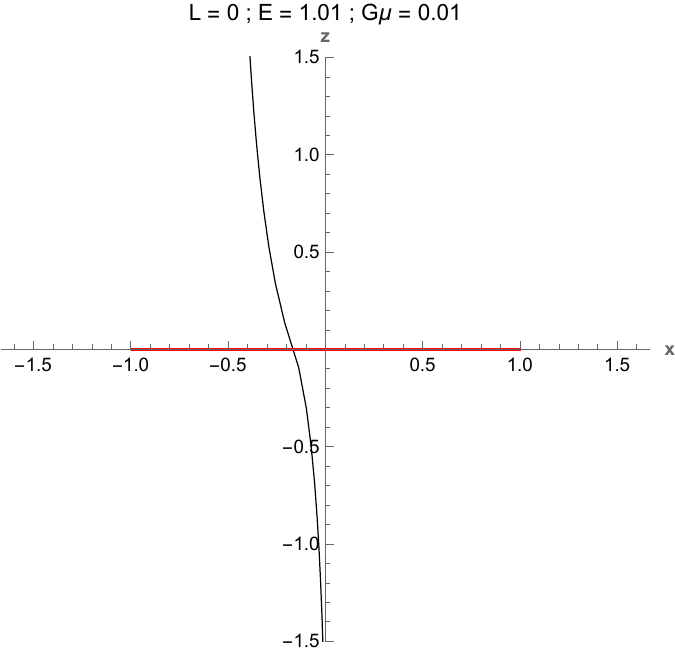}
        \caption{}
        \label{fig:scatter_z_XY}
    \end{subfigure}
    \hfill
    \begin{subfigure}[t]{0.3\textwidth}
        \centering
        \includegraphics[width=\textwidth]{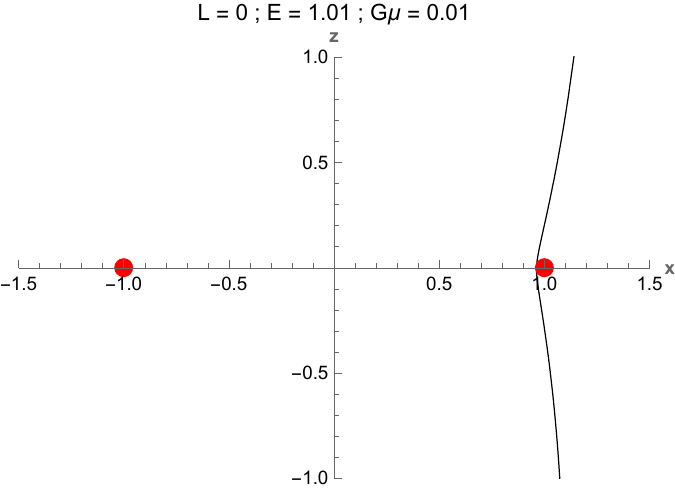}
        \caption{}
        \label{fig:scatter_z_RZ}
    \end{subfigure}
\caption{Unbound trajectories of massive test particles passing near the vorton, shown in projections onto the $xy$-plane (left) and $xz$-plane (right), for different initial velocities.}
\label{fig:Scattering}
\end{figure}

\subsubsection{Unbound Orbits}

We also examine unbound orbits, where test particles are deflected but not gravitationally captured. They are achieved by assigning an initial velocity in either the radial $r$ or vertical $z$ toward the vorton. As shown in Fig.~\ref{fig:Scattering}, particles experience a transient gain in angular momentum due to the frame-dragging effect as it traverses close to the vorton. This effect is clearly illustrated in Figs.~\ref{fig:scatter_over_XY}, \ref{fig:scatter_upover_XY} and \ref{fig:scatter_z_XY}. The deflection and temporary azimuthal motion become evident as the particles pass close to the vorton. In Fig.~\ref{fig:scatter_upover_RZ}, a particle passes directly through the vorton's central region, effectively threading through the ring-like structure, and eventually escapes, resembling a scattering event through a circular hoop.

\subsection{Null Geodesics}

In this work we primarily focus on the timelike orbital dynamics around vorton. For the null case ($\epsilon = 0$), we restrict our analysis to unbound trajectories analogous to the scattering orbits shown in Fig.~\ref{fig:Scattering}. In~\cite{putra2024gravitationalfieldlensingcircular} we have studied in great detail the gravitational lensing due to a circular chiral vorton. Here, we construct the null trajectories using the full three-dimensional weak-field geodesic equations given in Eqs.~\eqref{geodesicfinalr}-\eqref{geodesicfinalz}. 

\begin{figure}[H] 
\centering
    \begin{subfigure}[t]{0.3\textwidth}
        \centering
        \includegraphics[width=\textwidth]{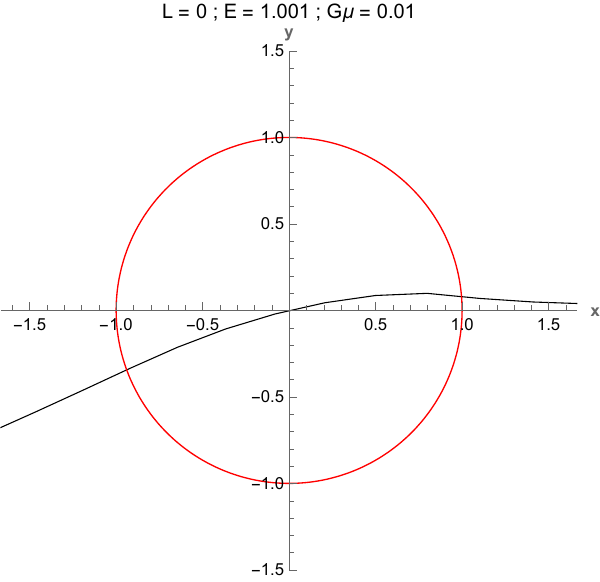}
        \caption{}
        \label{fig:null_scatter_upover_gmu0.01_XY}
    \end{subfigure}
    \hfill
    \begin{subfigure}[t]{0.3\textwidth}
        \centering
        \includegraphics[width=\textwidth]{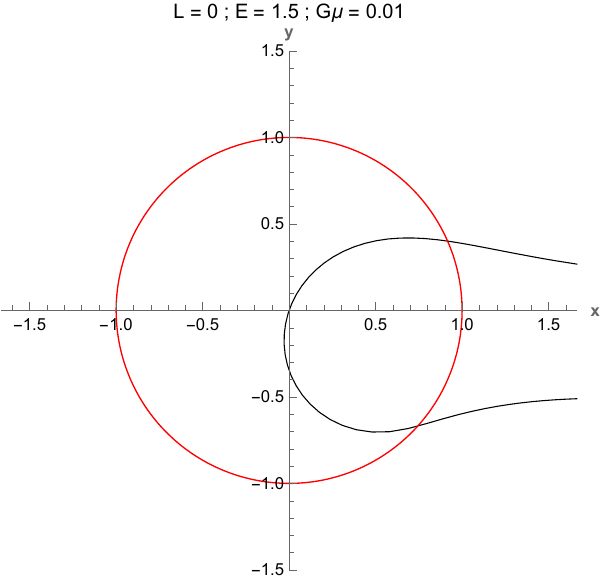}
        \caption{}
        \label{fig:null_scatter_upover_back_gmu0.01_XY}
    \end{subfigure}
    \hfill
   \begin{subfigure}[t]{0.3\textwidth}
        \centering
        \includegraphics[width=\textwidth]{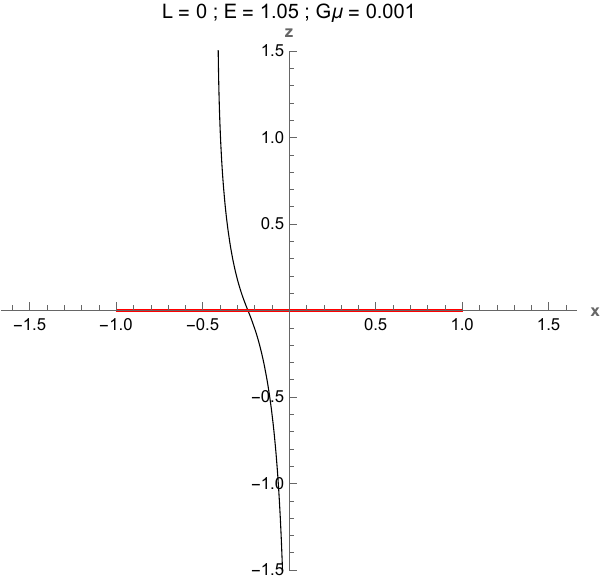}
        \caption{}
        \label{fig:null_scatter_fromz_gmu0.001_XZ}
    \end{subfigure}

    \medskip
    \begin{subfigure}[t]{0.3\textwidth}
        \centering
        \includegraphics[width=\textwidth]{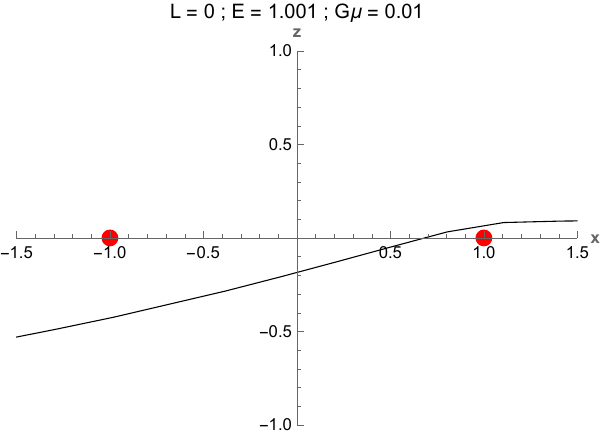}
        \caption{}
        \label{fig:null_scatter_upover_gmu0.01_XZ}
    \end{subfigure}
    \hfill
    \begin{subfigure}[t]{0.3\textwidth}
        \centering
        \includegraphics[width=\textwidth]{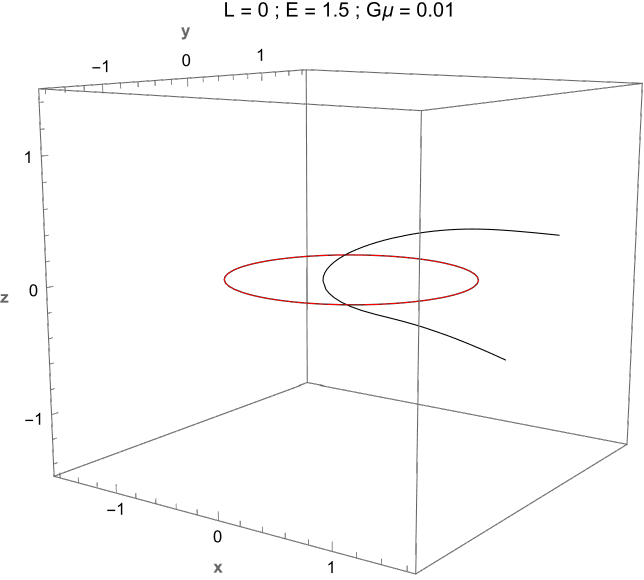}
        \caption{}
        \label{fig:null_scatter_upover_back_gmu0.01_3D}
    \end{subfigure}
    \hfill
   \begin{subfigure}[t]{0.3\textwidth}
        \centering
        \includegraphics[width=\textwidth]{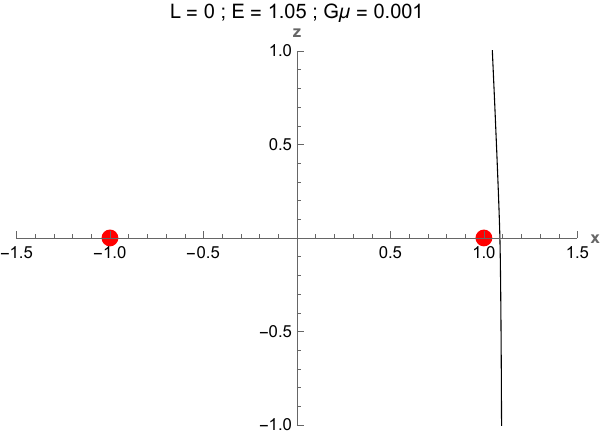}
        \caption{}
        \label{fig:null_scatter_fromz_gmu0.001_XZ1}
    \end{subfigure}
\caption{Unbound null trajectories passing through the interior of the vorton ($r<R$).}
\label{fig:null_scattering_upover}
\end{figure}

The null trajectories exhibit several interesting features. In some cases, a photon can pass through the vorton’s core and be strongly deflected back toward its source, as shown in Fig.~\ref{fig:null_scatter_upover_back_gmu0.01_3D}. This strong bending suggests the possibility of retrolensing, where light is redirected back toward the observer due to the strong gravitational field of the vorton \cite{Holz:2002uf}. Since retrolensing arises in the strong-deflection regime, a full field-theoretic treatment would be necessary for a quantitative analysis.

\begin{figure}[H] 
\centering
    \begin{subfigure}[t]{0.32\textwidth}
        \centering
        \includegraphics[width=\textwidth]{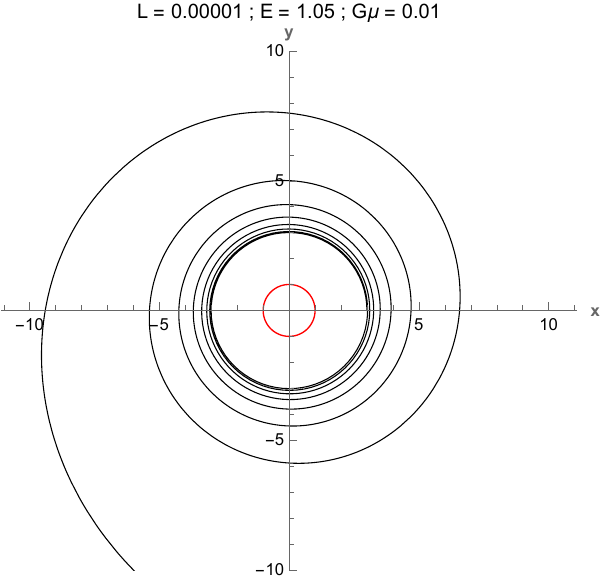}
        \caption{}
        \label{fig:null_ecape}
    \end{subfigure}
    \hfill
    \begin{subfigure}[t]{0.32\textwidth}
        \centering
        \includegraphics[width=\textwidth]{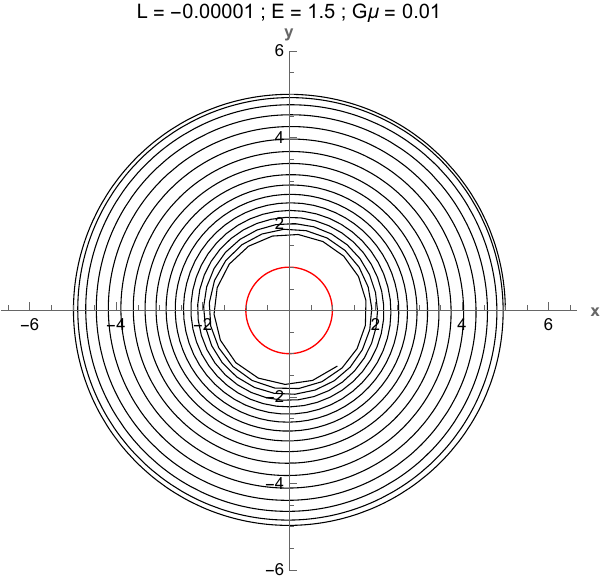}
        \caption{}
        \label{fig:null_terminating}
    \end{subfigure}
    \hfill
    \begin{subfigure}[t]{0.32\textwidth}
        \centering
        \includegraphics[width=\textwidth]{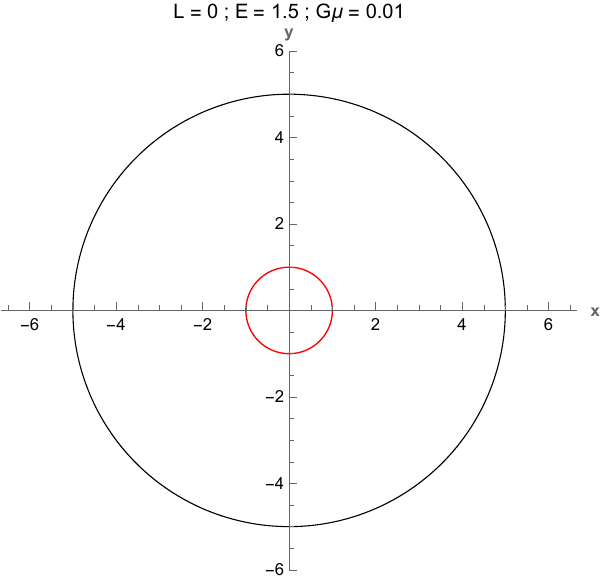}
        \caption{}
        \label{fig:null_circular}
    \end{subfigure}
\caption{Null geodesics around a chiral vorton with $G\mu = 0.01$, showing: (a) an escape orbit, (b) a terminating orbit, and (c) a perfectly circular orbit.}
\label{fig:null_escspcirc}
\end{figure}

Alternatively, as illustrated in Fig.~\ref{fig:null_scatter_upover_gmu0.01_XY}, a photon may thread through the vorton, acquire angular momentum from the frame-dragging region, and emerge below its initial trajectory. This suggests the formation of a double images when the vorton is viewed edge-on. Other cases with a nonzero initial velocity in the $z$ direction  can be seen in Figs.~\ref{fig:null_scatter_fromz_gmu0.001_XZ} and \ref{fig:null_scatter_fromz_gmu0.001_XZ1} when a photon gains angular momentum in the direction of the vorton's frame dragging. This gravitational lensing effect makes an observer perpendicular to the vorton's plane see a distorted background image. These behaviors are consistent with the findings of~\cite{putra2024gravitationalfieldlensingcircular}, which showed that a chiral vorton can distort background light and generate double images when viewed along its equatorial plane. Morevover, we also found that photons can undergo circular orbits when $L = 0$ (see Fig.~\ref{fig:null_circular}), whereas for $L\neq 0$ the trajectories become either terminating (Fig.~\ref{fig:null_terminating}) or unbound (Fig.~\ref{fig:null_ecape}).

\subsection{Chaotic Motion Around Vorton}

The trajectory solutions of the geodesic equations~\eqref{geodesicfinalr}-\eqref{geodesicfinalz} reveal that the dynamics of massive test-particles exhibit a high sensitivity to initial conditions, particularly when the motion departs from the equatorial plane, $z\neq 0$, as shown in Fig.~\ref{fig:crownorbit}. A significant contributing factor is the inseparability of the radial ($r$) and vertical ($z$) coordinates in the metric functions $\nu(r,z)$ (Eq.~\eqref{eq:numetric}) and $A(r,z)$ (Eq.~\eqref{eq:Ametric}), which prevents the complete separation of variables. This non-separability could lead to chaotic motion, similar to what happens to a Kerr black hole immersed in external fields~\cite{PhysRevD.109.064042_geod_motion_swirl_universe,Cao:2024pdb_chaotic_motion_swirl_univer}.

To probe the presence of chaos in our system, we employ the {\it Poincar$\text{\'e}$ surface of section} (PSOS) technique, a standard diagnostic in Hamiltonian dynamical systems. It is essentially a two-dimensional slice of the phase space of the solutions of the equations of motions. A regular motion corresponds to identifiable island-structures (Kolmogorov–Arnold–Moser or KAM tori), while chaotic motion is characterized by apparently random points scattered throughout region~\cite{Zelenka:2019nyp_KAMexp}. By analyzing how the patterns on the section depend on parameters, such as the string tension $G\mu$ and the initial conditions, we gain insight into the existence of integrable regimes and its transition to the chaotic ones around a vorton. 

\begin{figure}[H] 
\centering
    \begin{subfigure}[t]{0.32\textwidth}
        \centering
        \includegraphics[width=\textwidth]{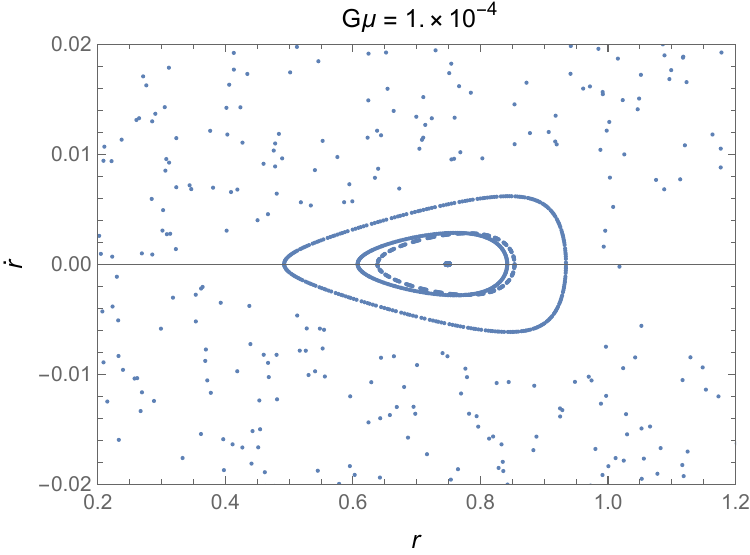}
        \caption{}
        \label{fig:PSOS_Crown_Gmu0.0001}
    \end{subfigure}
    \begin{subfigure}[t]{0.32\textwidth}
        \centering
        \includegraphics[width=\textwidth]{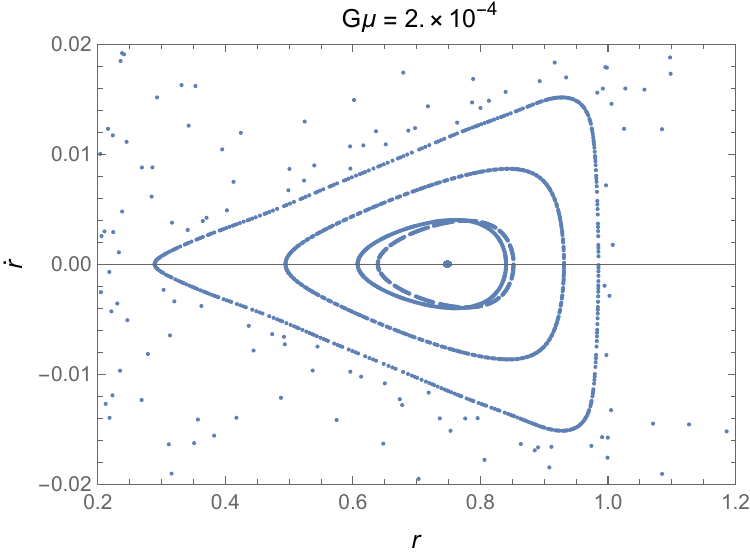}
        \caption{}
        \label{fig:PSOS_Crown_Gmu0.0002}
    \end{subfigure}
    \begin{subfigure}[t]{0.32\textwidth}
        \centering
        \includegraphics[width=\textwidth]{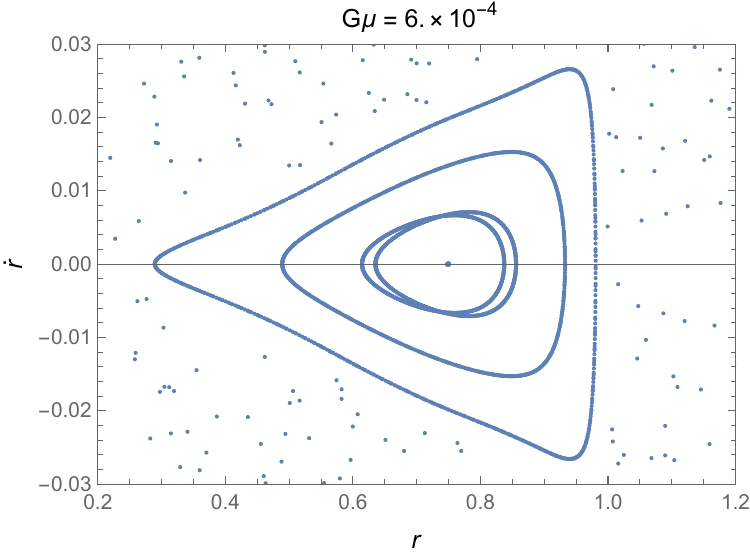}
        \caption{}
        \label{fig:PSOS_Crown_Gmu0.0006}
    \end{subfigure}
\caption{PSOS for a particle with initial $r = 0.6$–$1.1$ (step $0.1$), $z = 1$, $L = 0.001$, and $\dot{r} = \dot{z} = 0$, around a vorton with $R = 1$ and $G\mu = 1\times10^{-4}$ (left), $2\times10^{-4}$ (middle), $6\times10^{-4}$ (right).}    
\label{fig:PSOS_Crown}
\end{figure}

\begin{figure}[H] 
\centering
    \begin{subfigure}[t]{0.32\textwidth}
        \centering
        \includegraphics[width=\textwidth]{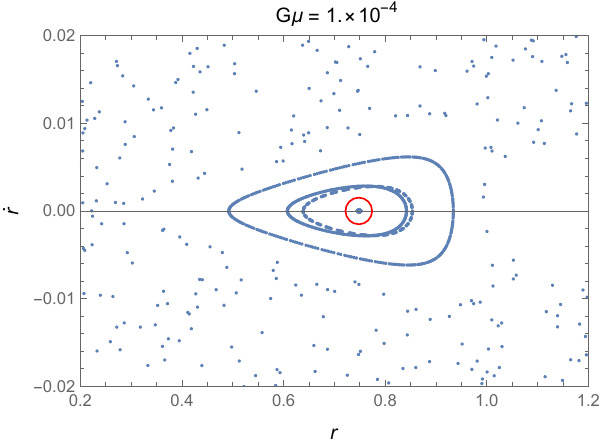}
        \caption{}
        \label{fig:PSOS_Crown_Gmu0.0001_2}
    \end{subfigure}
    \hfill
    \begin{subfigure}[t]{0.32\textwidth}
        \centering
        \includegraphics[width=\textwidth]{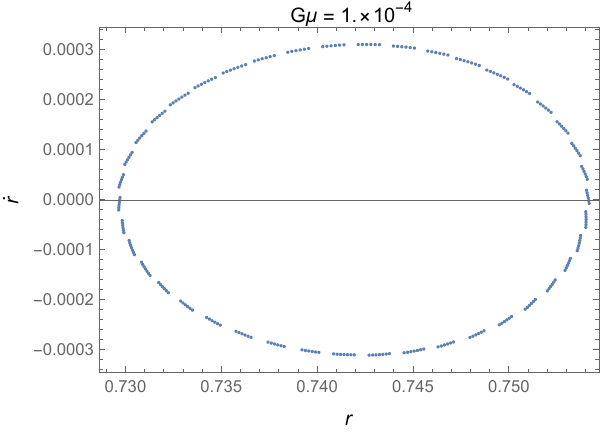}
        \caption{}
        \label{fig:PSOS_Crown_Gmu0.0001_Closeup}
    \end{subfigure}
    \hfill
    \begin{subfigure}[t]{0.25\textwidth}
        \centering
        \includegraphics[width=\textwidth]{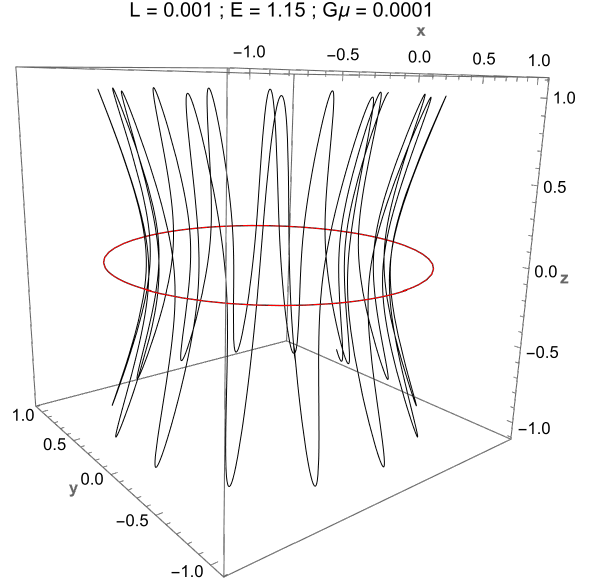}
        \caption{}
        \label{fig:PSOS_Crown_Gmu0.0001_Orbit}
    \end{subfigure}
\caption{PSOS of Fig.~\ref{fig:PSOS_Crown_Gmu0.0001} with a point highlighted in red. Middle: zoom on the highlighted point, showing a quasi-periodic orbit. Right: corresponding orbit trajectory.}    
\label{fig:PSOS_Closeup_orbit_View}
\end{figure}

In Fig.~\ref{fig:PSOS_Crown} we show the PSOS obtained by varying the initial radial position $r$ from $0.6$ to $1.1$ in increments of $0.1$. The phase-space structure shows a sequence of well-defined islands associated with quasi-periodic orbits. Increasing the tension $G\mu$ while keeping the same set of initial conditions widens the region of stability and width of the islands, as can be seen in Figs.~\ref{fig:PSOS_Crown_Gmu0.0001} and \ref{fig:PSOS_Crown_Gmu0.0006}. As can easily be seen, each island has a single point at the center, which would naively suggest the presence of a periodic orbit. However, closer inspection reveals that it is not the case. Fig.~\ref{fig:PSOS_Closeup_orbit_View} shows that the points actually correspond to quasi-periodic orbits rather than a closed periodic path.

\begin{figure}[H] 
\centering
    \begin{subfigure}[t]{0.32\textwidth}
        \centering
        \includegraphics[width=\textwidth]{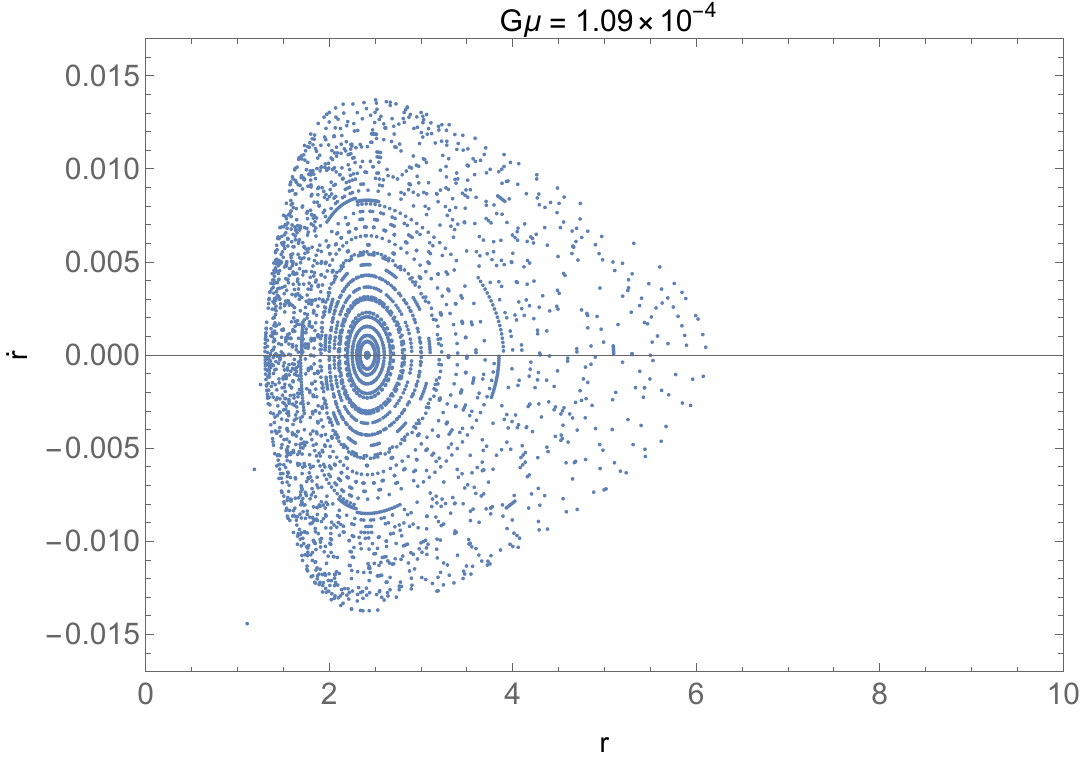}
        \caption{}
        \label{fig:PSOS_Planar_Gmu0.00010}
    \end{subfigure}
    \hfill
    \begin{subfigure}[t]{0.32\textwidth}
        \centering
        \includegraphics[width=\textwidth]{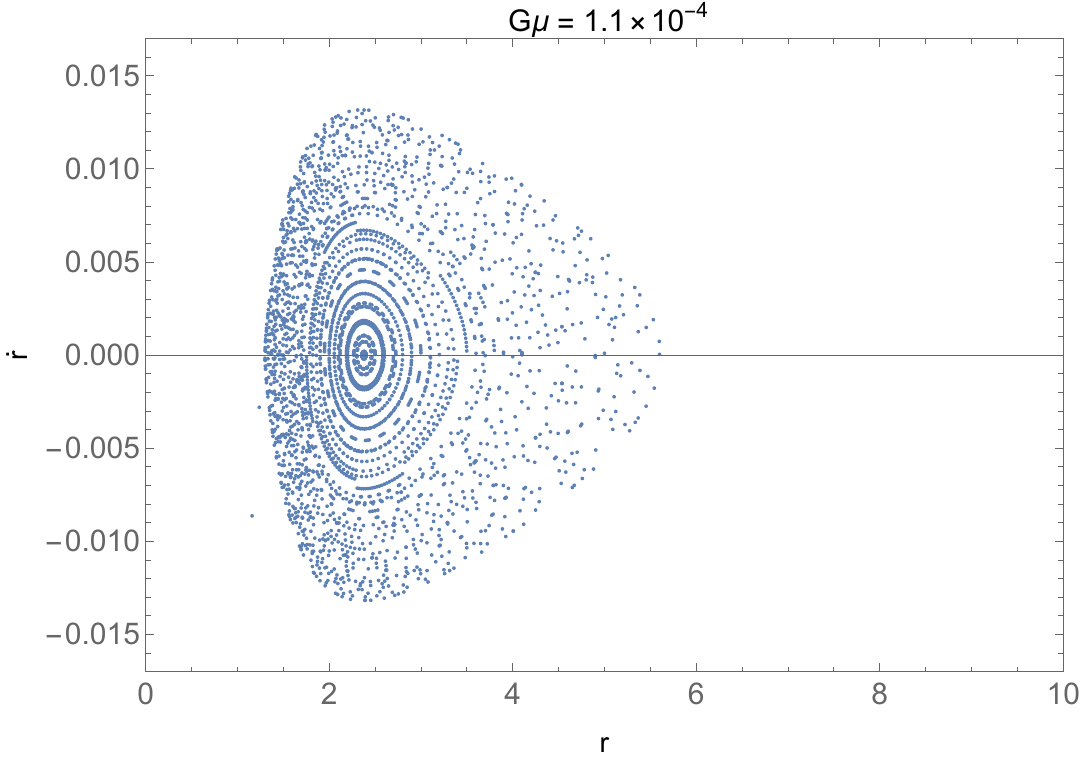}
        \caption{}
        \label{fig:PSOS_Planar_gmu0.00011}
    \end{subfigure}
    \begin{subfigure}[t]{0.32\textwidth}
        \centering
        \includegraphics[width=\textwidth]{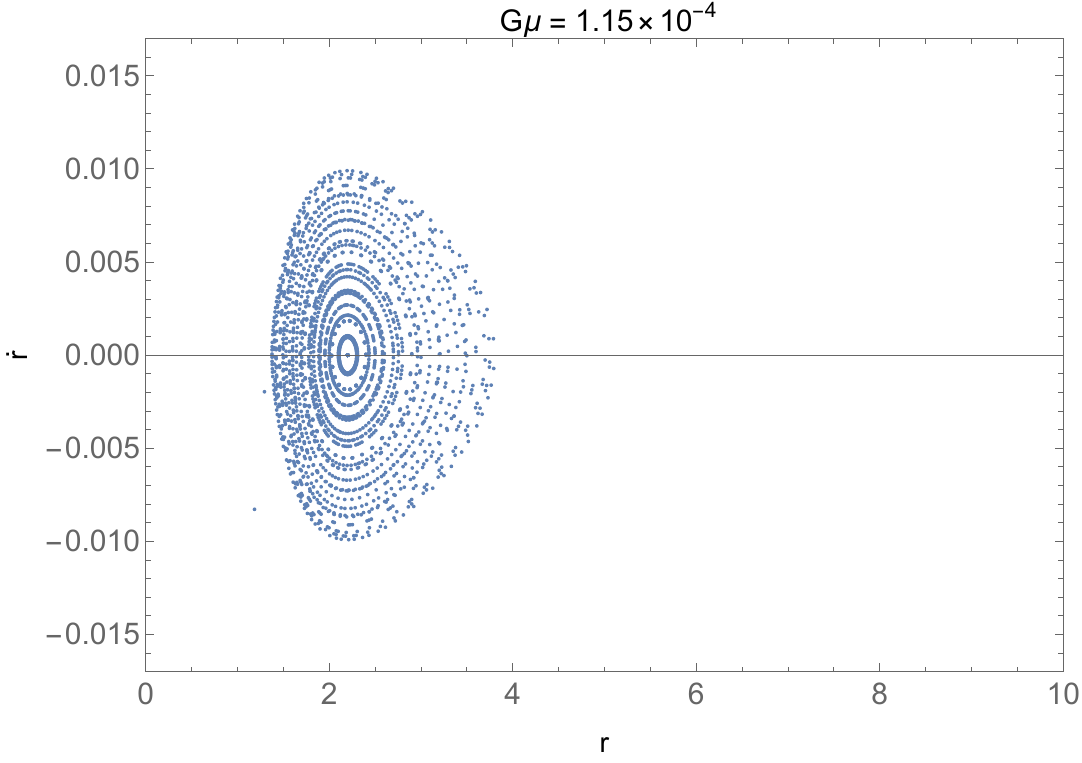}
        \caption{}
        \label{fig:PSOS_Planar_gmu0.000115}
    \end{subfigure}
    \medskip
    \begin{subfigure}[t]{0.32\textwidth}
        \centering
        \includegraphics[width=\textwidth]{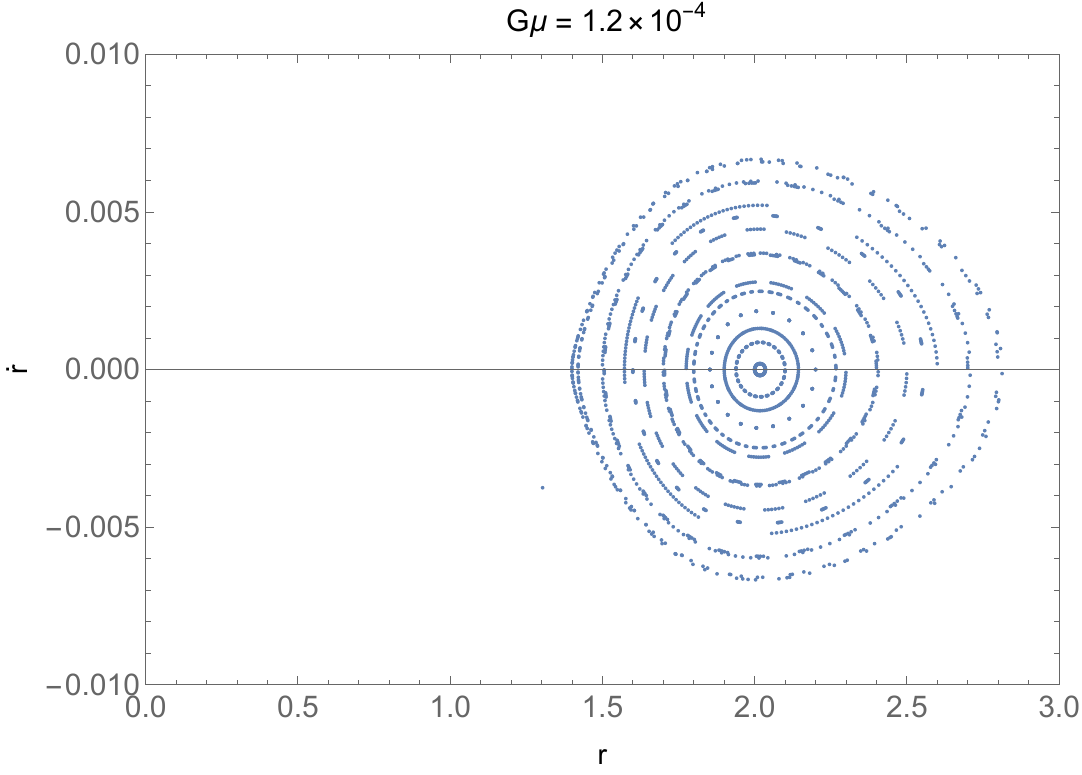}
        \caption{}
        \label{fig:PSOS_Planar_gmu0.00012}
    \end{subfigure}
    \hfill
    \begin{subfigure}[t]{0.32\textwidth}
        \centering
        \includegraphics[width=\textwidth]{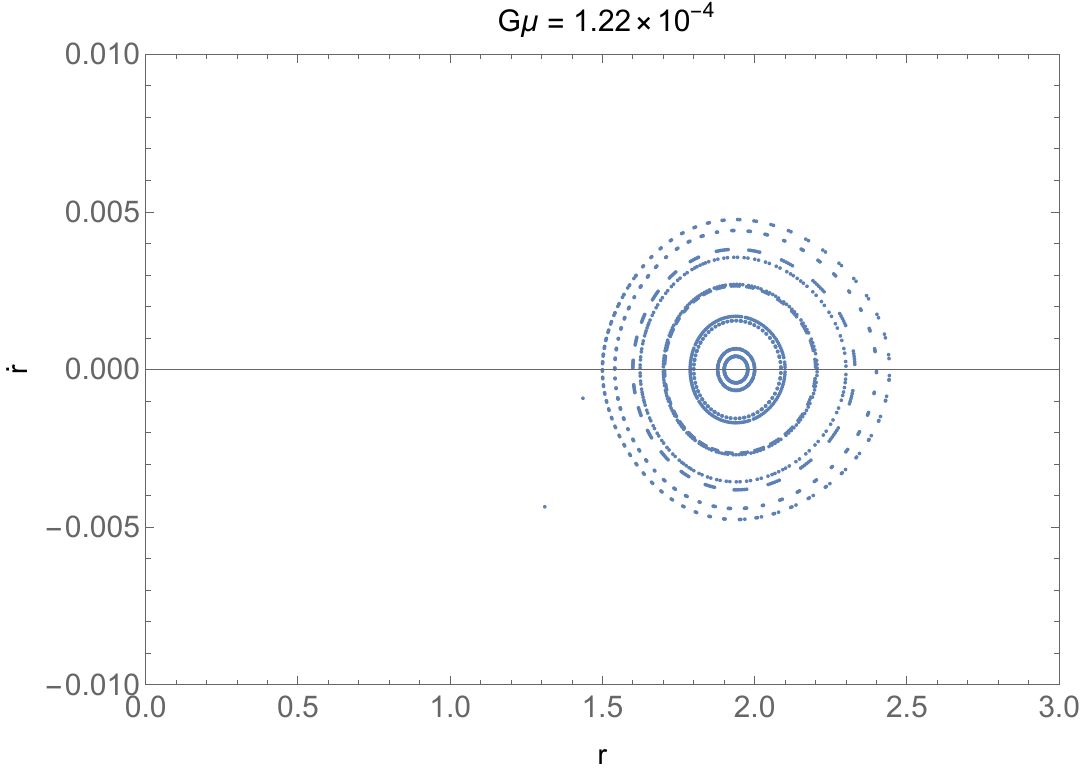}
        \caption{}
        \label{fig:PSOS_Planar_gmu0.000122}
    \end{subfigure}
    \hfill
    \begin{subfigure}[t]{0.32\textwidth}
        \centering
        \includegraphics[width=\textwidth]{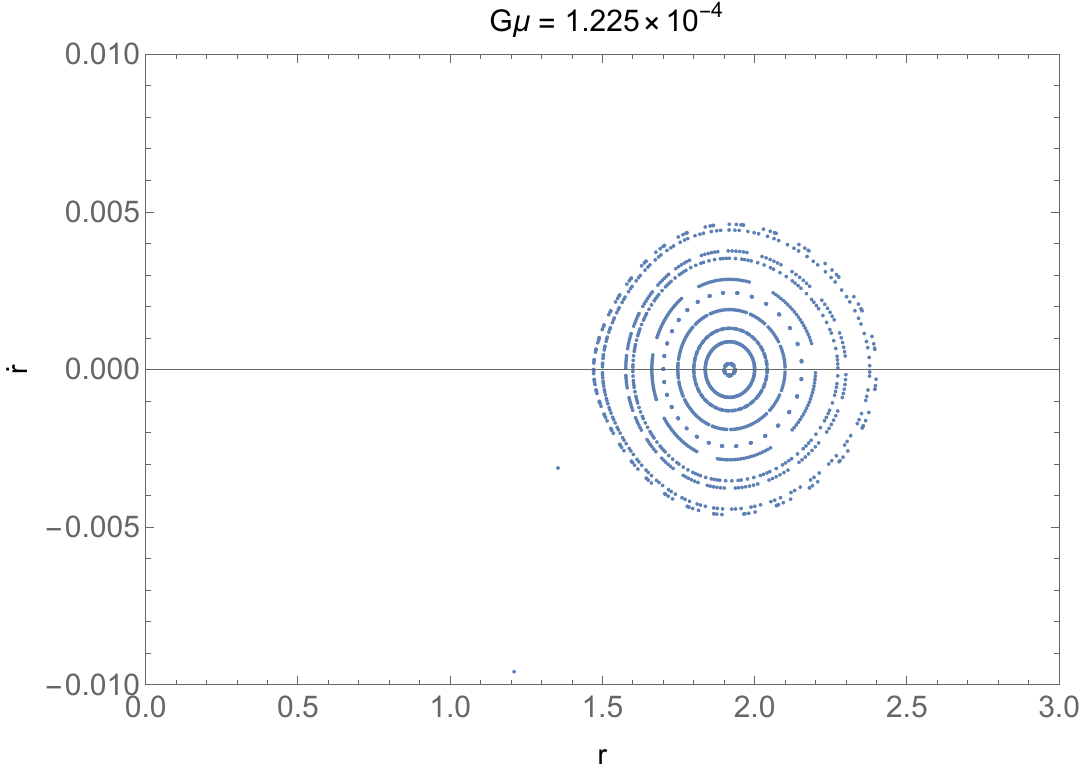}
        \caption{}
        \label{fig:PSOS_Planar_gmu0.0001225}
    \end{subfigure}
    \medskip
    \begin{subfigure}[t]{0.32\textwidth}
        \centering
        \includegraphics[width=\textwidth]{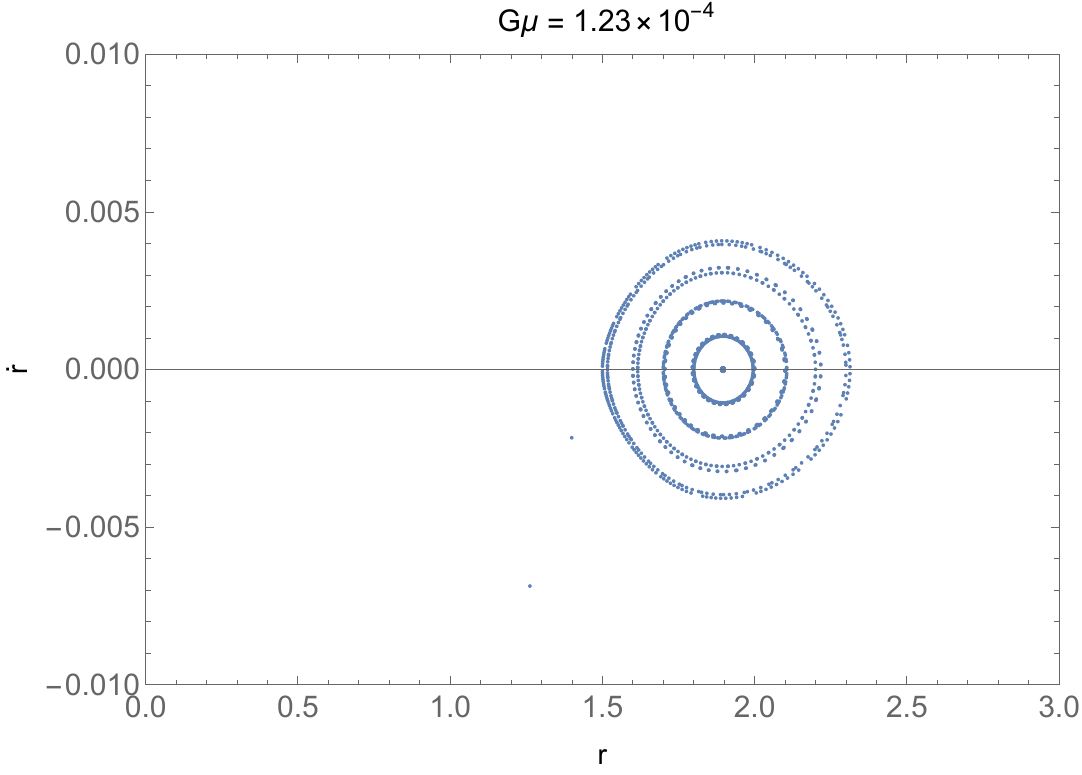}
        \caption{}
        \label{fig:PSOS_Planar_gmu0.000123}
    \end{subfigure}
    \hfill
    \begin{subfigure}[t]{0.32\textwidth}
        \centering
        \includegraphics[width=\textwidth]{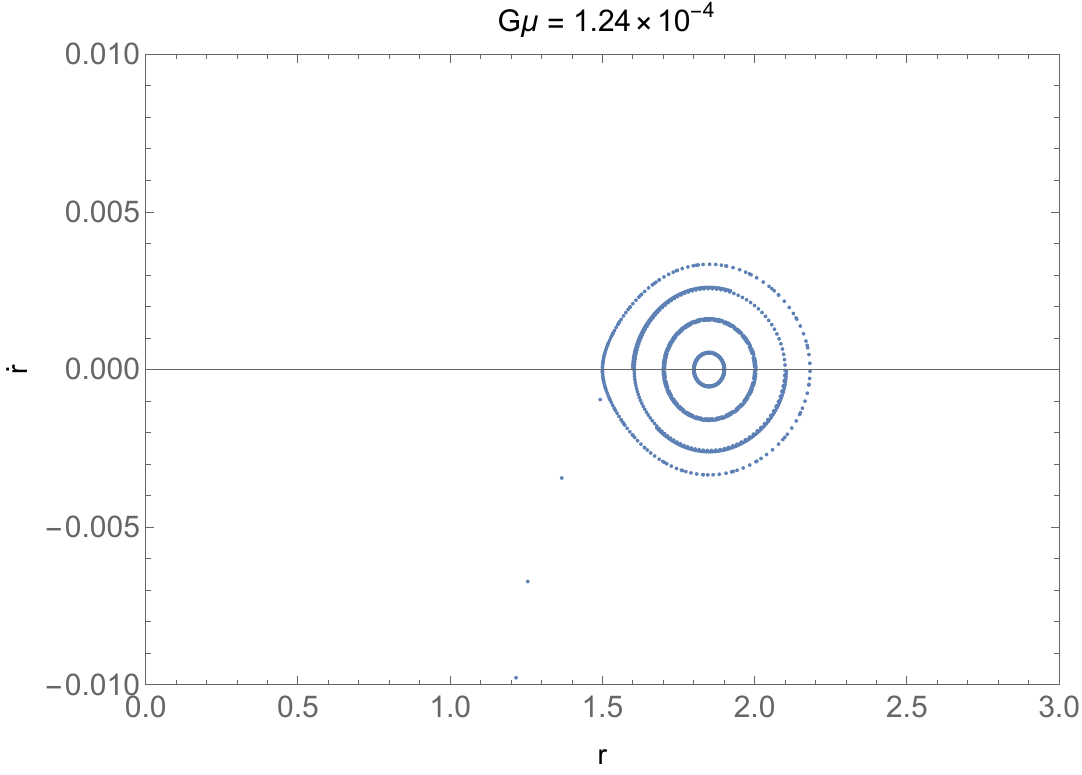}
        \caption{}
        \label{fig:PSOS_Planar_gmu0.000124}
    \end{subfigure}
    \hfill
    \begin{subfigure}[t]{0.32\textwidth}
        \centering
        \includegraphics[width=\textwidth]{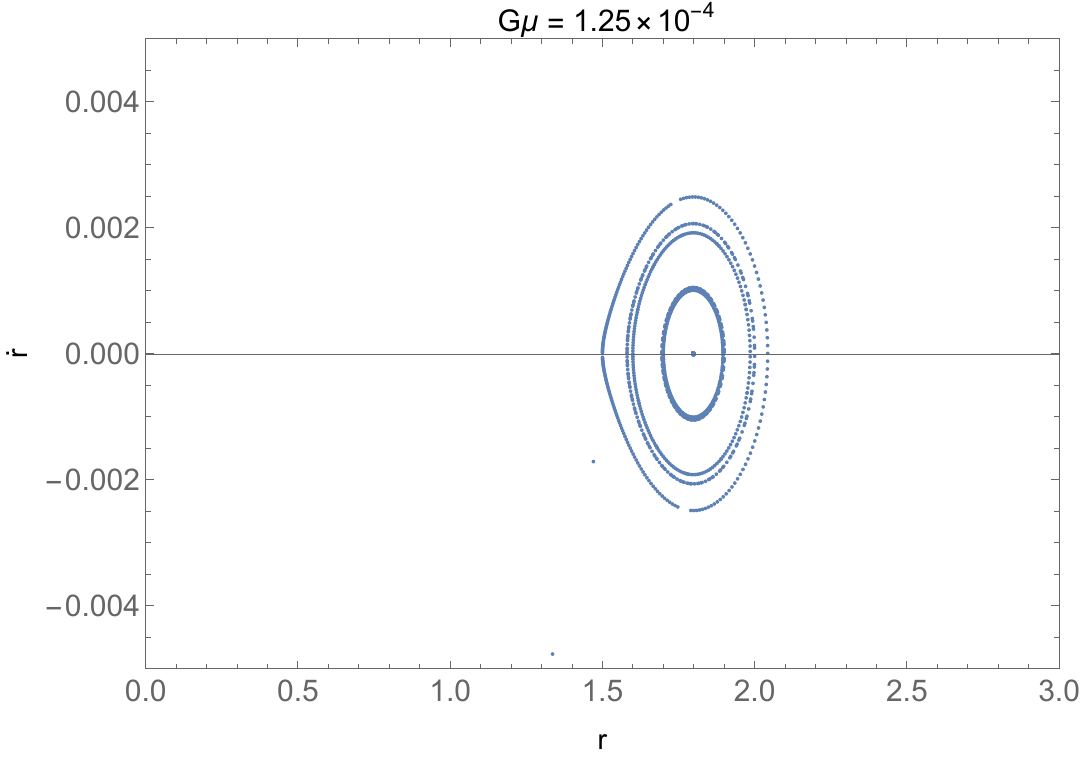}
        \caption{}
        \label{fig:PSOS_Planar_gmu0.000125}
    \end{subfigure}
\caption{PSOS for planar orbits ($z=0$) with $L=0.066$, $E=1.15$, sampled over $0 < r < 10$ ($\delta r=0.1$) for different $G\mu$ values.}
\label{fig:PSOS_Planar}
\end{figure}

In Fig.~\ref{fig:PSOS_Planar} we show a set of PSOSs for equatorial motion with $L = 0.066$. In contrast with the non-planar case, here the effect of the tension $G\mu$ is to progressively reduce the accessible domain around the vorton. The remaining stable region appears as a set of islands centered roughly in the interval $1 \lesssim r \lesssim 4$, corresponding to a potential well in which bound orbits are stable. As $G\mu$ increases, some of the invariant tori break up into chains of Birkhoff islands, signalling the onset of local chaos \cite{Chen:2016tmr_kammerge}, as illustrated in Figs.~\ref{fig:PSOS_Planar_gmu0.00012}-\ref{fig:PSOS_Planar_gmu0.0001225}. For slightly larger values of the tension, these chains reconnect into smooth tori again (see Figs.~\ref{fig:PSOS_Planar_gmu0.000123}-\ref{fig:PSOS_Planar_gmu0.000125}), indicating a partial restoration of regularity in phase space. This behavior reflects the sensitive dependence of the orbital structure on both the tension parameter and the choice of conserved quantities, which implies that the geodesics around vorton contains both integrable and chaotic sectors, depending on parameter regime.

\section{Gyroscope Spin Precession}

One notable prediction of general relativity is the precession of gyroscopes. A gyroscope, with its intrinsic angular momentum, tends to maintain the orientation of its spin axis. Under the influence of a gravitational field, however, this axis undergoes precession. The effect arises in two forms: {\it geodetic precession}, caused by the curvature of spacetime due to a central mass, and {\it Lense–Thirring precession}, induced by the rotation of that mass.

In the case of a vorton, both effects are expected to display nontrivial spatial dependence due to the ring topology and the associated gravitomagnetic field. To compute the precession rate, we will heavily follow the general formalism developed in~\cite{Chakraborty_2017_distinguishingKerrandNaked}. A test gyroscope attached to a stationary observer moving through a stationary spacetime possesses a general timelike Killing vector $K$ which can be in coordinate basis as
\begin{equation}
K = \partial_o + \omega \partial_c,
\end{equation}
where $\omega$ is the angular velocity of the observer, $\partial_0$ ($\partial_c$) is the timelike (spacelike) component, normalized so that $K^0=1$. Its corresponding dual co-vector is
\begin{equation}
   \tilde{K}=g_{0\nu}dx^{\nu}+\Omega g_{c\gamma}dx^{\gamma}. 
\end{equation}
The spin of the gyroscope would undergo Fermi-Walker transport along the four-velocity given by
\begin{equation} \label{timelikevectoru}
    u = \frac{1}{\sqrt{-K^2}}K.
\end{equation}

The general spin precession frequency $\Omega_p$ is then the rescaled vorticity field of the observer congruence and expressed as~\cite{Chakraborty_2017_distinguishingKerrandNaked,Straumann:2013spu}
\begin{equation}
    \begin{split}
        \tilde{\Omega}_p & = \frac{1}{2K^2}\ast \left(\tilde{K}\wedge d\tilde{K}\right), \\
        (\Omega_p)_\mu &=\frac{1}{2K^2}\eta_{\mu}\ ^{\alpha\beta\lambda} K_\alpha \partial_\beta K_\lambda,
    \end{split}
\end{equation}
where $\tilde{\Omega}_p$ is the co-vector of $\Omega_p$, operator $*$ represents the Hodge dual, and $\eta_{\mu}\ ^{\alpha\beta\lambda}$ is the volume-form components in the spacetime. Inserting $K$, $\Omega_p$ is obtained as~\cite{Chakraborty_2017_distinguishingKerrandNaked}
\begin{eqnarray} \label{gyroprecessGeneral}
     (\Omega_p)_l &=&\frac{\varepsilon_{ckl}}{2\sqrt{-g}\left(1+ 2\omega \frac{g_{0c}}{g_{00}} +\omega^2 \frac{g_{cc}}{g_{00}} \right)} \\
    & &\left[\left(g_{0c,k} - \frac{g_{0c}}{g_{00}}g_{00,k}   \right) + \omega\left(g_{cc,k} - \frac{g_{cc}}{g_{00}}g_{00,k}   \right) + \omega^2 \left(\frac{g_{0c}}{g_{00}}g_{cc,k} - \frac{g_{cc}}{g_{00}}g_{0c,k}   \right)    \right] \partial_l.
\end{eqnarray}
One can see that Eq.~\eqref{gyroprecessGeneral} is a general precession frequency applicable for any stationary non-static spacetime. In this work, we investigate the modulus of the precession frequency $|\vec{\Omega_p}|$, which for our case with coordinates $\{t,r,\phi,z\}$ is
\begin{equation} \label{modulusprecession}
     |\Vec{\Omega}_p| = \frac{1}{C}\sqrt{(\Omega_r)^2 + (\Omega_z)^2},
\end{equation}
where 
\begin{eqnarray} \label{componentsprecessvector}
        C &=& 2\sqrt{-g}\left(1+ 2\omega \frac{g_{t\phi}}{g_{tt}} +\omega^2 \frac{g_{\phi\phi}}{g_{tt}} \right),\\
       \Omega_r &=& -\sqrt{g_{rr}}\biggr[\left(g_{t\phi,z} - \frac{g_{t\phi}}{g_{tt}}g_{tt,z}   \right) + \omega\left(g_{\phi\phi,z} - \frac{g_{\phi\phi}}{g_{tt}}g_{tt,z}   \right) + \omega^2 \left(\frac{g_{t\phi}}{g_{tt}}g_{\phi\phi,z} - \frac{g_{\phi\phi}}{g_{tt}}g_{t\phi,z}   \right)    \biggr],\\
        \Omega_z &=& \sqrt{g_{zz}}\biggr[\left(g_{t\phi,r} - \frac{g_{t\phi}}{g_{tt}}g_{tt,r}   \right) + \omega\left(g_{\phi\phi,r} - \frac{g_{\phi\phi}}{g_{tt}}g_{tt,r}   \right) + \omega^2 \left(\frac{g_{t\phi}}{g_{tt}}g_{\phi\phi,r} - \frac{g_{\phi\phi}}{g_{tt}}g_{t\phi,r}   \right)    \biggr].
\end{eqnarray}

\subsection{Lense-Thirring Precession ($\omega=0$)}
Taking $\omega = 0$, the gyroscopes are fixed in space and values of the precession are taken. This effectively probes the precession values induced by the rotation of the vorton or more commonly known as the Lense-Thirring (LT) precession. Eqs.~\eqref{componentsprecessvector} become
\begin{eqnarray} \label{componentsprecessvectorLT}
    C &=& 2\sqrt{-g},\nonumber\\
    \Omega_r &=& -\sqrt{g_{rr}}\left(g_{t\phi,z} - \frac{g_{t\phi}}{g_{tt}}g_{tt,z}   \right),\nonumber\\
    \Omega_z &=& \sqrt{g_{zz}}\left(g_{t\phi,r} - \frac{g_{t\phi}}{g_{tt}}g_{tt,r}   \right).
\end{eqnarray}
Eq.~\eqref{componentsprecessvectorLT} can also be obtained by choosing a timelike killing vector field of $K = \partial_0$ so as to not introduce rotational movement to the observers attached to the gyroscopes. We can then insert the metric components and find that the Lense-thirring vector ($\Vec{\Omega}_p^{LT}$) components are
\begin{eqnarray}\label{LTprecessionwithmetric}
    \frac{\Omega^{LT}_r}{C} &=& \frac{e^{-\nu}r}{e^{4\nu}-r^2A^2}\left[A_{,z}\left(e^{4\nu}-r^2A^2 \right) - 4e^{4\nu}A \nu_{,z}  \right],\nonumber\\
    \frac{\Omega^{LT}_z}{C} &=& \frac{e^{-\nu}}{e^{4\nu}-r^2A^2}\left[A_{,r}\left(2r\nu_{,r} - 1 \right)   -A_{,r}\left(e^{4\nu}-r^2A^2 \right)  \right].
\end{eqnarray}
From Eq.~\eqref{modulusprecession}, this yields the magnitude of Lense-Thirring precession of gyroscopes around the vorton of
\begin{eqnarray}\label{LTprecessionmagnitude}
        \Omega^{LT}_p &=& \frac{e^{-\nu}}{2\left(e^{4\nu}-r^2A^2\right)}\bigg[ r^2\left\{A_{,z}\left(e^{4\nu}-r^2A^2 \right) - 4e^{4\nu}A \nu_{,z}  \right\}^2\nonumber\\
        &&+ \left\{-2e^{4\nu} A\left(2r\nu_{,r} - 1 \right)   +A_{,r}\left(e^{4\nu}+r^3 A^2 \right)\right\}^2 \bigg]^{\frac{1}{2}}.
\end{eqnarray}

\begin{figure}[H] 
\centering
    \begin{subfigure}[t]{0.38\textwidth}
        \centering
        \includegraphics[width=\textwidth]{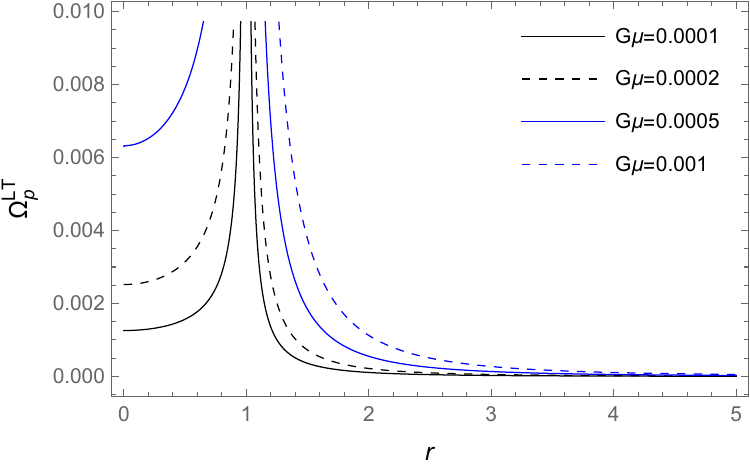}
        \caption{}
        \label{fig:LT_precess_gmueff_r}
    \end{subfigure}
    \begin{subfigure}[t]{0.38\textwidth}
        \centering
        \includegraphics[width=\textwidth]{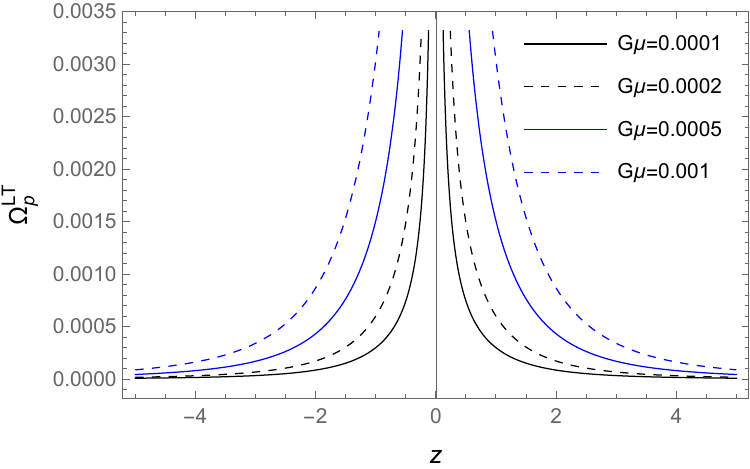}
        \caption{}
        \label{fig:LT_precess_gmueff_z}
    \end{subfigure}
    
    \medskip
    \begin{subfigure}[t]{0.38\textwidth}
        \centering
        \includegraphics[width=\textwidth]{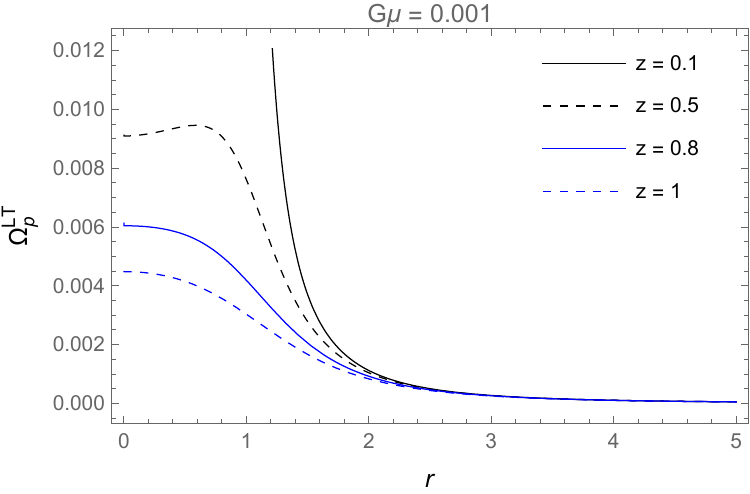}
        \caption{}
        \label{fig:LT_precess_zeff_gmu0001}
    \end{subfigure}
    \begin{subfigure}[t]{0.38\textwidth}
        \centering
        \includegraphics[width=\textwidth]{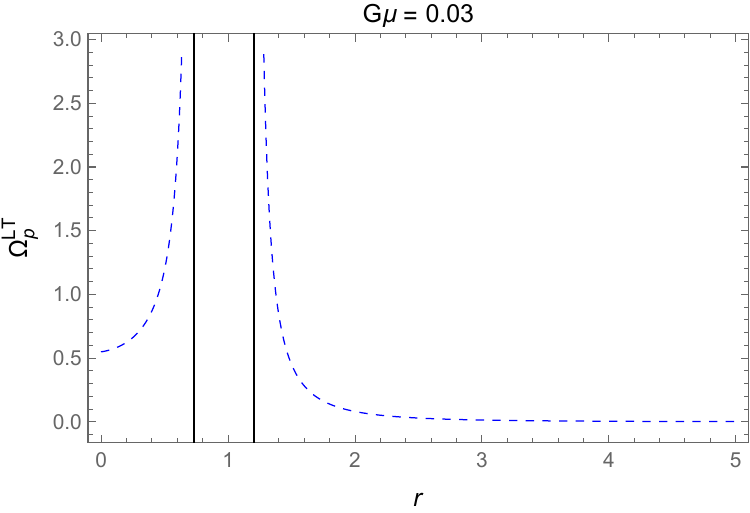}
        \caption{}
        \label{fig:LT_precess_ergoregion_gmu003}
    \end{subfigure}
\caption{Magnitude of LT precession frequency $\Omega_p^{LT}$. Panels (a) and (b) show its radial and $z$ dependence, respectively, for various $G\mu$. (c) Radial profile at fixed $z$ for $G\mu = 0.001$. (d) Ergoregion boundary (solid black) and $\Omega_p^{LT}$ (dashed blue) for $G\mu = 0.03$.}
\label{fig:LTprecessmagnitude}
\end{figure}

The magnitudes of $\Omega_p^{LT}$ shown in Fig.~\ref{fig:LTprecessmagnitude} exhibit a divergence at the boundaries of the ergoregion, as highlighted in Fig.~\ref{fig:LT_precess_ergoregion_gmu003}, and tend to zero as $r \to \infty$. The divergence of $\Omega_p^{LT}$ at $r=R$ indicates that an object transitioning from inside the vorton ring with a positive radial velocity would undergo a sudden increase in precession frequency. Although the weak-field, thin-string approximation used here does not allow a physical crossing of this region, a full-field treatment may yield a similar qualitative behavior. 

Inside the ring ($r < R$), Fig.~\ref{fig:LT_precess_gmueff_z} shows that $\Omega_p^{LT}$ attains a higher magnitude compared to the exterior region ($r > R$). Along the $z$ direction, the distribution shown in Fig.~\ref{fig:LT_precess_zeff_gmu0001} is more symmetric than the corresponding radial one. The ergoregion boundaries are determined by the condition that $g_{tt}$ vanishes,
\begin{equation}
    -e^{2\nu(r,z)} + e^{-2\nu(r,z)}r^2 A(r,z)^2 = 0. 
\end{equation}
For sufficiently small string tension ($G\mu \ll 1$), the ergoregion becomes extremely narrow and practically indistinguishable from the vorton core. 

\begin{equation}
    -e^{2\nu(r,z)} + e^{-2\nu(r,z)}r^2 A(r,z)^2 = 0. 
\end{equation}
The ergoregion is found to practically vanish when taking values of $G\mu \ll 1$ and becomes indistinguishable from the vorton core. 
\subsection{General Gyroscope Precession}

For the general precession case ($\omega \neq 0$), the four-velocity in Eq.~\eqref{timelikevectoru} must remain timelike, which requires
\begin{equation}
    K^2 = g_{00} + 2\omega g_{0c} + \omega^2g_{cc} < 0.
\end{equation}
This constraint restricts the allowable angular velocity $\omega$ to the interval
\begin{equation} \label{constrainsom}
    \omega_-(r,z) < \omega(r,z) < \omega_+(r,z),
\end{equation}
where 
\begin{equation} \label{omegapm}
    \omega_\pm = \frac{-g_{t\phi} \pm \sqrt{g_{t\phi}^2 - g_{tt}g_{\phi\phi}}}{g_{\phi\phi}}.
\end{equation}
Following Refs.~\cite{Wu_hairykerrprecession_2023,Chakraborty_2017_distinguishingKerrandNaked}, we introduce a parameter $k$ with the range of $0<k<1$ such that
\begin{equation}
    \omega = k\,\omega_+ + (1-k)\omega_- = \frac{-g_{t\phi} + (2k -1) \sqrt{g_{tt}^2 - g_{t\phi}g_{\phi\phi}}}{g_{\phi\phi}},
\end{equation}
which, after substituting the explicit metric components, yields
\begin{equation} \label{constraintomegafin}
    \omega = (2k - 1 )\frac{e^{2\nu}}{r} + A.
\end{equation}
This form guarantees that $\omega$ automatically satisfies the constraint in  Eq.~\eqref{constrainsom}. 

\begin{figure}[H] 
\centering
    \begin{subfigure}[t]{0.32\textwidth}
        \centering
        \includegraphics[width=\textwidth]{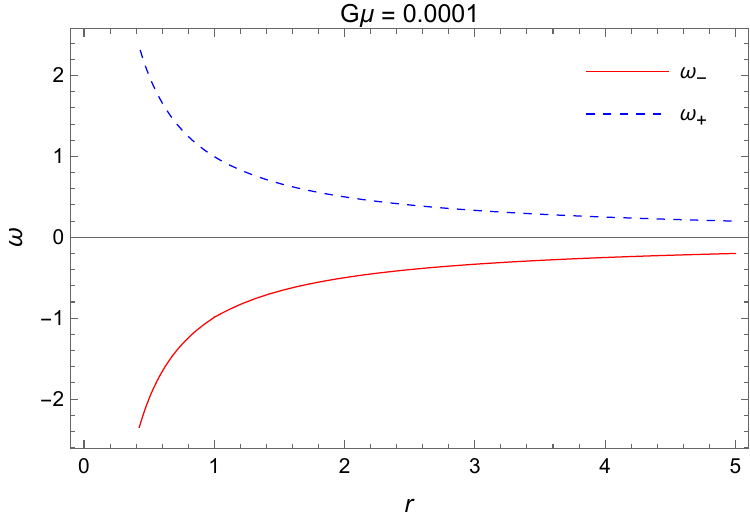}
        \caption{}
        \label{fig:omwminwmax_gmu_00001}
    \end{subfigure}
    \hfill
    \begin{subfigure}[t]{0.32\textwidth}
        \centering
        \includegraphics[width=\textwidth]{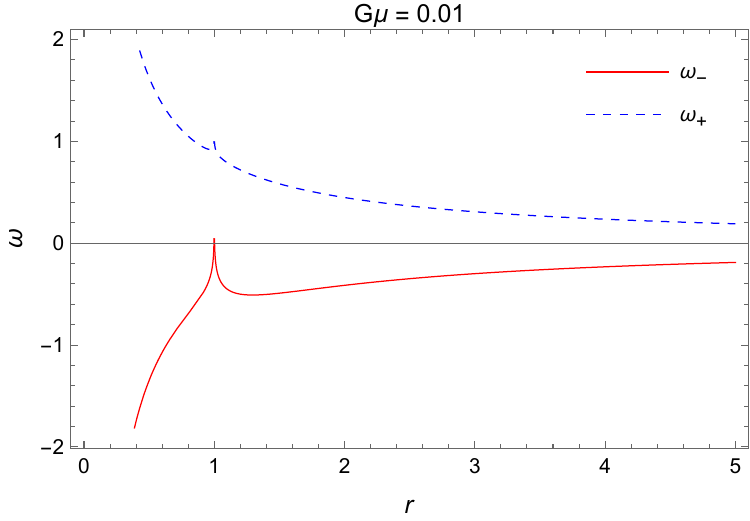}
        \caption{}
        \label{fig:omwminwmax_gmu_001}
    \end{subfigure}
    \hfill
    \begin{subfigure}[t]{0.32\textwidth}
        \centering
        \includegraphics[width=\textwidth]{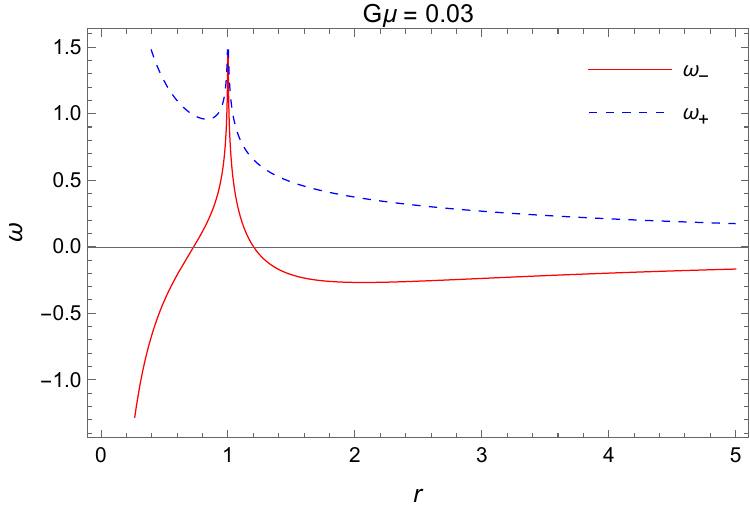}
        \caption{}
        \label{fig:omwminwmax_gmu_003}
    \end{subfigure}
\caption{Plots of the two orbital frequency branches, $\omega_-$ (red) and $\omega_+$ (dashed-blue), shown for different values of the string tension parameter $G\mu$, (a) $G\mu = 0.0001$, (b) $G\mu = 0.01$, (c) $G\mu = 0.03$ on the equatorial plane.}
\label{fig:valuesomegawminwmax}
\end{figure}

The values of $\omega_{\pm}$ are shown in Fig.~\ref{fig:valuesomegawminwmax} for three representative choices of the string tension parameter, $G\mu = 10^{-4},\  10^{-2},$ and $0.03$. The two branches $\omega_-$ and $\omega_+$ have opposite signs because they correspond to orbital motion in opposite directions (clockwise and couunterclockwise, respectively), which is clearly visible in Figs.~\ref{fig:omwminwmax_gmu_00001}, \ref{fig:omwminwmax_gmu_001}, and \ref{fig:omwminwmax_gmu_003}. As $G\mu$ increases, the gravitational field becomes stronger near the vorton and extends over a larger effective region, modifying the allowed values of $\omega_{\pm}$ in accordance with Eq.~\ref{omegapm}. This trend is evident in Figs.~\ref{fig:omwminwmax_gmu_001} and \ref{fig:omwminwmax_gmu_003}, where both $\omega_{\pm}$ increase for $G\mu = 0.01$ and $0.03$. For $G\mu = 0.03$, both branches become strictly positive and singular near the vorton core (within the ergoregion). This is due to the clockwise frame-dragging effect: maintaining a timelike trajectory forces $\omega_{\pm}$ to be positive as they counteract the dragging. The frequencies become singular near $R=1$, signaling the onset of spacelike behavior and indicating the breakdown of the thin-string approximation once an ergoregion is present.

\begin{figure}[H] 
\centering
    \begin{subfigure}[t]{0.32\textwidth}
        \centering
        \includegraphics[width=\textwidth]{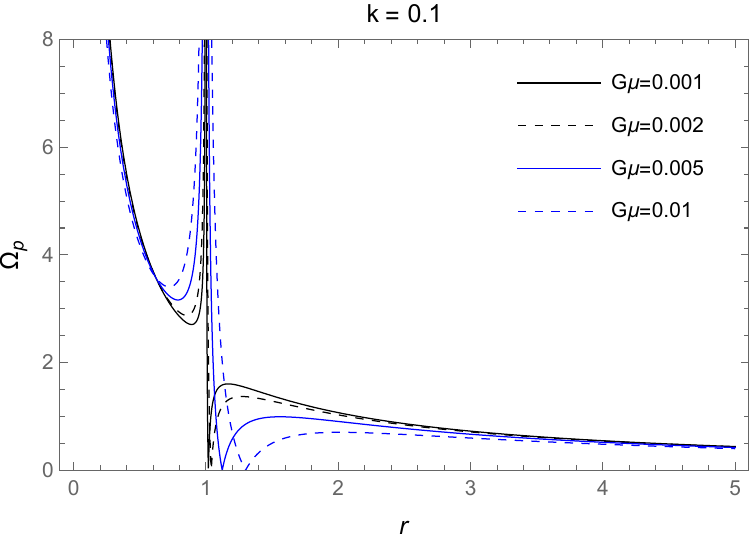}
        \caption{}
        \label{fig:gen_precess_gmueff_r_k0.1}
    \end{subfigure}
    \hfill
    \begin{subfigure}[t]{0.32\textwidth}
        \centering
        \includegraphics[width=\textwidth]{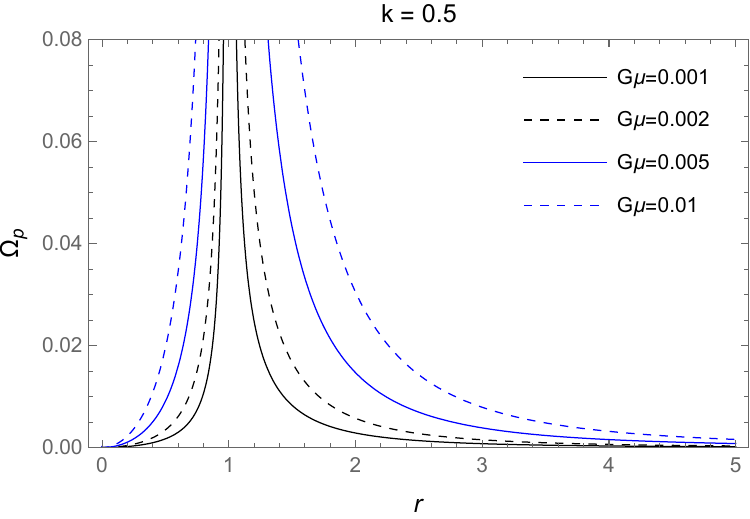}
        \caption{}
        \label{fig:gen_precess_gmueff_r_k0.5}
    \end{subfigure}
    \hfill
    \begin{subfigure}[t]{0.32\textwidth}
        \centering
        \includegraphics[width=\textwidth]{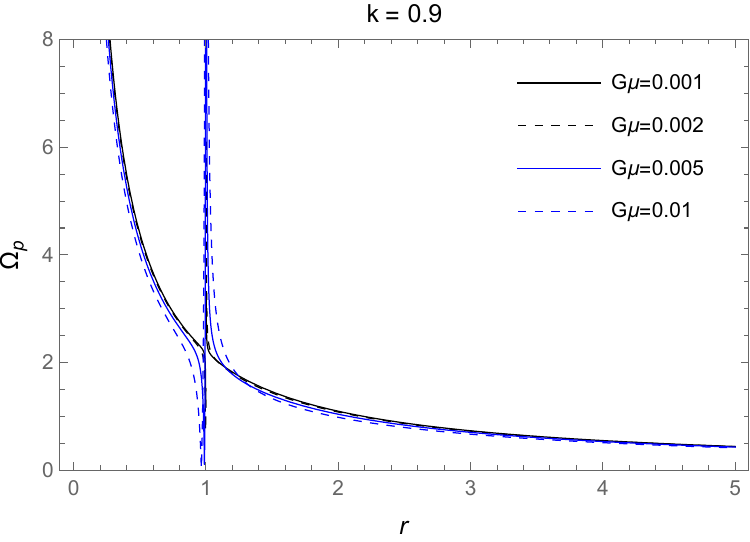}
        \caption{}
        \label{fig:gen_precess_gmueff_r_k0.9}
    \end{subfigure}
    \caption{Magnitude of $\Omega_p$ as a function of $r$ for several values of $G\mu$ $(0.001,0.002,0.005,0.01)$. Results are shown for (a) $k=0.1$, (b) $k=0.5$, and (c) $k=0.9$.}
\label{fig:gen_precess_gmueff_r}
\end{figure}

\begin{figure}[H] 
\centering
    \begin{subfigure}[t]{0.32\textwidth}
        \centering
        \includegraphics[width=\textwidth]{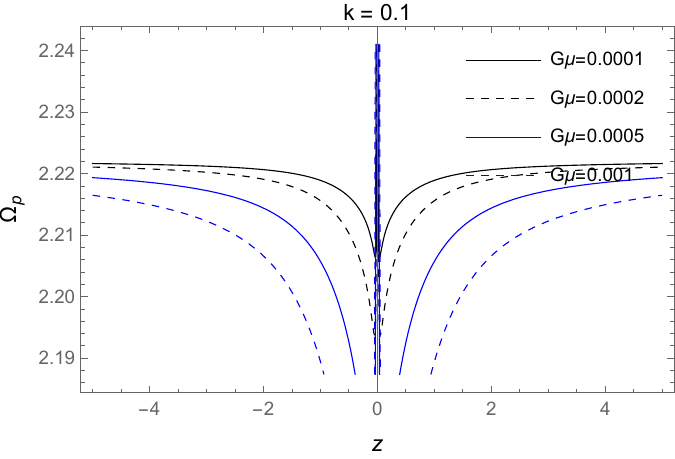}
        \caption{}
    \label{fig:gen_precess_gmueff_z_k0.1}
    \end{subfigure}
    \hfill
    \begin{subfigure}[t]{0.32\textwidth}
        \centering
        \includegraphics[width=\textwidth]{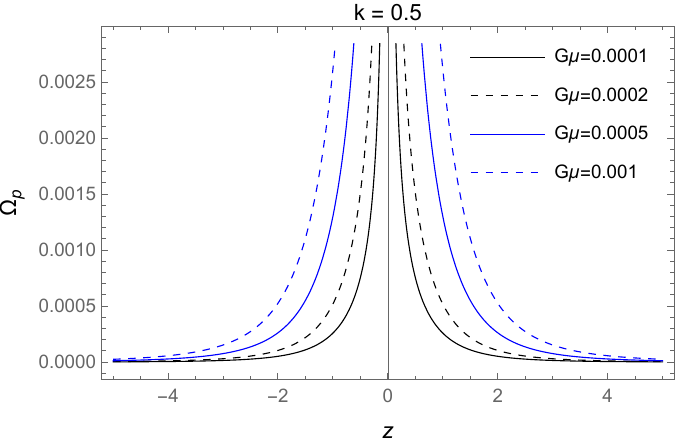}
        \caption{}
    \label{fig:gen_precess_gmueff_z_k0.5}
    \end{subfigure}
    \hfill
    \begin{subfigure}[t]{0.32\textwidth}
        \centering
        \includegraphics[width=\textwidth]{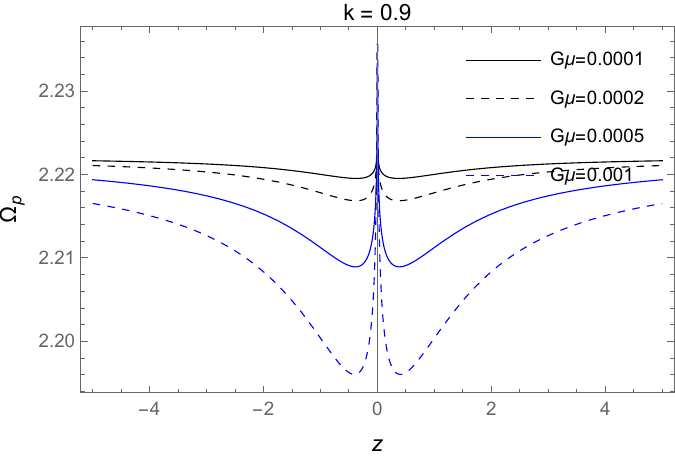}
        \caption{}
    \label{fig:gen_precess_gmueff_z_k0.9}
    \end{subfigure}
\caption{Magnitude of $\Omega_p$ as a function of $z$, shown for several values of $G\mu$ $(0.001, 0.002, 0.005, 0.01)$ with (a) $k=0.1$, (b) $k=0.5$, and (c) $k=0.9$.}
\label{fig:gen_precess_gmueff_z}
\end{figure}

\begin{figure}[H] 
\centering
    \begin{subfigure}[t]{0.38\textwidth}
        \centering
        \includegraphics[width=\textwidth]{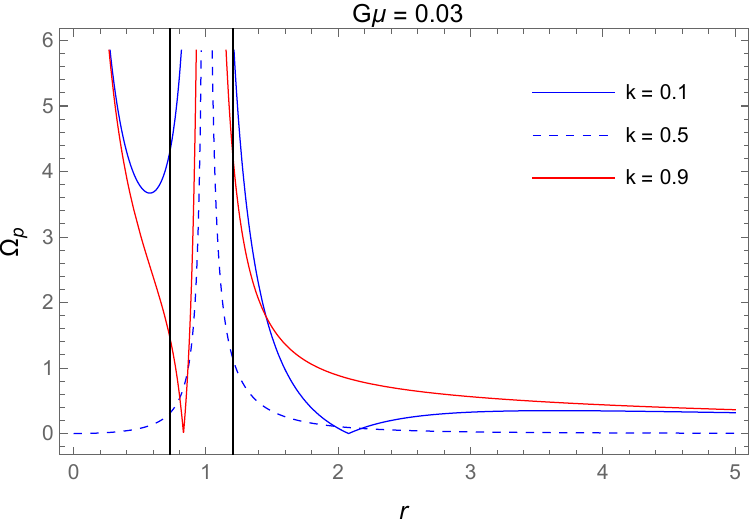}
        \caption{}
        \label{fig:gen_precess_gmueff_r_ergoregion_gmu0.03}
    \end{subfigure}
    \begin{subfigure}[t]{0.38\textwidth}
        \centering
        \includegraphics[width=\textwidth]{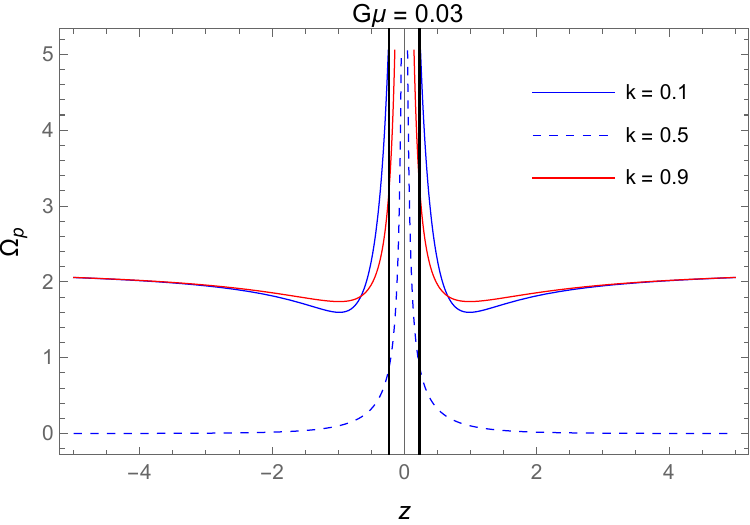}
        \caption{}
        \label{fig:gen_precess_gmueff_z_ergoregion_gmu0.03}
    \end{subfigure}
\caption{Magnitude of $\Omega_p$ with $G\mu = 0.03$, plotted as a function of (a) $r$ and (b) $z$ for three values of the parameter $k$ $(0.1, 0.5, 0.9)$. The solid black curve denotes the ergoregion boundary.}
\label{fig:gen_precess_ergoregion_gmu0.03}
\end{figure}

A special case occurs when $k = 1/2$, for which  $\omega(r,z) = A(r,z)$,
corresponding to $-g_{t\phi}/g_{tt}$, the angular frequency of the rotating frame of the vorton. Observers moving with this angular velocity are known as {\it Zero Angular Momentum Observers} (ZAMOs). For them, prograde and retrograde motions around the vorton become indistinguishable. The constraint on $\omega$ ensures that the the precession magnitude in Eq.~\eqref{modulusprecession} remains valid both inside and outside the ergoregion. We can then evaluate the general precession of a gyroscope moving in the vorton spacetime. Substituting the metric functions and the  
$\omega$ constraint from Eq.~\eqref{constraintomegafin} into Eqs.~\eqref{componentsprecessvector} yields
\begin{eqnarray} \label{genprecessionwithmetricwithomcinstraint}
        C &=& \frac{8e^{2\nu} \left(1-k\right) kr }{e^{4\nu} - r^2A^2}, \\
        \Omega_r &=& \frac{2e^{\nu}r}{e^{4\nu} - r^2A^2} \Big[ \left(1-2k+k^2 \right)r A_{,z} + 2e^{2\nu} \nu_{,z}(2k-1) \Big], \\ 
        \Omega_z &=& -\frac{2e^{\nu}}{e^{4\nu} - r^2A^2} \Big[\left(1-2k + k^2 \right) r^2 A_{,r} + e^{2\nu} \left(2k -1 \right)\left(2r\nu_{,r} -1 \right)\Big].
\end{eqnarray}
The magnitude of the general precession vector then becomes
\begin{eqnarray} \label{precessmagngeneralk}
    \Omega_p &=& \frac{e^{-\nu}}{4|k-1|k}\Big[\left\{ rA_{,z}(1-2k+2k^2) + 2e^{2\nu}\nu_{,z}(2k-1)  \right\}^2 \\ 
    & &+\frac{1}{r^2}\left\{r^2A_{,r}(1-2k+2k^2) + e^{2\nu} (2k-1)\left(2r\nu_{,r} -1 \right)    \right\}^2  \Big]^{\frac{1}{2}}.
\end{eqnarray}
For ZAMOs ($k=0.5$), it reduces to the particularly simple form
\begin{equation}
    \Omega_p = \frac{e^{-\nu}}{4}r^2\left( \left(A_{,z}\right)^2 + \left(A_{,r}\right)^2 \right)^{\frac{1}{2}}.
\end{equation}
A notable feature is that for ZAMOs the precession magnitude $\Omega_p$ closely resembles the Lense–Thirring precession profile with the distinction that it smoothly approaches zero both inside and outside the vorton. This behaviour is illustrated in Figs.~\ref{fig:gen_precess_gmueff_r} and \ref{fig:gen_precess_gmueff_z}. In contrast, for general values of $k$, the $\Omega_p$ can penetrate the ergoregion, as seen in Figs.~\ref{fig:gen_precess_ergoregion_gmu0.03}. This distinguishes the general precession behaviour from the Lense–Thirring case, where $\Omega^{LT}_p$ remains ill-defined within the ergoregion. 

\begin{figure}[H] 
\centering
    \begin{subfigure}[t]{0.32\textwidth}
        \centering
        \includegraphics[width=\textwidth]{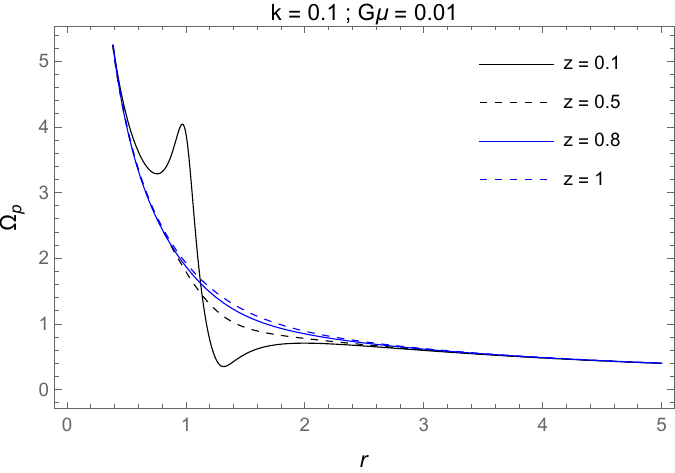}
        \caption{}
        \label{fig:gen_precess_zeff_gmu0.01_k0.1}
    \end{subfigure}
    \hfill
    \begin{subfigure}[t]{0.32\textwidth}
        \centering
        \includegraphics[width=\textwidth]{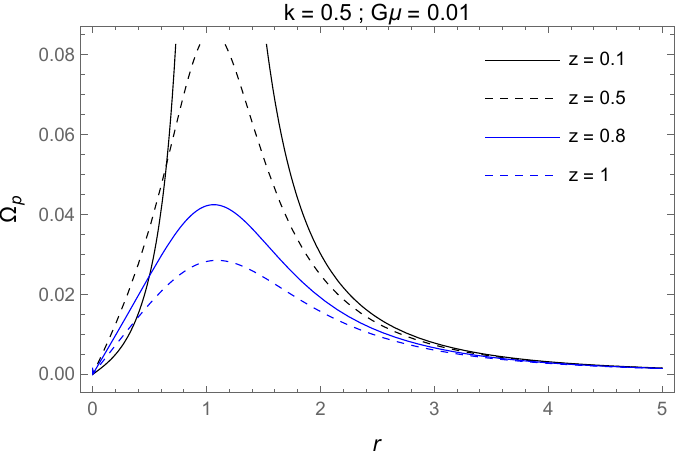}
        \caption{}
        \label{fig:gen_precess_zeff_gmu0.01_k0.5}
    \end{subfigure}
    \hfill
    \begin{subfigure}[t]{0.32\textwidth}
        \centering
        \includegraphics[width=\textwidth]{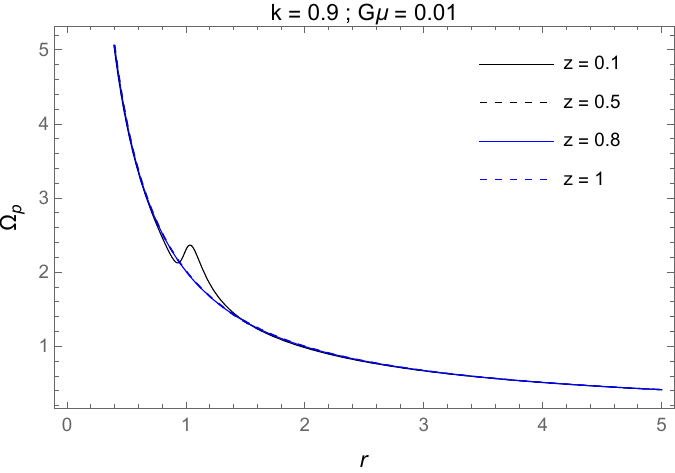}
        \caption{}
        \label{fig:gen_precess_zeff_gmu0.01_k0.9}
    \end{subfigure}
    \caption{Magnitude of $\Omega_p$ as a function of $r$ for several values of $z$ $(0.1,0.5,0.8,1)$, shown for (a) $k=0.1$, (b) $k=0.5$, and (c) $k=0.9$.}
\label{fig:gen_precess_zeff_gmu0.01}
\end{figure}

The behaviour of $\Omega_p$ on the equatorial plane ($z = 0$) exhibits a divergence at the vorton core ($r = R$). As in the Lense–Thirring case, this singularity arises from the thin-string, weak-field approximation and is expected to be regularized in a full-field treatment. However we see a distinct behavior here. In Figs.~\ref{fig:gen_precess_gmueff_r_k0.9} and \ref{fig:gen_precess_gmueff_z_k0.9} ($k=0.9$),  $\Omega_p$ decreases with increasing radius, then rises sharply near the core, and subsequently decreases again after crossing the vorton ring. Although the singularity prevents an actual transition from $r<R \to r>R$ on the equatorial plane, a similar pattern appears for trajectories slightly off the plane, as illustrated in Fig.~\ref{fig:gen_precess_zeff_gmu0.01}. For $k=0.1$, the behaviour is qualitatively different. Instead of the “dip–rise–dip” structure observed for $k=0.9$, the precession magnitude decreases and then exhibits a sharp increase resembling a localized pulse, as shown in Fig.~\ref{fig:gen_precess_zeff_gmu0.01_k0.1}. We also note that for $k = 0.1$, the radial profile of $\Omega_p$ closely resembles the general precession frequency in Kerr spacetime for $k = 0.9$, as discussed in Ref.~\cite{Chakraborty_2017_distinguishingKerrandNaked}.

\begin{figure}[H] 
\centering
    \begin{subfigure}[t]{0.32\textwidth}
        \centering
        \includegraphics[width=\textwidth]{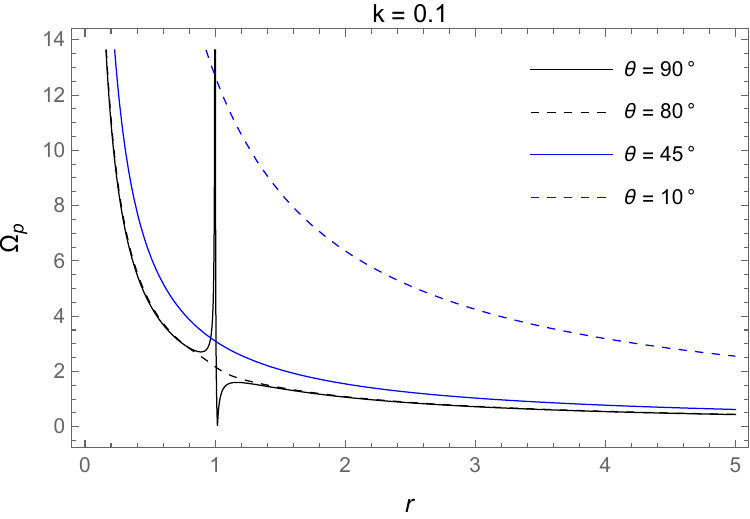}
        \caption{}
        \label{fig:gen_precess_sphericalk0.1}
    \end{subfigure}
    \hfill
    \begin{subfigure}[t]{0.32\textwidth}
        \centering
        \includegraphics[width=\textwidth]{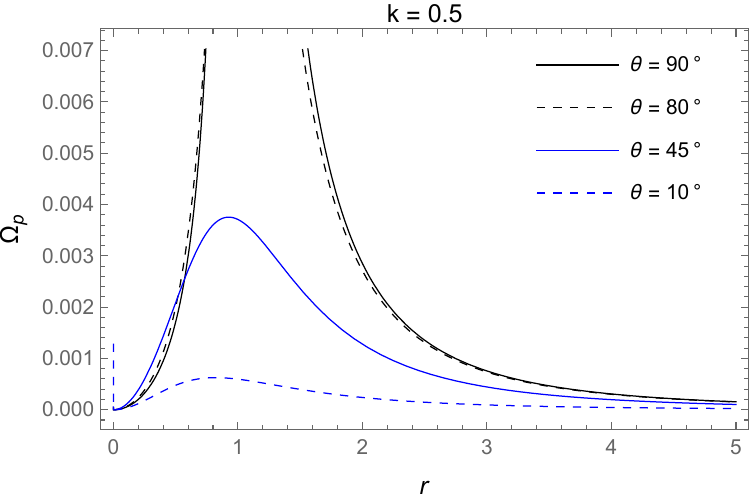}
        \caption{}
        \label{fig:gen_precess_sphericalk0.5}
    \end{subfigure}
    \hfill
    \begin{subfigure}[t]{0.32\textwidth}
        \centering
        \includegraphics[width=\textwidth]{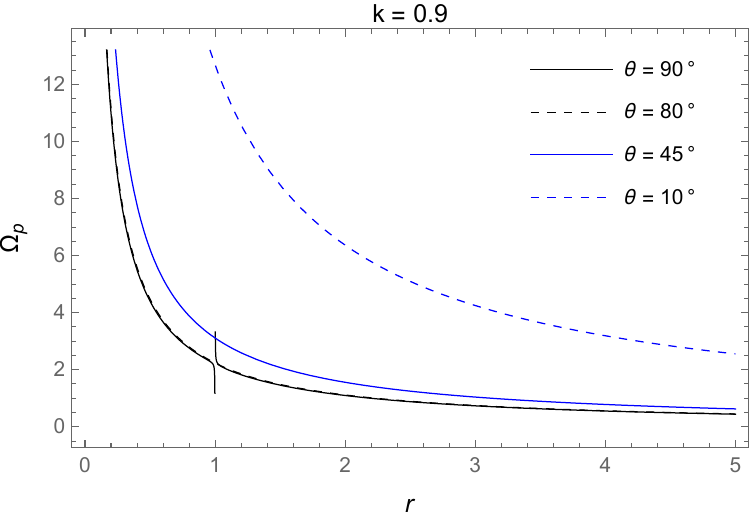}
        \caption{}
        \label{fig:gen_precess_sphericalk0.9}
    \end{subfigure}
\caption{Magnitude of $\Omega_p$ as a function of $r$ in spherical coordinates, shown for several polar angles $\theta$ ($10^\circ, 45^\circ, 80^\circ, 90^\circ$).}
\label{fig:gen_precess_spherical}
\end{figure}

As shown in Figs.~\ref{fig:gen_precess_gmueff_z_k0.1} and \ref{fig:gen_precess_gmueff_z_k0.9}, $\Omega_p$ is not asymptotically zero as $z \to \infty$. This behaviour arises because the orbital frequency $\omega$ retains a finite value at large $z$, allowing the gyroscope to continue precessing far from the vorton. To address this artefact, it is more appropriate to analyze the asymptotic behaviour using spherical coordinates: $r = \rho\sin{\theta},\ z = \rho\cos{\theta}$.
In these coordinates, Fig.~\ref{fig:gen_precess_spherical} shows that $\Omega_p$ consistently decays to zero for all values of $\theta$.

\section{Comparison with Kerr Spacetimes and Possible Signatures}

We now compare our results with those obtained for Kerr black holes and naked singularities in \cite{Chakraborty_2017_distinguishingKerrandNaked}. The qualitative behaviour of the precession frequency near the singular region differs markedly between the two systems. For trajectories slightly above the vorton plane, we find a distinctive two-minima structure in the general precession frequency. In contrast, the Kerr naked singularity analyzed in \cite{Chakraborty_2017_distinguishingKerrandNaked} (see their Fig.~5a) exhibits only a single minimum. These traits of spin-precession could be used as a diagnostic for distinguishing vortons from rotating naked singularities. In principle, such precession signatures could be detected through the timing behaviour of celestial radiation sources, most notably pulsars or X-ray binaries. However, identifying a vorton-induced precession feature requires a sufficiently long and well-characterized timing history to rule out intrinsic variability or other astrophysical sources of modulation.

Observationally, any compact object drifting through the field of a vorton would experience a shift in its spin-precession rate, leading to measurable alterations in the received signals. In binary pulsar systems, precession rates can be inferred from the temporal evolution of pulse profiles and calculated using the system parameters such as the mass of the stars and the eccentricity of their orbits \cite{PhysRevD.45.1840}. It is therefore essential to have an accurate baseline for the intrinsic precession expected from the binary dynamics. It allows one to distinguish precession arising from the system itself from additional precession induced by a nearby external source. 

A more ideal observational setting is provided by isolated pulsars, where the absence of a binary companion eliminates orbital contributions to the spin-precession signal. For example, the precession of the isolated pulsar PSR B1828–11 has been inferred from the distinctive evolution of its beam signature and timing behavior, hypothesized as arising from internal torques related to the star’s composition~\cite{Rezania_2003,Link_2001}. Recently, variations in the precession period of another isolated pulsar have been attributed to gradual changes in stellar deformation~\cite{Ashton_2017}. Furthermore, precession rates have also been measured in binary pulsars—such as the neutron-star systems PSR J1946+2052~\cite{Meng_2024} and PSR J1141–6545~\cite{Venkatraman_Krishnan_2019} based on the temporal evolution of the signals. These cases demonstrate that spin-precession signatures can indeed be extracted with sufficient precision to isolate effects arising from external perturbations once the intrinsic contributions of the system are properly accounted for.

\section{Conclusion}

In this work we have studied the timelike and null geodesics, as well as the spin-precession, around the weak-field spacetime of a circular chiral vorton. Using the weak-field metric, we derived the reduced geodesic equations, Eqs.~\eqref{geodesicfinalr}–\eqref{geodesicfinalz}, and used the corresponding effective potential to guide the selection of initial conditions and to characterize the qualitative behavior of particle motion. The effective potential exhibits a well around the vorton core and an infinite centrifugal barrier at $r = 0$, and it allows for the existence of circular orbits, suggesting that accretion-like structures may form around the loop. Frame-dragging effects were also examined, showing that prograde and retrograde trajectories experience different precession behaviour, consistent with the vorton’s circulating current.

Our analysis of bound trajectories revealed several classes of orbits, including: precessing bound orbits , circular orbits, and toroidal trajectories. These orbits typically remain confined near the vorton core, although their long-term stability requires a full-field treatment beyond the present weak-field approximation. 

A distinctive class of bound trajectories identified in this work is what we refer to as crown-like oscillatory orbits (Figs.~\ref{fig:crownorbit} and \ref{fig:crownorbitLeff}). These arise when a test particle oscillates between regions inside and outside the vorton loop without intersecting its own path during a single radial–vertical cycle. Unlike toroidal orbits, which fully encapsulate the vorton, crown orbits exhibit a characteristic multi-lobed structure in the $rz$-plane: the particle moves above and below the vorton plane while remaining radially confined to a finite interval around the ring. This behavior is enabled by the coupled $r$–$z$ dynamics of the spacetime, which prevents separation of variables and allows vertical oscillations to modulate the radial turning points. The dependence of these orbits on the angular momentum $L$ is also notable: even for $L=0$, frame-dragging induces sufficient azimuthal motion to sustain repeated oscillations, while small negative values of $L$ shift the phase and drift rate of the pattern. These crown-like trajectories therefore represent a generic feature of motion in the ring-like gravitational field of a vorton.

To investigate the non-integrability of the system, we constructed Poincaré surfaces of section (PSOS) for both equatorial and off-plane motion. The chaotic behavior emerge for orbits with $z\neq0$, as seen in Figs.~\ref{fig:PSOS_Crown}. This can be seen from the formation of Birkhoff island chains and scattered phase-space points. The equatorial motions (Figs.~\ref{fig:PSOS_Planar}), however, remains largely regular. The transition between regular and chaotic regimes is sensitive to both the initial conditions and the vorton tension $G\mu$. In particular, the crown orbits' trajectories are quasi-periodic and display moderate sensitivity to initial conditions. This behavior is consistent with the emergence of KAM-type structures in the PSOS. 

We further computed the Lense–Thirring precession frequency $\Omega_p^{LT}$ for stationary gyroscopes, and the general spin precession frequency $\Omega_p$ for a rotating gyroscope with an arbitrary angular velocity $\omega$ parameterized by a parameter $k$ ($0<k<1$). The LT frequency diverges at the boundary of the ergoregion, which is effectively at the vorton core for small at small $G\mu$. The general precession frequency, however, can extend smoothly through the ergoregion but still diverges at the ring core. Both frequencies exhibit distinctive off-plane features: local magnitude increase and decrease as a test gyroscope traverses the vicinity of the vorton. This phenomena of a certain disturbance or ``pulse" of the precession frequency (Fig.~\ref{fig:gen_precess_zeff_gmu0.01_k0.5} and \ref{fig:gen_precess_zeff_gmu0.01_k0.9}) would indicate that a spinning object would undergo a disturbance in its precession frequency as it approaches the vorton. 

The signatures presented here differ from those of Kerr black holes and share some qualitative similarities with Kerr naked singularities, but they also exhibit structures unique to the ring topology of a vorton. The observation of radiation emitting celestial bodies such as pulsars and binary star systems, for example, could provide a possible avenue for identifying or constraining vortons. Developing concrete observational diagnostics, however, requires extending this analysis beyond the weak-field approximation and incorporating realistic astrophysical environments. We leave these investigations for future work.

\section{Acknowledgement}
HSR is funded by Hibah PUTI Q1 UI No.~PKS-196/UN2.RST/HKP.05.00/2025.


\end{document}